\documentclass[lettersize,journal]{IEEEtran}
\usepackage{amsfonts}
\usepackage[ruled,linesnumbered]{algorithm2e}
\usepackage{textcomp}
\usepackage{stfloats}
\usepackage{url}
\usepackage{verbatim}
\usepackage{cite}
\usepackage{amsmath, amssymb, amsthm} 
\usepackage{wasysym} 
\usepackage{microtype} 

\usepackage{enumitem} 
\usepackage{url} 

\usepackage{booktabs} 
\usepackage{multirow} 
\usepackage{array} 
\usepackage{makecell} 
\usepackage{multirow} 

\usepackage{graphicx} 
\usepackage[caption=false,labelfont=sf,textfont=sf]{subfig}

\usepackage[capitalize]{cleveref} 
\usepackage{bm} 
\usepackage{cases} 
\usepackage{float} 
\usepackage{tabularx}
\newcolumntype{Y}{>{\centering\arraybackslash}X}
\usepackage[dvipsnames]{xcolor} 
\definecolor{blue}{RGB}{38,56,150} 

\newtheorem{theorem}{Theorem}

\begin{document}
	
\title{A Novel Truncated Norm Regularization Method for Multi-channel Color Image Denoising}

\author{Yiwen Shan, Dong Hu, and Zhi Wang,~\IEEEmembership{Member,~IEEE}
    \thanks{This work was supported by the Fundamental Research Funds for the Central Universities under Grant SWU-XDJH202303. \emph{(Corresponding author: Zhi Wang.)}}%
    \thanks{Yiwen Shan and Dong Hu are with the College of Computer and Information Science, Southwest University, Chongqing 400715, China (e-mail: yvinshan@foxmail.com; donghu@email.swu.edu.cn).}
    \thanks{Zhi Wang is with the College of Computer and Information Science, Southwest University, Chongqing 400715, China, and also with the Big Data and Intelligence Engineering School, Chongqing College of International Business and Economics, Chongqing 401520, China (e-mail: chiw@swu.edu.cn).}
}

\markboth{IEEE Transactions on Circuits and Systems for Video Technology}%
{Shell \MakeLowercase{\textit{et al.}}: A Sample Article Using IEEEtran.cls for IEEE Journals}

\IEEEpubid{}

\maketitle

\begin{abstract}
    Due to the high flexibility and remarkable performance, low-rank approximation has been widely studied for color image denoising. 
    However, existing methods usually ignore the cross-channel difference or the spatial variation of noise, which limits their capacity in the task of real world color image denoising. 
    To overcome these drawbacks, this paper proposes a double-weighted truncated nuclear norm minus truncated Frobenius norm minimization (DtNFM) model, and apply it to color image denoising through exploiting the nonlocal self-similarity prior. 
    The proposed DtNFM model has two merits. 
    First, it models and utilizes both the cross-channel difference and the spatial variation of noise. 
    This provides sufficient flexibility for handling the complex distribution of noise in real world images. 
    Second, the proposed DtNFM model provides a close approximation to the underlying clean matrix since it can treat different rank components flexibly. 
    To solve the DtNFM model, an efficient algorithm is devised through exploiting the framework of alternating directions method of multipliers (ADMM). 
    Meanwhile, the truncated nuclear norm minus truncated Frobenius norm regularized least squares subproblem is discussed in detail, and the results show that its global optimum can be directly obtained in closed form.
    Therefore, the DtNFM model can be efficiently solved by a single ADMM. 
    Rigorous mathematical derivation proves that the solution sequences generated by our proposed algorithm converge to a single critical point. 
    Extensive experiments on synthetic and real noise datasets demonstrate that the proposed method outperforms many state-of-the-art color image denoising methods. MATLAB code is available at https://github.com/wangzhi-swu/DtNFM.
\end{abstract}


\section{Introduction}
\IEEEPARstart{I}{mage} denoising serves as an indispensable process for manifold tasks in computer vision, such as semantic image segmentation \cite{ISeg_survey1, Seg_TCSVT} and image retrieval \cite{Peng1, Peng2}. 
It aims to recover the clean image ${X}$ from its corrupted observation ${Y}$, and such problem can be expressed as
\begin{equation}
    {Y} = {X} + {N}, \label{eq_ID}
\end{equation}
where ${N}$ is the white Gaussian noise. 
Since problem \eqref{eq_ID} is ill-posed, prior information should be exploited in order to regularize the solution space and improve the denoising performance. 
With them, image denoising can be formulated as the following well-posed problem: 
\begin{equation}
    \arg \min_{{X}} \mathcal{F}(X) + \lambda \!\cdot\! \Phi({X}), 
    \label{eq_ID_prior}
\end{equation}
where $\mathcal{F}(\cdot)$ is a loss function, $\Phi(\cdot)$ is a regularizer corresponding to some prior information, and $\lambda$ is a parameter. 
In the past decade, a large amount of works have been carried out on image denoising. 
State-of-the-art methods are mainly based on sparse representation \cite{SparRep_tcsvt_2014, RCSR, MCA_GSC, CAS}, deep learning \cite{FSLID, FFDNet, DnCNN, DRUNet, DeamNet, DSRDP}, and low-rank approximation \cite{EBD, DLRQP, SSLRDM, HLTA-GN, MCWNNM, MCWSNM, MCNNFNM}. 
{{The sparse representation-based methods estimate the denoised images by finding a sparse linear combination of bases. 
And the bases can be learned from the observation. 
However, it is difficult for those methods to handle the spatially variant noise. 
The deep learning-based methods learn a function from a large amount of data. 
The function, in which the prior information are modeled implicitly, relates the corrupted image to the denoised results. 
However, their generalization are usually not good since the dependence on the training data. 
}}%
%
\par
Due to the high efficiency and strong denoising capability, low-rank approximation-based methods have been widely studied. 
Along this line, lots of state-of-the-art methods are proposed based on the  nonlocal self-similarity (NSS) prior \cite{NL_means, NSP}. 
NSS indicates that there spread many similar structures across a natural image. 
Those similar structures can be gathered to construct a corrupted patch matrix, denoted as $\mathbf{Y}$. 
Since the gathered patches have similar structures, matrix $\mathbf{Y}$ is (or approximately) low rank. 
Hence its denoised version, denoted as $\mathbf{X}$, can be estimated by solving the following optimization problem: 
\begin{equation}
    \arg \min_{\mathbf{X}} \frac{1}{2\sigma^2_n} \Vert \mathbf{Y}-\mathbf{X} \Vert_F^2 + \lambda \!\cdot\! \mathbf{Rank}(\mathbf{X}). \label{eq_rank_min}
\end{equation}
However, problem \eqref{eq_rank_min} is NP-hard \cite{rank_NP_hard} since $\mathbf{Rank}(\!\cdot\!)$ is nonconvex and discontinuous. 
Inspired by compressed sensing \cite{CS1}, $\mathbf{Rank}(\!\cdot\!)$ is relaxed to the nuclear norm, leading to the famous nuclear norm minimization (NNM) problem: 
\begin{equation}
    \arg \min_{\mathbf{X}} \frac{1}{2\sigma^2_n} \Vert \mathbf{Y}-\mathbf{X} \Vert_F^2 + \lambda \Vert \mathbf{X}\Vert_{*}, \label{eq_NNM}
\end{equation}
where $\Vert \mathbf{X}\Vert_{*} = \sum_{i}\sigma_i(\mathbf{X})$, and $\sigma_i(\mathbf{X})$ is the $i$th leading singular value. 
In theory, Fazel et al. \cite{Convex_Envelop} proved that nuclear norm is the best approximation of the $\mathbf{Rank}(\!\cdot\!)$ in all convex functions. 
Moreover, NNM-based models can be solved efficiently \cite{SVT, FPCA, APGL}. 
Although having theoretical supports and efficient solvers, NNM still has drawbacks. 
Lots of experiments on image denoising \cite{WNNM, WNNM-2014, LRQA} demonstrate that the results produced by NNM-based models deviate from the optimal solutions severely. 
This is attributed to the nuclear norm treating all singular values equally. 
Consequently, the dominant singular values are over-shrunk during the optimization process. 
In a nutshell, nuclear norm cannot approximate the rank function with sufficient accuracy, especially in handling inverse problems in image processing. 
Therefore, a variety of nonconvex surrogates of the rank function have been studied \cite{PSSV, NNFN, wang1, wang2, wang3, wang4}. 
In theory, nonconvex surrogates can approximate the rank function better and make the denoising models produce better performance. 
The famous weighted nuclear norm minimization (WNNM) model \cite{WNNM} assigns different weights to different singular values. 
In this way, the larger singular values would be shrunk less, and hence the major information would be protected better. 
This practice makes WNNM outperforms NNM on grayscale image denoising. 
More importantly, solving WNNM is highly efficient since the global optimum can be obtained in closed-form. 
However, the performance of WNNM degrades sharply when the noise becomes stronger. 
That is attributed to the leading singular values still being over-shrunk during the optimization. 
To further improve the denoising capacity of WNNM, Xie et al. \cite{WSNM} proposed the weighted Schatten $p$-norm minimization (WSNM) model. 
WSNM provides more flexible treatments for different rank components. 
Hence the estimated matrix can be closer to the optimal solution of the rank minimization problem in \eqref{eq_rank_min}. 
However, solving WSNM is computationally expensive, since it no longer allow closed-form solutions and have to be solved by numerical algorithms \cite{GST}. 
\par 
When denoising color images, the following two challenges have to be considered. 
First, the data in different color channels has correlations. 
Second, the noise has not only cross-channel difference but also spatial variation in a single channel. 
Therefore, the naive strategy that applies grayscale denoisers to each color channel independently cannot obtain satisfactory results since it neglects all those points aforementioned. 
Three modified strategies can be considered to make low-rank approximation methods feasible for color image denoising. 
The first is to transform the color image from RGB space into a new color space where channels demonstrate less correlation, and then denoise each channel independently \cite{CBM3D}. 
However, the transformation may complicate the distribution of noise \cite{MCWNNM}. 
In addition, the correlation among color channels cannot be fully exploited. 
The second strategy is to encode color images using more complex data structures, for example, the quaternion \cite{QWNNM-dn, LRQA, DLRQP}, in order to preserve the inherent correlation among color channels. 
However, existing quaternion-based color image denoising methods mostly ignore the cross-channel difference and spatial variation of noise. 
Moreover, the algorithms converge slowly when solving quaternion-based models. 
The third strategy is to introduce mutuality between RGB channels and denoise them jointly. 
Along this line, WNNM and WSNM are respectively extended to color image denoising in \cite{MCWNNM} and \cite{MCWSNM}. 
Their key idea is introducing a weight matrix to characterize the cross-channel difference of noise. 
However, the extended models inherit the drawbacks from their grayscale version. 
In addition, both of them ignore the spatial variation of noise. 
\par 
\begin{table}
    \centering
    \label{tab_drawbacks_aforementioned}
    \caption{Comparisons among low-rank approximation models on (A): flexibility to dealing with the cross-channel difference of noise; (B): flexibility to dealing with spatially variant noise; (C): efficient solver (can be solved via a single iterative algorithm); (D): theoretical convergence guarantee.}
    \begin{tabularx}{\linewidth}{p{1.8cm}<{\centering}| p{4.45cm}<{\centering}| YYYY}
        \hline
        \!\!\!Model & Expression & A & B & C & D \\
        \hline
        \!\!\!LRQA \cite{LRQA} & $ \min_{\dot{\mathbf{X}}} \frac{1}{2}\Vert \dot{\mathbf{Y}} - \dot{\mathbf{X}} \Vert_F^2 + \lambda \Psi(\dot{\mathbf{X}}) $ & $\times$ & $\times$ & \checked & $\times$ \\ %
        \!\!\!MCWNNM \cite{MCWNNM} & $\min_{\mathbf{X}} \Vert \mathbf{W}(\mathbf{Y}-\mathbf{X}) \Vert_F^2 + \Vert \mathbf{X} \Vert_{w,*}$ & \checked & $\times$ & \checked & \checked \\
        \!\!\!MCWSNM \cite{MCWSNM} & $\min_{\mathbf{X}} \Vert \mathbf{W}(\mathbf{Y}-\mathbf{X}) \Vert_F^2 + \Vert \mathbf{X} \Vert_{w,Sp}$ & \checked & $\times$ & $\times$ & \checked \\
        \!\!\!NNFNM \cite{MCNNFNM} & $\min_{\mathbf{X}} \Vert \mathbf{W}(\mathbf{Y}-\mathbf{X}) \Vert_F^2 + \lambda \Vert \mathbf{X} \Vert_{*-F}$ & \checked & $\times$ & \checked & \checked \\
        \!\!\!DtNFM (Ours) & $\min_{\bm{\mathbf{X}}} \Vert \mathbf{C}(\bm{\mathbf{Y}} - \bm{\mathbf{X}}) \mathbf{S} \Vert_F^2 + \lambda \Vert \mathbf{X} \Vert_{t,*-F}$ & \checked & \checked & \checked & \checked \\
        \hline
    \end{tabularx}
\end{table}
In summary, as shown in Table \ref{tab_drawbacks_aforementioned}, none of the aforementioned low-rank approximation models obtain (A-B) flexibility to dealing with noise, (C) efficient solver, and (D) theoretical convergence guarantee simultaneously. 
To fill in this blank, this paper proposes a new low-rank approximation model, called double-weighted truncated nuclear norm minus truncated Frobenius norm minimization (DtNFM), and apply it to color image denoising through exploiting the framework of NSS prior. 
The proposed DtNFM model has two merits. 
First, it is flexible in dealing with the noise. 
Concretely, two weight matrices are designed to respectively model the cross-channel difference and spatial variation of noise. 
And a heuristic scheme is proposed for adaptively determining the trade-off between two weight matrices. 
Second, the proposed DtNFM model is flexible in treating different rank components, which guarantees the underlying clean matrix can be estimated with sufficient accuracy. 
The novelty and contributions of this work are summarized as follows: 
\begin{itemize}
    \item The DtNFM model is proposed and applied to color image denoising. 
    The proposed model possesses sufficient flexibility to deal with the cross-channel difference and spatial variation of noise. 
    Meanwhile, the original clean matrix can be estimated with more accuracy, since its low-rankness obtains a closer approximation. 
    \item To solve the DtNFM model, an efficient algorithm is developed through exploiting the framework of ADMM. 
    Meanwhile, the truncated nuclear norm minus truncated Frobenius norm regularized least squares subproblem is discussed in detail. It is shown that the global optimum can be directly obtained in closed-form rather than transforming the corresponding nonconvex optimization problem into two convex problems \cite{TL12}. Leveraging on it, the DtNFM model can be efficiently solved by a single ADMM, without nesting other iterative algorithms. 
    \item Theoretical convergence guarantee is obtained, even though the optimization problem resulting from the DtNFM model is nonconvex. 
    Concretely, we prove that the solution sequences generated by our algorithm converge to a single critical point, which indicates our algorithm is capable to solve the DtNFM model. 
    \item To validate the denoising capacity of DtNFM, extensive experiments are carried out on 1) spatially invariant, 2) spatially variant, and 3) real-world noise removal. 
    The results demonstrate that the proposed DtNFM method outperforms state-of-the-art color image denoising methods. 
\end{itemize}
\par 
The rest of this paper is organized as follows. 
Section II presents the notations and some related works. 
In Section III, the DtNFM model is proposed. Then, an efficient algorithm is devised and its convergence is analyzed. 
In Section IV, the experimental results are reported and analyzed. 
Finally, some conclusions are drawn in Section V. 
\section{Notations and Related Works}
In this section, we summarize the notations used in this paper. 
Then, we introduce some famous low-rank approximation-based color image denoising methods and the ADMM framework. 
\subsection{Notations}
In this paper, vectors and matrices are respectively denoted by lowercase boldface and uppercase boldface. 
$\mathbf{I}$ denotes the identity matrix. 
For a vector $\mathbf{x}=[x_i]\in \mathbb{R}^{m}$, $\mathrm{Diag}(\mathbf{x})$ creates a $m\times m$ diagonal matrix with $x_i$ as the $i$th element of the main diagonal. 
And $\Vert \mathbf{x} \Vert_2 = \sqrt{ \sum_{i=1}^{n}x_i^2 }$ is the vector $\ell_2$ norm. 
Given $\mathbf{X}, \mathbf{Y} \in \mathbb{R}^{m\times n}$, $\langle \mathbf{X}, \mathbf{Y} \rangle = \sum_{j=1}^{n}\sum_{i=1}^{m}\mathbf{X}_{ij}\times\mathbf{Y}_{ij}$ is their inner product, 
$\mathbf{X} \!\oslash\! \mathbf{Y} \in \mathbb{R}^{m\times n}$ is the Hadamard division, 
and $[\mathbf{X}; \mathbf{Y}] \in \mathbb{R}^{2m\times n}$ vertically appends matrix $\mathbf{Y}$ to $\mathbf{X}$. 
$\Vert \mathbf{X} \Vert_* = \sum_i \sigma_i(\mathbf{X})$ is its nuclear norm, where $\sigma_i(\mathbf{X})$ is the $i$th leading singular value. 
$\Vert \mathbf{X} \Vert_F = \sqrt{\sum_i \sigma_i(\mathbf{X})^2}$ is the Frobenius norm. 
\subsection{Color Image Denoising Methods Based on Low-rank Approximation}
The multi-channel weighted nuclear norm minimization (MCWNNM) \cite{MCWNNM} can be formulated as 
\begin{equation}
    \arg \min_{\mathbf{X} \in\mathbb{R}^{m\times n}} \Vert \mathbf{W}(\mathbf{Y}-\mathbf{X}) \Vert_F^2 + \Vert \mathbf{X} \Vert_{w,*},
    \label{eq_MCWNNM}
\end{equation}
where $\mathbf{X}, \mathbf{Y}$ are respectively the estimated patch matrix and the corrupted observation. 
The weight matrix $\mathbf{W} = \mathrm{Diag} ([\sigma_r^{-1}\mathbf{1}; \sigma_g^{-1}\mathbf{1}; \sigma_b^{-1}\mathbf{1}]) \in \mathbb{R}^{m\times m}$ is used to model the cross-channel difference of noise, with $\sigma_{c}$ ($c\in \lbrace r,g,b \rbrace$) as the noise standard deviation in channel $c$. 
Since $\mathbf{W}_{ii} \propto \sigma_c^{-1}$, the stronger noise will result in the data in this channel providing less contribution to the denoised result. 
The MCWNNM model pioneers the use of weight matrix, and verifies the effectiveness of the joint denoising strategy. 
The $\Vert \mathbf{X} \Vert_{w,*}=\sum_i w_i \sigma_i(\mathbf{X})$ in \eqref{eq_MCWNNM} is the weighted nuclear norm, where $\lbrace w_i\rbrace$ are determined by some prior knowledge. 
Benefited from the weights, the larger singular values are shrunk less during the optimization of MCWNNM. 
Hence the major information can be preserved better. 
More importantly, it was proved in \cite{WNNM-2014} that the proximal operator of weighted nuclear norm, formulated as
\begin{equation}
    \arg \min_{\mathbf{X} \in\mathbb{R}^{m\times n}} \frac{1}{2} \Vert \mathbf{X} - \mathbf{Y} \Vert_F^2 + \Vert \mathbf{X} \Vert_{w,*},
    \label{eq_prox_wnnm}
\end{equation}
has a cheap closed-form solution. 
Therefore, the problem in \eqref{eq_MCWNNM} can be efficiently solved by a single ADMM algorithm, without nesting other iterative algorithms. 
Nevertheless, MCWNNM inherits the drawback of WNNM that the leading singular values might not be protected well enough, especially in the case of reducing severe noise. 
%
\par
{{%
To alleviate this problem, the MCWSNM model \cite{MCWSNM} is proposed, which can be formulated as
\begin{equation}
    \arg \min_{\mathbf{X} \in\mathbb{R}^{m\times n}} \Vert \mathbf{W}(\mathbf{Y}-\mathbf{X}) \Vert_F^2 + \Vert \mathbf{X} \Vert_{w,Sp}^{p},
    \label{eq_MCWSNM}
\end{equation}
where $\Vert \mathbf{X} \Vert_{Sp}^{p} = \sum_i w_i \sigma(\mathbf{X})^p$ is the weighted Schatten $p$-norm. 
The parameter $p\in (0,1]$ provides more flexibility to control the shrinkage on singular values. 
Hence the over-shrinking problem of MCWNNM can be alleviate. 
However, the proximal operator of the weighted Schatten $p$-norm has to be solved by iterative algorithms, such as the GST \cite{GST}. 
Consequently, to solve the problem in \eqref{eq_MCWSNM}, one has to nest the GST algorithm into the ADMM, which is inefficient. 
The nuclear norm minus Frobinius norm (NNFN) minimization \cite{MCNNFNM} combines the advantages of MCWNNM and MCWSNM, since the NNFN not only allows closed-form proximal operator, but also have sufficient flexibility on shrinking the singular values. 
However, the aforementioned methods are not adequate to handle the spatially variant noise, since the weight matrix they use can only model the cross-channel difference of noise. 
%
\par
Recently, lots of quaternion-based methods are proposed for color image denoising \cite{LRQA, DLRQP}. 
Performing the low-rank approximation in quaternion domain, those methods are able to preserve more correlation among RGB channels. 
However, modeling the mathematical characters of the noise, such as the cross-channel difference and spatial variation, becomes difficult in the quaternion domain. 
}}
%
%
\subsection{Alternating Direction Method of Multipliers (ADMM)}
\label{sec_admm}
The ADMM is an efficient algorithm to solve a wide variety of problems arising in statistics and machine learning. 
Consider the general regularized optimization problem: 
\begin{equation}
	\arg\min_{\mathbf{x}} \mathcal{F}(\mathbf{x}) + \lambda \Phi(\mathbf{x}), \label{ADMM_0}
\end{equation}
where $\mathbf{x} \in \mathbb{R}^{n}$, 
$\mathcal{F}: \mathbb{R}^{n} \rightarrow \mathbb{R} \cup \lbrace+\infty \rbrace$ is convex, 
$\Phi: \mathbb{R}^{n} \rightarrow \mathbb{R} \cup \lbrace+\infty \rbrace$ is a generic regularizer, 
$\lambda>0$ is a parameter. 
Problem \eqref{ADMM_0} can be rewritten as
\begin{align}
	\min_{\mathbf{x}, \mathbf{z}} \; \mathcal{F}(\mathbf{x}) + \lambda \Phi(\mathbf{z}) \quad
	\mathrm{s.t.} \ \ \mathbf{x} - \mathbf{z} = \mathbf{0},
	\label{eq_ADMM_1}
\end{align}
where $\mathbf{z} \in \mathbb{R}^{n}$ is the auxiliary variable, $\mathbf{0} \in \mathbb{R}^{n}$ is a vector of zeros. 
The augmented Lagrangian for problem (\ref{eq_ADMM_1}) is
\begin{equation}
	\mathcal{L}_{\rho} (\mathbf{x}, \mathbf{z}, \mathbf{y}) = \mathcal{F}(\mathbf{x}) + \lambda \Phi(\mathbf{z}) + \mathbf{y}^\top (\mathbf{x} - \mathbf{z}) + \frac{\rho}{2} \Vert \mathbf{x} - \mathbf{z} \Vert_2^2, 
\end{equation}
where $\mathbf{y} \in \mathbb{R}^{n}$ is the Lagrangian multiplier, and $\rho>0$ is the penalty parameter. 
The ADMM minimizes the augmented Lagrangian with respect to $\mathbf{x}$ and $\mathbf{z}$ in an alternative manner: 
\begin{numcases}{}
	\mathbf{x}_{k+1} = \arg \min_{\mathbf{x}}\limits \ \mathcal{L}_{\rho} (\mathbf{x}, \mathbf{z}_k, \mathbf{y}_k), \label{eq_ADMM_x} \\
	\mathbf{z}_{k+1} = \arg \min_{\mathbf{z}}\limits \ \mathcal{L}_{\rho} (\mathbf{x}_{k+1}, \mathbf{z}, \mathbf{y}_k),
\end{numcases}
where $k \in \mathbb{N}$ denotes the iteration. 
Then, the Lagrangian multiplier is updated in a feedback manner: 
\begin{equation}
	\mathbf{y}_{k+1} = \mathbf{y}_k + \rho (\mathbf{x}_{k+1} - \mathbf{z}_{k+1}). \label{eq_ADMM_y}
\end{equation}
Steps \eqref{eq_ADMM_x} $\sim$ \eqref{eq_ADMM_y} are performed until convergence. 
Albeit the great success of ADMM on convex problems, it is still a great challenge to extend the ADMM to solving nonconvex problems. 
In particular, the convergence of the ADMM remains as an open issue when the objective becomes nonconvex. 
Nevertheless, lots of works have shown that ADMM works very well for various nonconvex problems, obtaining satisfactory performance with high efficiency \cite{ADMM_subspace_clustering, LiuT2023, LatLRR, Xie2, Xie3}. 
\begin{figure*}[t]
\centering
\includegraphics[width=.7\linewidth]{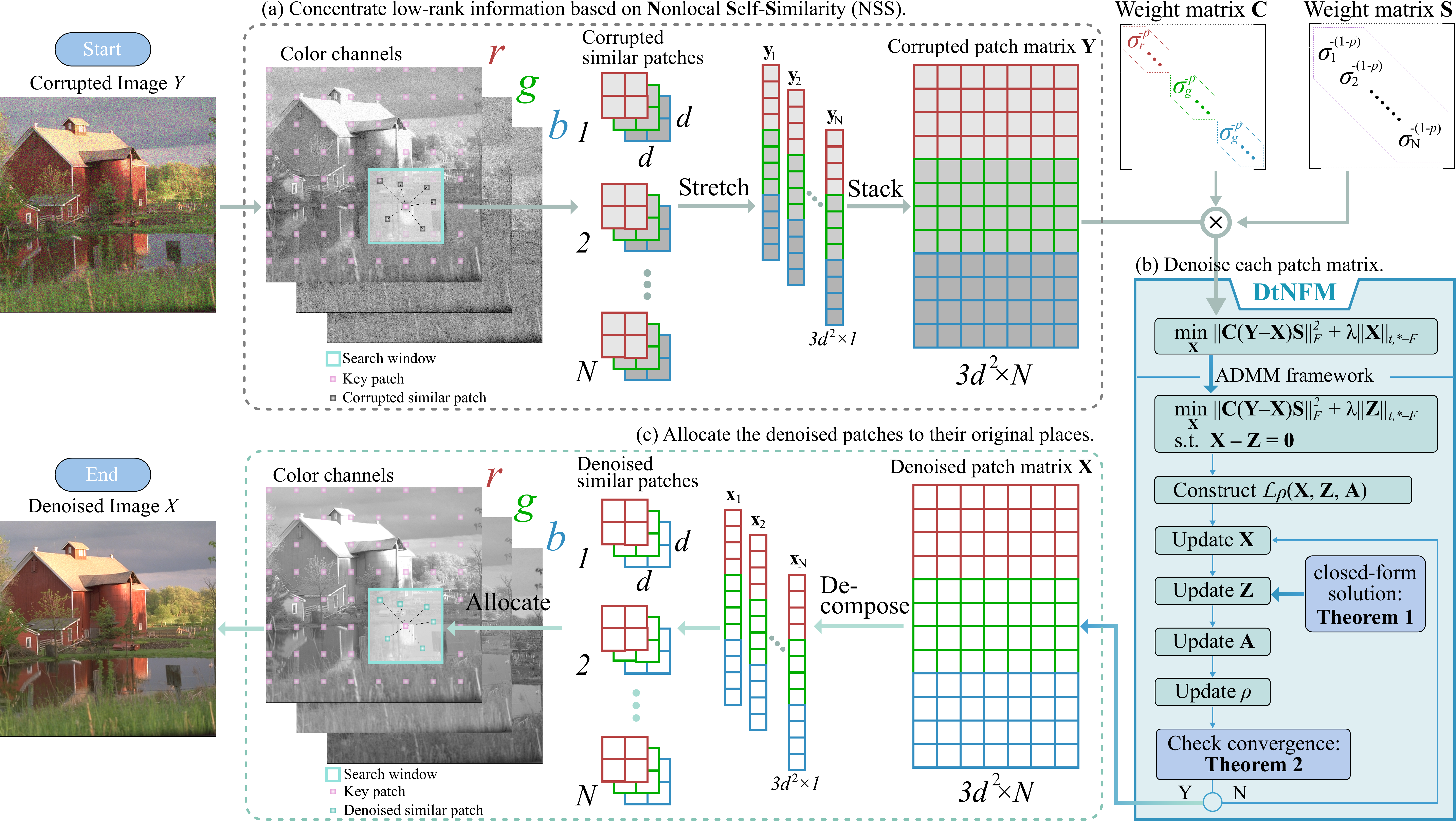}
\caption{{{An overview of our DtNFM method. 
(a) Group similar patches. For each key patch, the similar patches are found via the $k$NN algorithm. Those similar patch compose a patch matrix $\mathbf{Y}$. Using the data in $\mathbf{Y}$, the weight matrices $\mathbf{C}$ and $\mathbf{S}$ are constructed. (b) Denoise each patch matrix. The matrices $\mathbf{Y}, \mathbf{C}$ and $\mathbf{S}$ are used to formulate the DtNFM model in \eqref{eq_DtNFM}. The generated optimization problem is solved by the proposed Algorithm \ref{alg_admm}. At each iteration, the variables $\mathbf{X}$, $\mathbf{Z}$, $\mathbf{A}$, and $\rho$ are updated in an alternating manner. (c) Generate the denoised image. The patch matrix $\mathbf{X}$ outputed from DtNFM model is decomposed to patches. Those denoised patches are settled to their original places. After processing all of the $P$ patch matrices, the denoised image is obtained. 
}}}
\label{fig_flowchart}
\end{figure*}
\section{The Proposed Model}
In this section, we first decompose the color image denoising problem from the whole image to each patch matrix through exploiting the NSS prior. 
Then, the DtNFM model is proposed to characterize the denoising problem of each patch matrix. 
After that, a closed-form proximal operator is proposed for the truncated nuclear norm minus truncated Frobenius norm (tNF) regularizer. 
Leveraging on it, an efficient algorithm is developed to solve the DtNFM model. 
Finally, the convergence guarantee of our algorithm is provided. 
\subsection{Problem Formulation}
The proposed DtNFM method is composed of two parts: 1) patch grouping via exploiting the NSS prior and 2) low-rank approximation by solving the problem resulted by the DtNFM model. 
The detailed processes are shown in Fig. \ref{fig_flowchart}. 
Given a corrupted color image $Y \in \mathbb{R}^{H\times W\times 3}$, $P$ key patches are assigned across it. 
The key patches are spaced apart with an equal distance $s$, and each sizes $d\times d\times 3$ pixels. 
And hence the number of key patches $P = \lceil (H-d)/s \rceil\times\lceil (W-d)/s \rceil$. 
For each key patch, its $N$ nearest neighbors, i.e., the most similar patches, are extracted from a search window around it. 
Those extracted similar patches are then stretched to $N$ column vectors $\mathbf{y}_i\in \mathbb{R}^{3d^2}, i \in \lbrace 1,\cdots, N \rbrace$. 
The $N$ vectors are stacked horizontally to form a corrupted patch matrix $\mathbf{Y} \in \mathbb{R}^{3d^2 \times N}$. 
Note that the relationship between key patches and patch matrices is one-to-one. 
Therefore, $P$ corrupted patch matrices should be constructed in total. 
Then, the proposed DtNFM model will operate on each of them, estimating its clean version $\mathbf{X} \in \mathbb{R}^{3d^2\times N}$. 
After that, $\mathbf{X}$ will be decomposed to $N$ vectors $\mathbf{x}_i \in \mathbb{R}^{3d^2}, i \in \lbrace 1,\cdots,N \rbrace$. 
For each vector, it will be reshaped back to a denoised patch and then settled to its original place. 
After processing all of the $P$ denoised patch matrices, the denoised image can be obtained. %
To obtain better results, the procedure above should be repeated $\theta$ iterations. 
The complete procedure of the proposed color image denoising method is summarized in Algorithm \ref{alg_framework}. 
\begin{algorithm}[t]
    \caption{Color image denoising via the DtNFM method.}
    \label{alg_framework}
    \KwIn{Corrupted image $Y \in \mathbb{R}^{H\times W\times 3}$, noise standard deviations $[\sigma_{r\_0}; \sigma_{g\_0}; \sigma_{b\_0}]$\;}
    \KwOut{Denoised image $X \in \mathbb{R}^{H\times W\times 3}$\;}
    Initialize $X_0 = Y_0 = Y, \theta, s, N$\;
    \For{$l = 1:\theta$}{
        Iterative regularization $Y_{l} = X_{l-1} + \delta (Y_0 - X_{l-1})$\;
        Assign $P$ key patches in $X_{i-1}$\;
        \For{Each key patch}{
            Extract $N$ most similar neighbors to form the patch matrix $\mathbf{Y}$\;
            Construct the weight matrices $\mathbf{C}$ and $\mathbf{S}$\;
            $\mathbf{X} = \mathrm{DtNFM\_model}(\mathbf{Y}, \mathbf{C}, \mathbf{S})$\;
        }
        Aggregate all of $P$ denoised patch matrices to generate $X_{l}$\;
    }
\end{algorithm}
\par
A problem left over from Algorithm \ref{alg_framework} is how to estimate the clean patch matrix $\mathbf{X} \in \mathbb{R}^{3d^2\times N}$ with high accuracy and efficiency. 
To this end, we propose the DtNFM model, which can be formulated as 
\begin{equation}
    \arg \min_{\mathbf{X}} \Vert \mathbf{C}(\mathbf{Y} - \mathbf{X}) \mathbf{S} \Vert_F^2 + \lambda \Vert \mathbf{X} \Vert_{t,*-F},
    \label{eq_DtNFM}
\end{equation}
where $\mathbf{C} \in \mathbb{R}^{3d^2\times 3d^2}$ is the weight matrix to model the cross-channel difference of noise, $\mathbf{S} \in \mathbb{R}^{N\times N}$ is to model the spatial variation of noise. 
{{%
The truncated nuclear norm minus truncated Frobenius norm (tNF) is defined as
\begin{equation}
    \!\!\!\!\Vert \mathbf{X} \Vert_{t,*-F} = \!\Big(\sum_{i=t+1}^{\min(3d^2, N)} \sigma_i(\mathbf{X}) \!\Big) - \!\alpha \Big(\sum_{i=t+1}^{\min(3d^2, N)} \sigma_i(\mathbf{X})^{2} \Big)^{\frac{1}{2}},
    \label{eq_tnf}
\end{equation}
where $t\in \mathbb{N}$ and $\alpha \in [0, +\infty)$ are parameters.}} %
Compared with other low-rank approximation based models, the proposed DtNFM model has the following three advantages simultaneously. 
\par
\textit{$\bullet$ It is able to deal with the spatially variant noise,} 
since two weight matrices are used to characterize the noise. 
The matrix $\mathbf{C}$ is to characterize the noise standard deviations in three color channels, making the DtNFM model able to handle the cross-channel difference of noise. 
Meanwhile, the DtNFM model can model the noise standard deviations on each of the $N$ neighbor patches using the matrix $\mathbf{S}$. 
Hence it has sufficient flexibility in handling the noise variation among patches. 
\begin{figure}[tb]
\centering
\subfloat[]{
    \includegraphics[width=0.48\linewidth]{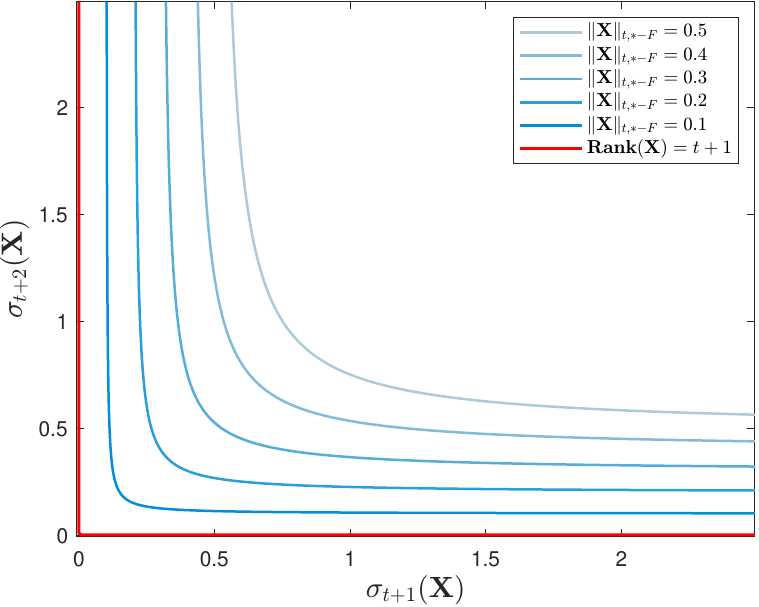}
}
\subfloat[]{
    \includegraphics[width=0.48\linewidth]{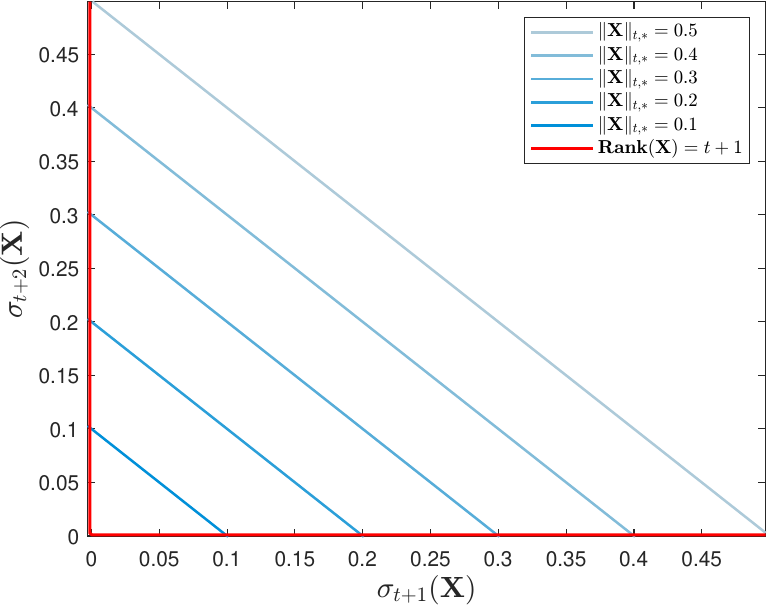}
}%
\caption{{%
    Contours of the tNF and the truncated nuclear norm. 
    Assume the matrix $\mathbf{X} \in \mathbb{R}^{n\times(t+2)}$ with $n \ge t+2$. 
    As the $\Vert \mathbf{X} \Vert_{t,*-F}$ in (a) is minimized from 0.5 to 0.1, its curve approaches to the axes on which $\mathbf{Rank}(\mathbf{X}) = t+1$ holds. 
    Thus the rank of matrix $\mathbf{X} \in \mathbb{R}^{n\times(t+2)}$ will be regularized down to $t+1$. 
    And the low-rankness of $\mathbf{X}$ will be promoted by the tNF minimization. 
    In contrast, the truncated nuclear norm behaves worse than the tNF on promoting the low-rankness. 
}}
\label{fig_contour_L12}
\end{figure}
\par {{%
\textit{$\bullet$ It can estimate the underlying low-rank matrix with high accuracy,} since the tNF regularizer can give a close approximation to the rank function. 
Recap that the estimated patch matrix $\mathbf{X} \in \mathbb{R}^{3d^2 \times N}$ should be of low rank since the $N$ patches constituting it have similar structures.
As shown in Fig. \ref{fig_contour_L12}, compared with other regularizers, tNF has more capacity on promoting the low-rankness and finding low-rank matrices \cite{TL12}.
Furthermore, the tNF has sufficient flexibility on approximating the rank function since its shrinkage on different singular values can be tuned flexibly. 
%
\par
\textit{$\bullet$ It can be solved efficiently by a single iterative algorithm}, 
since we proved that the proximal operator associated with tNF allows cheap closed-form solution. 
In previous works \cite{TL12, NNFN}, such proximal operator has no analytical solution but has to be solved by inefficient iterative algorithms, such as the difference of convex algorithms (DCA) \cite{DCA}. 
This drawback restricts the application of many tNF-based models. 
To this end, we mathematically proved that the global optimum of the proximal operator of tNF can be easily obtained in closed-form. 
Hence we can directly obtain the solution of problem \eqref{eq_tnf_operator} in a single step. 
This offers an efficient algorithm to solve the problem in \eqref{eq_DtNFM}. 
}}%
\begin{figure}[tb]
    \centering
    \includegraphics[width=0.4\textwidth]{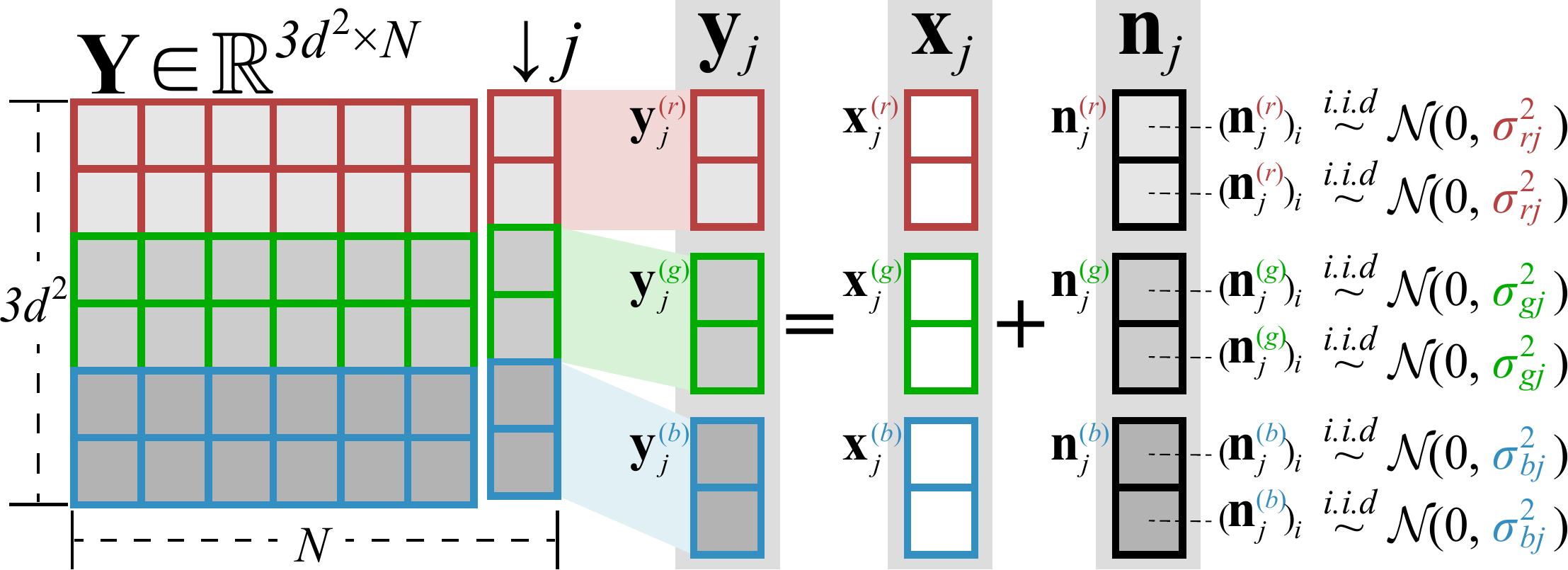}
    \caption{An illustration of the definitions of $\mathbf{y}_{j}, \mathbf{x}_{j}, \mathbf{x}_{j}^{(c)}$, and $\mathbf{n}_{j}$.}
    \label{fig_MAP_noise}
\end{figure}
\par
In the rest of this section, we elaborate the construction of weight matrices $\mathbf{C}$ and $\mathbf{S}$. 
Given a corrupted patch matrix $\mathbf{Y} \in \mathbb{R}^{3d^2\times N}$, as shown in Fig. \ref{fig_MAP_noise}, we denote the $j$th column as $\mathbf{y}_{j} = [\mathbf{y}_{j}^{(r)}; \mathbf{y}_{j}^{(g)}; \mathbf{y}_{j}^{(b)}] \in \mathbb{R}^{3d^2}$, where $\mathbf{y}_{j}^{(c)} \in \mathbb{R}^{d^2}$ is the observed data of channel $c \in \lbrace r, g, b \rbrace$. 
Correspondingly, define $\mathbf{x}_{j} = [\mathbf{x}_{j}^{(r)}; \mathbf{x}_{j}^{(g)}; \mathbf{x}_{j}^{(b)}], \mathbf{n}_{j} = [\mathbf{n}_{j}^{(r)}; \mathbf{n}_{j}^{(g)}; \mathbf{n}_{j}^{(b)}] \in \mathbb{R}^{3d^2}$ as the underlying ground truth data and the noise, respectively. 
Based on model \eqref{eq_ID}, we have $\mathbf{y}_{j} = \mathbf{x}_{j} + \mathbf{n}_{j}$ and $\mathbf{y}_{j}^{(c)} = \mathbf{x}_{j}^{(c)} + \mathbf{n}_{j}^{(c)}$. 
\par
Given the matrix $\mathbf{Y}$, the original clean matrix can be estimated through exploiting the framework of maximum a-posteriori: 
\begin{align}
    \mathbf{X}^{*} = \arg \max_{\mathbf{X}} \mathrm{P}(\mathbf{X}|\mathbf{Y})
    = \arg \max_{\mathbf{X}} \ \ln \mathrm{P}(\mathbf{Y} | \mathbf{X}) \!+\! \ln \mathrm{P}(\mathbf{X}). 
    \label{eq_map}
\end{align}
The likelihood term $\mathrm{P}(\mathbf{Y} | \mathbf{X})$ is determined by the statistics of noise. 
According to \cite{Assume_AWGN}, we assume the noise is not only independent among RGB channels, but also independent among the similar patches. 
In other words, we assume $(\mathbf{n}_j^{(c)})_i \stackrel{i.i.d}{\sim} \mathcal{N}(0, \sigma_{cj}^2)$ with $i \in \lbrace1,\cdots, d^2\rbrace$, where $\sigma_{cj}$ is the noise standard deviation in $j$th patch in channel $c$. 
And we define
\begin{equation}
    \sigma_{cj} = \sigma_c^p \cdot \sigma_j^{1-p}, \label{eq_sigma_cs}
\end{equation}
where $\sigma_c, \sigma_j$ are the noise standard deviations in color channel $c$ and $j$th patch, respectively, and $p \in [0,1]$ serves as a relative weight. 
$\sigma_j$ is defined as
\begin{equation}
    \sigma_j = \Big( \Big| \frac{\sigma_{r\_0}^2 + \sigma_{g\_0}^2 + \sigma_{b\_0}^2}{3} - \frac{1}{3d^2} \Vert \mathbf{y}_j - \hat{\mathbf{x}}_{j} \Vert_2^2 \Big| \Big)^{\frac{1}{2}}, 
\end{equation}
{{%
where $\sigma_{c\_0}$ is the original noise standard deviation in channel $c\in \lbrace r,g,b \rbrace$. 
In this paper, $\sigma_{c\_0}$ is given or can be estimated by off-the-shelf methods \cite{Noi_est}. 
$\hat{\mathbf{x}}_{j} \in \mathbb{R}^{3d^2}$ is the $j$th denoised patch output from the previous iteration of Algorithm 1, and is used to estimate the ground truth $\mathbf{x}_{j}$ in Fig. \ref{fig_MAP_noise}. 
Before Algorithm 1 iterates, $\hat{\mathbf{x}}_{j}$ is initialized to $\mathbf{y}_{j}$. 
And $\sigma_c$ is defined as
\begin{equation}
    \sigma_{c} = \Big( \Big| \sigma_{c\_0}^2 - \frac{1}{d^2} \Vert \mathbf{y}_{j}^{(c)} - \hat{\mathbf{x}}_{j}^{(c)} \Vert_2^2 \Big| \Big)^{\frac{1}{2}}. 
    \label{eq_sigma_c}
\end{equation}
where $\hat{\mathbf{x}}_{j}^{(c)}\in \mathbb{R}^{d^2}$ is the part of $\hat{\mathbf{x}}_j \in \mathbb{R}^{3d^2}$ in channel $c$. 
According to Fig. \ref{fig_MAP_noise}, we have $\hat{\mathbf{x}}_j = [\hat{\mathbf{x}}_{j}^{(r)}; \hat{\mathbf{x}}_{j}^{(g)}; \hat{\mathbf{x}}_{j}^{(b)}]$. 
}}
And the relative weight $p$ in \eqref{eq_sigma_cs} is determined by 
\begin{equation}
    p = \frac{v_{\mathbf{c}} + \epsilon}{v_{\mathbf{c}} + v_{\mathbf{s}} + 2\epsilon},
\end{equation}
where $v_{\mathbf{c}}, v_{\mathbf{s}} \in \mathbb{R}$ are respectively the coefficient of variation of $[\sigma_r; \sigma_g; \sigma_b]$ and $[\sigma_1, \sigma_2, \cdots, \sigma_N]$, 
and $\epsilon>0$ is a small value. 
\par
After determining $\sigma_{cj}$, we can determine $\mathrm{P}(\mathbf{Y} | \mathbf{X})$ based on the Gaussian probability density function: 
\begin{equation}
\!\!\mathrm{P}(\mathbf{Y} | \mathbf{X}) = \prod_{j=1}^{N} \prod_{c \in \lbrace r,g,b \rbrace}\!\! \frac{1}{\sqrt{2\pi}\sigma_{cj}} \exp\Big(-\frac{\Vert \mathbf{y}_{j}^{(c)} - \hat{\mathbf{x}}_{j}^{(c)} \Vert_2^2 }{2\sigma_{cj}^2} \Big),
\end{equation}
For the estimated patch matrix $\mathbf{X}$, it is expected to be of low-rank, i.e., the prior probability $\mathrm{P}(\mathbf{X}) \propto \exp(-\mathbf{Rank}(\mathbf{X}))$. 
However, the function $\mathbf{Rank}(\mathbf{X})$ is discontinuous, which would make the optimization problem \eqref{eq_map} be NP-hard. 
Therefore, the tNF ($\Vert \mathbf{X} \Vert_{t, *-F}$) is used to give a close approximation to the $\mathbf{Rank}(\mathbf{X})$. 
Then, the prior probability term is defined as
\begin{equation}
    \mathrm{P}(\mathbf{X}) = \exp(-\frac{\lambda}{2} \Vert \mathbf{X} \Vert_{t, *-F} ),
\end{equation}
where $\lambda > 0$ is a parameter. 
Finally, the problem \eqref{eq_map} can be deduced as
\begin{align}
& \arg \max_{\mathbf{X}} \ln\mathrm{P}(\mathbf{Y} | \mathbf{X}) + \ln\mathrm{P}(\mathbf{X}) \notag \\
=& \arg \max_{\mathbf{X}} \sum_{j=1}^{N} \sum_{c \in \lbrace r,g,b \rbrace} \!\!\! \big(- \frac{1}{2\sigma_{cj}^2} \Vert \mathbf{y}_j^{(c)} - \hat{\mathbf{x}}_j^{(c)} \Vert_2^2 \big) \!-\! \frac{\lambda}{2} \Vert \mathbf{X} \Vert_{t, *-F} \notag \\
=& \arg \min_{\mathbf{X}} \sum_{j=1}^{N} {\sigma_j^{-2(1-p)}} \!\!\!\! \sum_{c \in \lbrace r,g,b \rbrace} \!\!\!\! {\sigma_c^{-2p}} \Vert \mathbf{y}_j^{(c)} \!-\! \hat{\mathbf{x}}_j^{(c)}\! \Vert_2^2 \!+\! \lambda \Vert \mathbf{X} \Vert_{t, *-F} \notag \\
=& \arg \min_{\mathbf{X}} \sum_{j=1}^{N} {\sigma_j^{-2(1-p)}} \Vert \mathbf{C}(\mathbf{y}_j - \hat{\mathbf{x}}_{j}) \Vert_2^2 + \lambda \Vert \mathbf{X} \Vert_{t, *-F} \notag \\
=& \arg \min_{\mathbf{X}} \Vert \mathbf{C} (\mathbf{Y} - \mathbf{X}) \mathbf{S} \Vert_F^2 + \lambda \Vert \mathbf{X} \Vert_{t, *-F}, 
\end{align}
where 
$\mathbf{C} = \mathrm{Diag}([\sigma_r^{-p} \mathbf{1}; \sigma_g^{-p} \mathbf{1}; \sigma_b^{-p} \mathbf{1}]) \in \mathbb{R}^{3d^2 \times 3d^2}$, 
and $\mathbf{S} = \mathrm{Diag}([\sigma_1^{-(1-p)}, \cdots, \sigma_N^{-(1-p)}]) \in \mathbb{R}^{N \times N}$. 
Intuitively, as $\sigma_c$ becomes larger, the data in color channel $c$ will make less contribution to the estimation of $\mathbf{X}$. 
In the same manner, a larger $\sigma_j$ will reduce the contribution of $j$th noisy patch when estimating $\mathbf{X}$. 
%
%
%
\subsection{Optimization}
\label{sec_opt}
To solve the proposed DtNFM model, an efficient and effective algorithm is proposed via exploiting the framework of the ADMM. 
The original problem \eqref{eq_DtNFM} can be rewritten as
\begin{align}
    &\arg \min_{\mathbf{X}, \mathbf{Z}} \Vert \mathbf{C}(\bm{\mathbf{Y}} - \bm{\mathbf{X}}) \mathbf{S} \Vert_F^2 + \lambda \Vert \bm{\mathbf{Z}} \Vert_{t,*-F}, \notag \\
    &\ \mathrm{s.t.} \ \ \mathbf{X} - \mathbf{Z} = \mathbf{0}, 
    \label{eq_DtNFM_2}
\end{align}
where $\mathbf{Z}, \mathbf{0} \in \mathbb{R}^{3d^2\times N}$ are the auxiliary variable and a matrix of zeros, respectively. 
The augmented Lagrangian of problem \eqref{eq_DtNFM_2} is 
\begin{align}
    \mathcal{L}_{\rho} (\mathbf{X}, \mathbf{Z}, \mathbf{A}) =& \Vert \mathbf{C}(\bm{\mathbf{Y}} - \bm{\mathbf{X}}) \mathbf{S} \Vert_F^2 + \lambda \Vert \bm{\mathbf{Z}} \Vert_{t,*-F} \notag \\
    &+ \langle \mathbf{A}, \mathbf{X} - \mathbf{Z} \rangle + \frac{\rho}{2} \Vert \mathbf{X} - \mathbf{Z} \Vert_F^2, 
\end{align}
where $\mathbf{A} \in \mathbb{R}^{3d^2\times N}$ is the Lagrangian multiplier, and $\rho>0$ is the penalty parameter. 
The proposed algorithm solves the following subproblems alternatively until convergence.
\begin{numcases}{}
    \mathbf{X}_{k+1} = \arg \min_{\mathbf{X}} \mathcal{L}_{\rho_{k}} (\mathbf{X}, \mathbf{Z}_k, \mathbf{A}_k), \label{eq_upX_ori} \\
    \mathbf{Z}_{k+1} = \arg \min_{\mathbf{Z}} \mathcal{L}_{\rho_{k}} (\mathbf{X}_{k+1}, \mathbf{Z}, \mathbf{A}_k), \label{eq_upZ_ori} \\
    \mathbf{A}_{k+1} = \mathbf{A}_{k} + \rho_{k}(\mathbf{X}_{k+1}- \mathbf{Z}_{k+1}), \label{upD_close} \\
    \ \rho_{k+1} =   \mu\!\cdot\!\rho_{k}, \label{uprho_close}
\end{numcases}
where $k\in \mathbb{N}$ is the iteration number and $\mu>1$ is a scalar. 
To drive the algorithm to convergence, equation \eqref{uprho_close} is used to make the $\rho_k \rightarrow +\infty$ as $k\rightarrow \infty$. 
In the rest of this section, we detail the subproblems \eqref{eq_upX_ori} and \eqref{eq_upZ_ori}. 
\par
%
%
For subproblem \eqref{eq_upX_ori}, we have
\begin{align}
    \mathbf{X}_{k+1} =& \arg \min_{\mathbf{X}} \Vert \mathbf{C} (\mathbf{Y} - \mathbf{X}) \mathbf{S} \Vert_F^2 + \lambda \Vert \mathbf{Z}_{k} \Vert_{t,*-F} \notag \\
    &+ \langle \mathbf{A}_{k}, \mathbf{X} - \mathbf{Z}_{k} \rangle + \frac{\rho_k}{2}  \Vert \mathbf{X} - \mathbf{Z}_{k} \Vert_F^2 \notag \\
    =& \arg \min_{\mathbf{X}} \Vert \mathbf{C} ( \mathbf{Y}-\mathbf{X}) \mathbf{S}\Vert_F^2 \!+\! \frac{\rho_k}{2} \Vert \mathbf{X} \!-\! \mathbf{Z}_k \!+\! \frac{1}{\rho_{k}} \mathbf{A}_k \Vert_F^2.
    \label{upX_mid}
\end{align}
This is a standard least squares problem, which has a closed-form solution:
\begin{equation}
    \mathbf{X}_{k+1} = (2 \mathbf{C}^{\top}\!\mathbf{C} \mathbf{Y} \mathbf{S}^{\top}\!\mathbf{S} +  \rho_k\mathbf{Z}_k - \mathbf{A}_k) \!\oslash\! ( 2 \mathbf{C}^{\top}\!\mathbf{C} \mathbf{1} \mathbf{S}^{\top}\!\mathbf{S} + \rho_{k} \mathbf{1} ),
    \label{upX_close}
\end{equation}
where $\mathbf{1}\in \mathbb{R}^{3d^2\times N}$ is a matrix of ones. 
%
\par
Subproblem \eqref{eq_upZ_ori} can be equivalently rewritten as
\begin{equation}
    \mathbf{Z}_{k+1} \!=\! \arg \min_{\mathbf{Z}} \frac{1}{2} \Vert \mathbf{Z} - (\mathbf{X}_{k+1} + \frac{1}{\rho_k} \mathbf{A}_k) \Vert_F^2 \!+\! \frac{\lambda}{\rho_k} \Vert \mathbf{Z} \Vert_{t,*-F}. \label{upB}
\end{equation}
Problem \eqref{upB} is nonconvex, which is rather difficult to solve by traditional optimization techniques. 
To address this issue, we propose the following theorem to show that the global optimum of problem \eqref{upB} can be obtained in closed-form. 
\begin{theorem}
Assume that $\tau>0$ and $\mathbf{B} \in \mathbb{R}^{m\times n}$ admits SVD as $\mathbf{U}_{\mathbf{B}} \mathrm{Diag}(\bm{\sigma}(\mathbf{B})) \mathbf{V}_{\mathbf{B}}^{\top}$, without loss of generality, let $m \ge n$.
Then, the closed-form solution to
\begin{equation}
    \arg \min_{\mathbf{Z}} \frac{1}{2} \Vert \mathbf{Z} - \mathbf{B} \Vert_F^2 + \tau \Vert \mathbf{Z} \Vert_{t, *-F}, \label{eq_tnf_operator}
\end{equation}
is given by
\begin{equation}
	\mathbf{Z}^{*} = \mathbf{U}_{\mathbf{B}} \mathrm{Diag}(\varrho^*) \mathbf{V}_{\mathbf{B}}^{\top},
	\label{operator_solution}
\end{equation}
where
\begin{equation}
    \varrho^*_i \!= \!\left\{
    \begin{array}{ll}
        \!\!\sigma_{i}(\mathbf{B}), &\!\!\!\!\! \mathrm{if}\ \ 0 \le i < t+1, \\
        \!\!\!\bigl( 1 + \frac{\alpha \tau}{\Vert \mathcal{S}_{\tau} (\mathbf{r}) \Vert_{2}} \bigr)  \mathcal{S}_{\tau} (\sigma_{i}(\mathbf{B})), &\!\!\!\!\! \mathrm{if}\ \ t+1 \leq i \leq n,
    \end{array}
    \right. \label{operator_solution_2}
\end{equation}
where $\mathbf{r} = [0, \cdots, 0, \sigma_{t+1}(\mathbf{B}), \cdots, \sigma_{n}(\mathbf{B})]^{\top}$ and $\mathcal{S}_{\tau}(\mathbf{r})_{i} = \max( \mathbf{r}_{i} - \tau, 0)$ be the soft shrinkage \cite{SVT}.
\end{theorem}

%
%
%
\textit{Proof: }
Consider $\mathbf{Z}$ admits SVD as $\mathbf{U}_{\mathbf{Z}} \mathbf{\Sigma}_{\mathbf{Z}} \mathbf{V}_{\mathbf{Z}}^{\top}$.  
The first term of problem \eqref{eq_tnf_operator} can be rewritten as
\begin{equation}
    \frac{1}{2} \Vert \mathbf{Z} - \mathbf{B} \Vert_F^2 = \frac{1}{2} \left( \Vert \mathbf{Z} \Vert_F^2 - 2\langle \mathbf{Z}, \mathbf{B} \rangle + \Vert \mathbf{B} \Vert_F^2 \right).
    \label{operator_loss}
\end{equation}
According to Von Neumann's trace inequality, we have
\begin{align}
    \langle \mathbf{Z}, \mathbf{B} \rangle =& \mathrm{tr}(\mathbf{Z}^{\top}\mathbf{B}) 
    = \mathrm{tr}( \mathbf{V}_{\mathbf{Z}} \mathbf{\Sigma}_{\mathbf{Z}} \mathbf{U}_{\mathbf{Z}}^{\top} \mathbf{B} ) 
    = \mathrm{tr}( \mathbf{\Sigma}_{\mathbf{Z}} \mathbf{U}_{\mathbf{Z}}^{\top} \mathbf{B} \mathbf{V}_{\mathbf{Z}} ) \notag \\
    \le& \sum_{i=1}^{n} \sigma_{i}(\mathbf{Z}) \!\cdot\! \sigma_{i}(\mathbf{B}) \!\cdot\! \sigma_{i}(\mathbf{U}_{\mathbf{Z}}^{\top} \mathbf{V}_{\mathbf{Z}} ) \label{von_neumann} 
    = \sum_{i=1}^{n} \sigma_{i}(\mathbf{Z}) \!\cdot\! \sigma_{i}(\mathbf{B}).
\end{align}
The equality of (\ref{von_neumann}) occurs if and only if 
\begin{equation}
	\mathbf{U}_{\mathbf{Z}} = \mathbf{U}_{\mathbf{B}}, \mathbf{V}_{\mathbf{Z}} = \mathbf{V}_{\mathbf{B}}.
\end{equation}
Therefore, problem \eqref{eq_tnf_operator} can be rewritten as follows:
\begin{align}
    &\arg \min_{\mathbf{Z}} \frac{1}{2} \Vert \mathbf{Z} - \mathbf{B} \Vert_F^2 + \tau \Vert \mathbf{Z} \Vert_{t,*-F} \notag \\
    =& \arg \min_{\mathbf{Z}} \frac{1}{2} \Vert \mathbf{Z} \Vert_F^2 - \sum_{i=1}^{n} \sigma_{i}(\mathbf{Z}) \sigma_{i}(\mathbf{B}) + \frac{1}{2} \Vert \mathbf{B} \Vert_F^2 \notag \\ 
    &+ \tau \biggl( \sum_{i=t+1}^{n} \sigma_{i}(\mathbf{Z}) - \alpha \bigl( \sum_{i=t+1}^{n} \sigma_{i}(\mathbf{Z})^{2} \bigr)^{\frac{1}{2}} \biggr) \notag \\
    =& \arg \min_{\mathbf{Z}} \sum_{i=1}^{t}\! \left( \frac{1}{2} \sigma_{i}(\mathbf{Z})^2 - \sigma_{i}(\mathbf{Z}) \sigma_{i}(\mathbf{B}) \!\right)\! + \!\sum_{i=t+1}^{n}\! \biggl( \frac{1}{2} \sigma_{i}(\mathbf{Z})^2 \notag \\ 
    &- \sigma_{i}(\mathbf{Z}) \sigma_{i}(\mathbf{B}) + \tau \sigma_{i}(\mathbf{Z}) \!\biggr) \!- \alpha \tau\! \biggl( \sum_{i=t+1}^{n} \sigma_{i}(\mathbf{Z})^{2} \! \biggr)^{\frac{1}{2}}. 
    \label{linear_combination}
\end{align}
Therefore, the original problem \eqref{eq_tnf_operator} has been equivalently transformed into the combination of independent quadratic equations for each 
$\sigma_{i}(\mathbf{Z})$. Let $F(\bm{\sigma}(\mathbf{Z}))$ denote the objective function of (\ref{linear_combination}). The minimum of $F$, denoted as $\varrho^*_i$, is given by 
\begin{equation}
	\frac{\partial F}{\partial {\sigma_{i}}(\mathbf{Z})} = 0.
	\label{first-order}
\end{equation}
When $0 \le i < t+1$, it is trivial to obtain
\begin{equation}
	\varrho^*_i = \sigma_{i}(\mathbf{B}).
	\label{0--t}
\end{equation}
When $t+1 \le i \le n$, equation (\ref{first-order}) is expressed as
\begin{equation}
    \Bigl( 1 - \frac{\alpha \tau}{\sqrt{\sum_{i=t+1}^{n} (\varrho_{i}^{*})^{2}}} \Bigr) \varrho_{i}^{*} = \sigma_{i}(\mathbf{B}) - \tau.
    \label{t+1--T}
\end{equation}
And the solution of (\ref{t+1--T}) is
\begin{equation}
    \varrho^*_i = \Bigl( 1 + \frac{\alpha \tau}{\Vert \mathcal{S}_{\tau} (\mathbf{r}) \Vert_{2}} \Bigr) \!\cdot\! \mathcal{S}_{\tau} (\sigma_{i}(\mathbf{B})), \label{eq_sol_t_n}
\end{equation}
where $\mathbf{r} = [0, \ldots, 0, \sigma_{t+1}(\mathbf{B}), \ldots, \sigma_{T}(\mathbf{B})]^{\top} \in \mathbb{R}^{n}$ and $(\mathcal{S}_{\tau} (\mathbf{r}))_{i} = \max(\mathbf{r}_{i} - \tau, 0)$. 
Combining (\ref{0--t}) and (\ref{eq_sol_t_n}), we have 
\begin{equation}
	\varrho^*_i \!= \!\left\{
	\begin{array}{ll}
	\!\!\sigma_{i}(\mathbf{B}), &\!\!\!\!\! \mathrm{if}\ \ 0 \le i < t+1, \\
	\!\!\!\bigl( 1 + \frac{\alpha \tau}{\Vert \mathcal{S}_{\tau} (\mathbf{r}) \Vert_{2}} \bigr)  \mathcal{S}_{\tau} (\sigma_{i}(\mathbf{B})), &\!\!\!\!\! \mathrm{if}\ \ t+1 \le i \le n.
	\end{array}
	\right.
\end{equation}
And the optimal solution of problem (\ref{eq_tnf_operator}) is 
\begin{equation}
	\mathbf{Z}^{*} = \mathbf{U}_{\mathbf{B}} \mathrm{Diag}(\varrho^*) \mathbf{V}_{\mathbf{B}}^{\top}.
\end{equation}
$\hfill\blacksquare$ \par 

According to Theorem 1, the global optimum of subproblem \eqref{eq_upZ_ori} is given by \eqref{operator_solution} and \eqref{operator_solution_2}, where $\mathbf{B} = \mathbf{X}_{k+1} + \rho_k^{-1} \mathbf{A}_k$ and $\tau = \lambda/\rho_k$. 
Up to now, the global optima of both subproblem \eqref{eq_upX_ori} and \eqref{eq_upZ_ori} have been obtained in closed-form. 
\par
The algorithm would be terminated when the iteration number exceeds a threshold $K$ or the following stopping criteria hold simultaneously: 
(a) $\Vert \mathbf{X}_{k+1} - \mathbf{Z}_{k+1} \Vert_F \le \epsilon$, (b) $\Vert \mathbf{X}_{k+1} - \mathbf{X}_{k} \Vert_F \le \epsilon$, (c) $\Vert \mathbf{Z}_{k+1} - \mathbf{Z}_{k} \Vert_F \le \epsilon$, 
where $\epsilon > 0$ is a small tolerance. 
This stopping criterion is devised based on the convergence guarantee given by Theorem 2. 
Finally, the complete algorithm for solving problem \eqref{eq_DtNFM} is summarized in Algorithm \ref{alg_admm}. 
\begin{algorithm}[t] 
    \caption{Solving problem \eqref{eq_DtNFM} via ADMM.}
    \label{alg_admm}
    \KwIn{corrupted matrix $\mathbf{Y} \in \mathbb{R}^{3d^2\times N}$, weight matrix $\mathbf{C} \in \mathbb{R}^{3d^2\times 3d^2}, \mathbf{S} \in \mathbb{R}^{N\times N}$\;}%
    \KwOut{Denoised matrix $\mathbf{X}\in \mathbb{R}^{3d^2\times N}$\;}
    Initialize $\mathbf{X}_{0} = \mathbf{Z}_{0} = \mathbf{A}_{0} = \mathbf{0}, \rho_{0}, t, \lambda, \mu, \epsilon, k=0, K$\;
    \While{Stopping criterion is not satisfied}{
        $\mathbf{X}_{k+1} \!\!\leftarrow\!\! (2 \mathbf{C}^{2} \mathbf{Y} \mathbf{S}^{2} + \rho_k\mathbf{Z}_k - \mathbf{A}_k) \!\oslash\! ( 2 \mathbf{C}^{2} \mathbf{1} \mathbf{S}^{2} \!+\! \rho_{k} \mathbf{1} )$\;
        $\mathbf{Z}_{k+1} \!\leftarrow\arg \min_{\mathbf{Z}} \frac{1}{2} \Vert \mathbf{Z} - (\mathbf{X}_{k+1} + \frac{1}{\rho_{k}}\mathbf{A}_{k}) \Vert_F^2 + \lambda / \rho_{k} \Vert \mathbf{Z} \Vert_{t, *-F}$\;
        $\mathbf{A}_{k+1} \leftarrow \mathbf{A}_{k} + \rho_k(\mathbf{X}_{k+1}- \mathbf{Z}_{k+1})$\;
        $\rho_{k+1} \; \leftarrow \mu\cdot\rho_{k}$\;
        Check the stopping criterion:
        $((k\leftarrow k+1) > K)||(\Vert \mathbf{X}_{k+1} - \mathbf{Z}_{k+1} \Vert_F \le \epsilon)\&\&$
        $(\Vert \mathbf{Z}_{k+1} - \mathbf{Z}_{k}\Vert_F \le \epsilon) \&\&(\Vert \mathbf{X}_{k+1} - \mathbf{X}_{k}\Vert_F \le \epsilon)$\;
    }
\end{algorithm}
%
%
%
\subsection{Convergence and Complexity Analysis}
\label{sec_convergence}
In this section, we give the following theorem to establish the weak convergence result of Algorithm \ref{alg_admm}, which facilitates the construction of a reasonable termination condition.
The detailed proof can be found in the supplementary material. 
%
\begin{theorem}
The sequences $\lbrace \mathbf{X}_k \rbrace$ and $\lbrace \mathbf{Z}_k \rbrace$ generated from Algorithm \ref{alg_admm} satisfy
\begin{align}
    &(a) \lim_{k \rightarrow \infty}  \Vert \mathbf{X}_{k+1} - \mathbf{Z}_{k+1} \Vert_F = 0.\notag\\
    &(b) \lim_{k \rightarrow \infty} \Vert \mathbf{X}_{k+1} - \mathbf{X}_k \Vert_F = 0. \notag \\
      &(c) \lim_{k \rightarrow \infty} \Vert \mathbf{Z}_{k+1} - \mathbf{Z}_k \Vert_F = 0. \notag
\end{align}
\end{theorem}

Theorem 2 guarantees that $\lbrace \mathbf{X}_{k} \rbrace$ and $\lbrace \mathbf{Z}_{k} \rbrace$ converge to a single critical point, i.e., $\lim_{k\rightarrow \infty} \mathbf{X}_{k} = \lim_{k\rightarrow \infty} \mathbf{Z}_{k} = \mathbf{X}^{*}$. 
This converged solution proves that the proposed Algorithm \ref{alg_admm} is capable of solving the problem resulted by the DtNFM model in \eqref{eq_DtNFM}. 
%
\par
The time complexity of Algorithm \ref{alg_admm} is discussed in brief.
In step 3, updating $\mathbf{X} \in \mathbb{R}^{3d^2\times N}$ takes $\mathcal{O}(d^4N)$ time. 
In step 4, the complexity is $\mathcal{O}(d^2N^2 + d^4N )$ since the SVD of $\mathbf{X}_{k+1} + \rho_{k}^{-1} \mathbf{A}_{k} \in \mathbb{R}^{3d^2\times N}$ costs $\mathcal{O}(d^2N^2)$ time. 
The complexity of step 5 is $\mathcal{O} (d^2N)$. 
Therefore, the complexity of Algorithm \ref{alg_admm} is $\mathcal{O}((d^4N + d^2N^2)K)$. 
In Algorithm \ref{alg_framework}, step 6 costs $\mathcal{O}(S^2\log S)$ time, where $S\in \mathbb{N}_{+}$ is the side length of the search window. 
In step 7, constructing $\mathbf{C}\in \mathbb{R}^{3d^2\times 3d^2}$ and $\mathbf{S}\in \mathbb{R}^{N\times N}$ costs $\mathcal{O}(d^4 + N^2)$ time. 
The dominant cost lies in step 8, which is actually the Algorithm \ref{alg_admm}. 
In summary, the complexity of Algorithm \ref{alg_framework} is $\mathcal{O}((d^4N + d^2N^2)KP\theta)$. 
\begin{figure}[tb]
\centering
\captionsetup[subfigure]{captionskip=0pt, farskip=0pt}
\begin{minipage}{0.4\linewidth}
    \subfloat[]{
        \includegraphics[width=\linewidth]{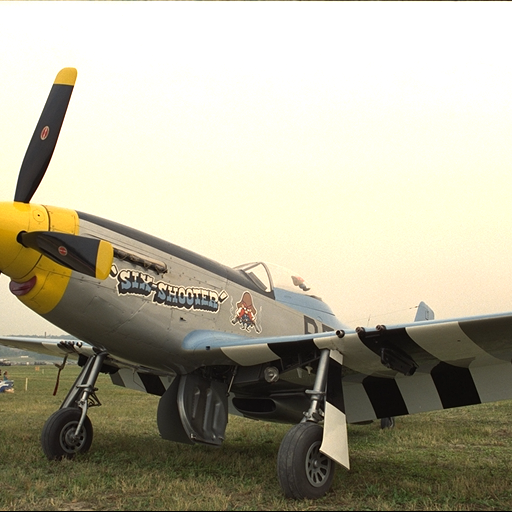}
        \label{fig_20_GT}
    }
\end{minipage}
\begin{minipage}{0.41\linewidth}
    \subfloat[]{
        \includegraphics[width=0.5\linewidth]{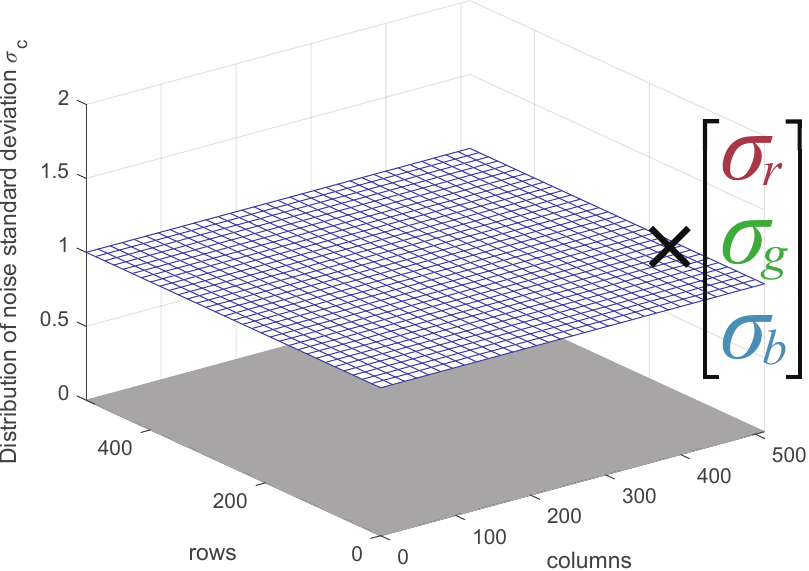} \label{fig_nmap_235}
    }
    \subfloat[]{
        \includegraphics[width=0.5\linewidth]{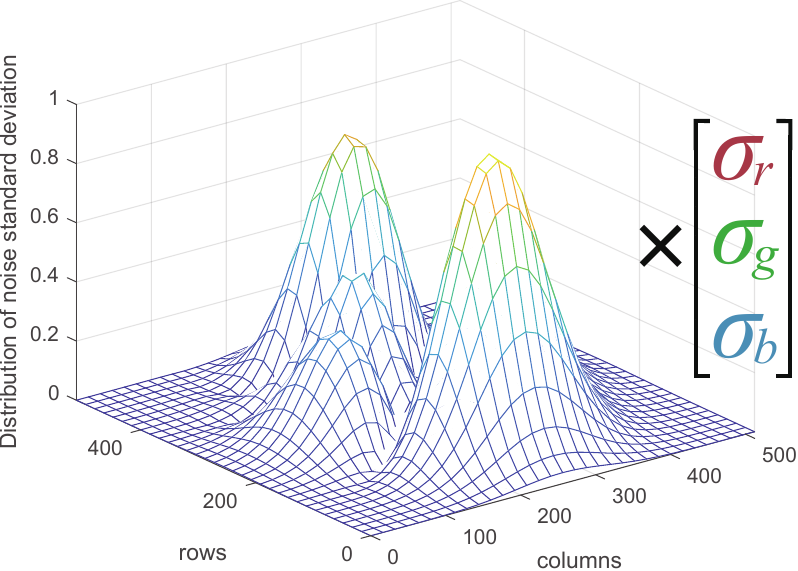} \label{fig_peak}
    }
    \\
    \subfloat[]{
        \includegraphics[width=0.5\linewidth]{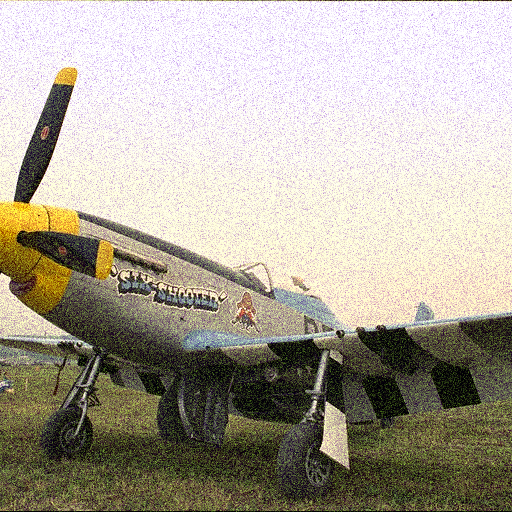}
    }
    \subfloat[]{
        \includegraphics[width=0.5\linewidth]{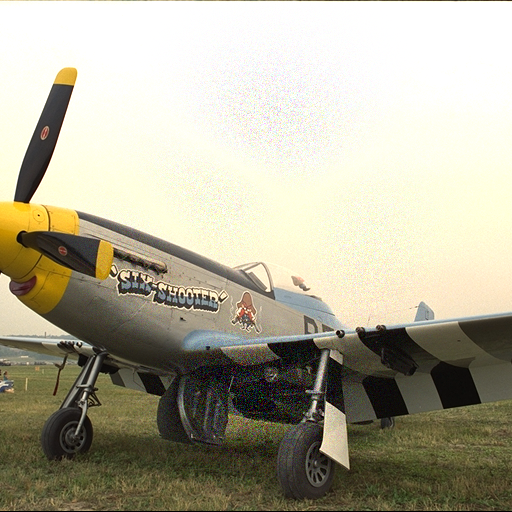}
    }
\end{minipage}
\caption{Illustrations of the spatially invariant noise and the spatially variant noise. (a) the ground truth image ``kodim20''. (b) The surface of the standard deviation of spatially invariant noise. (c) The surface of the standard deviation of spatially variant noise, which is returned by MATLAB code ``abs(peaks(512))''. A point $(x,y,z)$ means the pixel at $x$th row and $y$th column has Gaussian noise with standard deviations being $z\times [\sigma_{r\_0};\sigma_{g\_0};\sigma_{b\_0}]$, where $z \in [0,1]$. Specifically, $z \equiv 1$ in the subfigure (b). (d) The image corrupted by the noise in subfigure (b) with $[\sigma_{r\_0};\sigma_{g\_0};\sigma_{b\_0}] = [30; 10; 50]$. (e) The image corrupted by the noise in subfigure (c) with $[\sigma_{r\_0};\sigma_{g\_0};\sigma_{b\_0}] = [30; 35; 40]$.}
\label{fig_spatial_var}
\end{figure}
\subsection{{{Differences From Existing Methods}}}
{{
In this section, we discuss the difference between the proposed DtNFM and three highly related methods---EBD \cite{EBD}, SSLRDM \cite{SSLRDM}, and DLRQP \cite{DLRQP}. 
All of those methods, together with our DtNFM, are based on low-rank approximation, and exploit the NSS prior for image denoising. 
\par
First, the proposed DtNFM method is able to handle the spatially variant noise, since a dedicated weight matrix is devised to model the spatial variation of the noise. 
In contrast, the EBD, SSLRDM and DLRQP can only deal with the spatially invariant noise. 
Second, the DtNFM method can deal with the difference of the noise among RGB channels. 
While the DLRQP cannot model this cross-channel difference of noise. And the EBD and SSLRDM are not developed for color image denoising. 
Third, the algorithm for the DtNFM model possess theoretical convergence guarantee. 
In contrast, it is difficult to prove the convergence of the algorithm for the DLRQP model, since it incorporates a deep learning denoiser which is effectively a black box. 
In summary, the proposed DtNFM method has significant difference with those related methods. 
}}
\section{Experimental Results and Analysis}
\label{sec_exp}
To test the performance of the proposed DtNFM method, three kinds of experiments are conducted, i.e., spatially invariant noise removal, spatially variant noise removal, and real-world noise removal. 
Nine state-of-the-art methods are compared, including (1) CBM3D \cite{CBM3D}, (2) DRUNet \cite{DRUNet}, (3) Restormer \cite{Restormer}, (4) HLTA-GN \cite{HLTA-GN}, (5) NGmeet \cite{NGmeet}, (6) DLRQP \cite{DLRQP}, (7) MCWNNM \cite{MCWNNM}, (8) MCWSNM \cite{MCWSNM}, and (9) NNFNM \cite{MCNNFNM}. 
The parameters of compared methods are either tuned to the optimum or kept the same as the original codes. 
{{%
The experiments are implemented on a laptop with 2.1GHz CPU, 16GB RAM, and Nvidia GeForce MX350 GPU. 
The Restormer \cite{Restormer} is trained on a server with 2.1GHz CPU, 64G RAM, and Tesla T4 GPU. 
}}
\begin{table*}[t] 
\centering\scriptsize
\caption{ PSNR and SSIM results for all competing methods under $[\sigma_{r\_0}; \sigma_{g\_0}; \sigma_{b\_0}] = [20;35;5]$. Running time is in seconds. }
\begin{tabularx}{\textwidth}{p{0.2cm}<{\centering}YYYYY YYYYY}
\toprule
& CBM3D & DRUNet & \!{{Restormer}} & \!{{HLTA-GN}} & \!{{NGMeet}} & \!{{DLRQP}} & MCWNNM & MCWSNM & NNFNM & DtNFM\\
\# & PSNR SSIM & PSNR SSIM & PSNR SSIM & PSNR SSIM & PSNR SSIM & PSNR SSIM & PSNR SSIM & PSNR SSIM & PSNR SSIM & PSNR SSIM\\
\hline
1 & 28.66 0.8267 & 29.25 0.8537 & 25.59 0.7935 & 28.42 0.8139 & 28.13 0.8013 & 28.67 0.8266 & 31.08 0.8999 & 31.11 0.903 & 31.29 0.9057 & \textbf{31.46 0.9147}\\
2 & 31.53 0.7683 & 32.81 0.8334 & 25.02 0.6168 & 31.59 0.7980 & 31.47 0.7824 & 31.06 0.7788 & 33.76 0.8540 & 34.02 0.8682 & 34.05 0.8678 & \textbf{34.22 0.8761}\\
3 & 33.16 0.8454 & 34.75 0.9142 & 23.57 0.6603 & 33.08 0.8632 & 33.83 0.8922 & 32.35 0.8644 & 35.46 0.9004 & 35.96 0.9223 & 36.33 0.9252 & \textbf{36.15 0.9275}\\
4 & 31.71 0.8017 & 33.04 0.8561 & 23.76 0.6480 & 32.08 0.8308 & 31.98 0.8213 & 31.10 0.7966 & 34.32 0.8795 & 34.60 0.8922 & 34.71 0.8941 & \textbf{34.82 0.9001}\\
5 & 29.17 0.8657 & 30.64 0.8994 & 27.67 0.8359 & 27.73 0.8457 & 29.10 0.8623 & 29.49 0.8730 & 31.12 0.9159 & 31.23 0.9227 & 31.29 0.9251 & \textbf{31.35 0.9289}\\
6 & 29.98 0.8251 & 30.22 0.8606 & 25.80 0.7579 & 29.83 0.8313 & 29.73 0.8207 & 29.32 0.8159 & 32.29 0.8925 & 32.45 0.9016 & 32.54 0.9057 & \textbf{32.75 0.9122}\\
7 & 32.30 0.8749 & 34.96 \textbf{0.9439} & 24.25 0.7330 & 32.31 0.8944 & 33.34 0.9228 & 30.05 0.8767 & 34.64 0.9229 & 35.02 0.9401 & 35.00 0.9378 & \textbf{35.05} 0.9428\\
8 & 29.27 0.8766 & 30.45 0.9027 & 28.01 0.8590 & 28.72 0.8793 & 29.28 0.8749 & 29.10 0.8780 & 31.19 0.9214 & 31.28 0.9269 & \textbf{31.28} 0.9243 & 31.11 \textbf{0.9300}\\
9 & 32.69 0.8445 & 34.66 0.9117 & 23.24 0.6733 & 32.72 0.8662 & 33.36 0.8909 & 31.35 0.8622 & 34.86 0.8980 & 35.27 0.9164 & 35.33 0.9136 & \textbf{35.37 0.9187}\\
10 & 32.54 0.8382 & 34.61 0.9093 & 23.46 0.6792 & 32.64 0.8638 & 33.12 0.8747 & 30.86 0.8347 & 34.67 0.8954 & 35.07 0.9115 & 35.12 0.9099 & \textbf{35.15 0.9153}\\
11 & 30.54 0.8066 & 31.43 0.8453 & 25.10 0.7028 & 30.17 0.8170 & 30.32 0.8020 & 29.79 0.7960 & 32.53 0.8770 & 32.70 0.8873 & 32.76 0.8906 & \textbf{32.94 0.8969}\\
12 & 32.52 0.8082 & 33.31 0.8646 & 23.65 0.6495 & 33.02 0.8373 & 32.67 0.8268 & 31.57 0.8070 & 34.74 0.8769 & 35.14 0.8941 & 35.22 0.8955 & \textbf{35.29 0.9015}\\
13 & 27.44 0.8082 & 27.84 0.8201 & 26.16 0.7758 & 26.25 0.7722 & 26.82 0.7622 & 26.20 0.7820 & 29.25 0.8875 & 29.27 0.8875 & 29.41 0.8940 & \textbf{29.76 0.9053}\\
14 & 29.46 0.8023 & 30.50 0.8348 & 25.72 0.7403 & 28.49 0.7966 & 29.28 0.7895 & 29.92 0.8137 & 31.78 0.8840 & 31.87 0.8885 & 31.95 0.8913 & \textbf{32.14 0.8982}\\
15 & 32.05 0.8120 & 31.64 0.8668 & 25.93 0.7222 & 32.29 0.8438 & 32.38 0.8488 & 30.82 0.8246 & 34.31 0.8829 & 34.64 0.9006 & 34.64 0.8986 & \textbf{34.68 0.9066}\\
16 & 31.41 0.8159 & 32.42 0.8648 & 23.47 0.6504 & 31.66 0.8304 & 31.38 0.8241 & 31.23 0.8175 & 33.86 0.8861 & 34.11 0.8996 & 34.35 0.9040 & \textbf{34.51 0.9098}\\
17 & 31.83 0.8301 & 32.77 0.8820 & 24.59 0.7039 & 31.25 0.8539 & 32.08 0.8566 & 31.20 0.8405 & 34.08 0.8961 & 34.39 0.9100 & 34.48 0.9107 & \textbf{34.57 0.9147}\\
18 & 29.33 0.8109 & 30.06 0.8438 & 25.99 0.7210 & 27.45 0.7688 & 28.86 0.7972 & 29.05 0.8089 & 31.16 0.8795 & 31.26 0.8856 & 31.35 0.8888 & \textbf{31.60 0.8948}\\
19 & 31.00 0.8147 & 31.97 0.8591 & 25.23 0.6764 & 31.11 0.8311 & 31.08 0.8256 & 31.08 0.8251 & 33.23 0.8863 & 33.41 0.8969 & 33.49 0.8983 & \textbf{33.60 0.9059}\\
20 & 32.35 0.8248 & 29.80 0.8960 & 26.58 0.8014 & 32.62 0.8659 & 32.61 0.8653 & 30.50 0.8473 & 34.27 0.9008 & \textbf{34.69} 0.9162 & 34.29 0.9197 & 34.25 \textbf{0.9229}\\
21 & 30.34 0.8376 & 31.34 0.8904 & 24.46 0.7154 & 30.08 0.8501 & 30.14 0.8619 & 28.18 0.8264 & 32.48 0.8974 & 32.67 0.9106 & 32.62 0.9096 & \textbf{33.01 0.9162}\\
22 & 30.25 0.7829 & 30.89 0.8227 & 24.20 0.6366 & 30.34 0.8036 & 30.18 0.7846 & 30.23 0.7865 & 32.31 0.8645 & 32.44 0.8733 & 32.56 0.8761 & \textbf{32.66 0.8842}\\
23 & 33.26 0.8413 & 35.34 0.9185 & 23.37 0.6489 & 33.83 0.8819 & 34.45 0.9013 & 31.47 0.8665 & 35.53 0.9006 & 36.06 0.9216 & 35.97 0.9182 & \textbf{36.14 0.9241}\\
24 & 29.57 0.8358 & 30.49 0.8867 & 26.48 0.7600 & 28.64 0.8410 & 29.66 0.8439 & 29.71 0.8446 & 31.73 0.8996 & 31.90 0.9093 & 31.65 0.9072 & \textbf{31.97 0.9174}\\
\hline
Avg. & 30.93 0.8249 & 31.88 0.8742 & 25.05 0.7151 & 30.68 0.8367 & 31.05 0.8389 & 30.18 0.8289 & 33.11 0.8916 & 33.36 0.9036 & 33.40 0.9047 & \textbf{33.52 0.9110}\\
Time & 7.69 & 2.99 & \textbf{1.54} & 325.54 & 529.19 & 847.17 & 706.41 & 867.08 & 570.86 & 582.13\\
\bottomrule
	\end{tabularx}
	\label{tab_syn_235}
\end{table*}
\begin{table*}[ptb] 
\centering\scriptsize
\caption{ PSNR and SSIM results for all competing methods under $[\sigma_{r\_0}; \sigma_{g\_0}; \sigma_{b\_0}] = [30;10;50]$. Running time is in seconds.}
\begin{tabularx}{\textwidth}{p{0.3cm}<{\centering}YYYYY YYYYY}
    \toprule
    & CBM3D & DRUNet & \!{{Restormer}} & \!{{HLTA-GN}} & \!{{NGMeet}} & \!{{DLRQP}} & MCWNNM & MCWSNM & NNFNM & DtNFM\\
    \# & PSNR SSIM & PSNR SSIM & PSNR SSIM & PSNR SSIM & PSNR SSIM & PSNR SSIM & PSNR SSIM & PSNR SSIM & PSNR SSIM & PSNR SSIM\\
    \hline
    1 & 26.90 0.7496 & 27.11 0.7662 & 23.81 0.6980 & 27.60 0.7835 & 26.44 0.7174 & 26.52 0.7395 & 28.81 0.8340 & 28.86 0.8316 & 29.01 0.8388 & \textbf{29.55 0.8579}\\
    2 & 30.27 0.7276 & 29.29 0.7622 & 26.45 0.6791 & 30.75 0.7721 & 30.40 0.7526 & 29.15 0.7227 & 31.81 0.7990 & 31.95 0.8063 & 32.07 0.8091 & \textbf{32.42 0.8223}\\
    3 & 31.56 0.8038 & 31.50 0.8683 & 24.29 0.6574 & 31.97 0.8332 & 31.94 0.8510 & 30.41 0.8208 & 33.89 0.8752 & 34.07 0.8886 & 34.37 0.8928 & \textbf{34.77 0.9030}\\
    4 & 30.22 0.7546 & 30.76 0.8033 & 24.43 0.6194 & 31.02 0.7993 & 30.37 0.7775 & 29.11 0.7432 & 32.26 0.8322 & 32.39 0.8381 & 32.60 0.8434 & \textbf{32.96 0.8556}\\
    5 & 27.14 0.8010 & 27.96 0.8279 & 24.93 0.7672 & 26.47 0.8066 & 26.92 0.7878 & 26.87 0.7960 & 29.01 0.8703 & 29.13 0.8723 & 29.07 0.8756 & \textbf{29.89 0.8943}\\
    6 & 28.27 0.7650 & 28.39 0.7899 & 23.67 0.6486 & 28.89 0.8018 & 27.96 0.7503 & 27.74 0.7425 & 30.23 0.8388 & 30.31 0.8413 & 30.40 0.8520 & \textbf{31.03 0.8675}\\
    7 & 30.61 0.8438 & 31.74 0.9111 & 24.44 0.6878 & 30.69 0.8714 & 31.18 0.8898 & 28.79 0.8472 & 32.81 0.9048 & 33.02 0.9173 & 32.93 0.9121 & \textbf{33.62 0.9262}\\
    8 & 27.31 0.8340 & 28.23 0.8601 & 24.45 0.7823 & 27.54 0.8543 & 27.54 0.8378 & 26.44 0.8180 & 29.07 0.8838 & 29.25 0.8867 & 29.10 0.8884 & \textbf{29.87 0.9003}\\
    9 & 31.20 0.8139 & 32.52 0.8794 & 23.90 0.5937 & 31.63 0.8418 & 31.59 0.8610 & 29.38 0.8185 & 33.23 0.8729 & 33.47 0.8873 & 33.57 0.8854 & \textbf{34.04 0.8982}\\
    10 & 30.96 0.7976 & 32.44 0.8683 & 23.91 0.6015 & 31.43 0.8362 & 31.26 0.8364 & 28.87 0.7815 & 33.00 0.8649 & 33.19 0.8767 & 33.24 0.8758 & \textbf{33.88 0.8921}\\
    11 & 28.89 0.7478 & 28.86 0.7643 & 24.06 0.6278 & 29.20 0.7868 & 28.44 0.7341 & 28.30 0.7308 & 30.60 0.8239 & 30.72 0.8249 & 30.71 0.8339 & \textbf{31.28 0.8454}\\
    12 & 31.25 0.7695 & 31.42 0.7997 & 23.27 0.5678 & 32.13 0.8145 & 31.16 0.7887 & 29.75 0.7615 & 33.03 0.8379 & 33.16 0.8452 & 33.28 0.8515 & \textbf{33.70 0.8623}\\
    13 & 25.45 0.7100 & 25.51 0.6984 & 23.58 0.6942 & 25.05 0.7228 & 24.68 0.6427 & 25.38 0.7233 & 27.07 0.8094 & 27.25 0.8049 & 27.19 0.8152 & \textbf{27.88 0.8360}\\
    14 & 27.68 0.7307 & 28.09 0.7490 & 24.54 0.6744 & 27.44 0.7592 & 27.61 0.7167 & 27.41 0.7169 & 29.57 0.8193 & 29.61 0.8155 & 29.71 0.8241 & \textbf{30.28 0.8414}\\
    15 & 30.62 0.7746 & 29.92 0.8271 & 25.85 0.7191 & 31.24 0.8136 & 30.79 0.8095 & 29.23 0.7871 & 32.48 0.8446 & 32.73 0.8567 & 32.72 0.8577 & \textbf{33.14 0.8706}\\
    16 & 29.91 0.7607 & 30.30 0.7953 & 23.32 0.5782 & 30.73 0.8008 & 29.77 0.7631 & 29.35 0.7433 & 32.02 0.8402 & 32.04 0.8439 & 32.38 0.8554 & \textbf{32.81 0.8690}\\
    17 & 30.27 0.7899 & 29.98 0.8276 & 24.62 0.6892 & 30.03 0.8240 & 30.43 0.8212 & 28.23 0.7687 & 32.28 0.8636 & 32.42 0.8696 & 32.51 0.8731 & \textbf{33.05 0.8855}\\
    18 & 27.54 0.7386 & 27.60 0.7664 & 24.67 0.6958 & 26.30 0.7210 & 27.07 0.7231 & 27.01 0.7209 & 29.33 0.8202 & 29.42 0.8230 & 29.47 0.8287 & \textbf{30.12 0.8483}\\
    19 & 29.51 0.7655 & 30.00 0.7961 & 24.05 0.6207 & 30.20 0.8064 & 29.75 0.7820 & 28.22 0.7420 & 31.54 0.8432 & 31.56 0.8458 & 31.74 0.8526 & \textbf{32.15 0.8650}\\
    20 & 30.91 0.7973 & 28.00 0.8574 & 24.74 0.7716 & 31.44 0.8386 & 31.26 0.8387 & 28.90 0.8116 & 32.47 0.8585 & 32.85 0.8740 & 32.30 0.8837 & \textbf{32.92 0.8873}\\
    21 & 28.62 0.7954 & 29.23 0.8409 & 23.80 0.6379 & 29.01 0.8201 & 28.34 0.8128 & 26.40 0.7621 & 30.42 0.8561 & 30.60 0.8665 & 30.54 0.8667 & \textbf{31.20 0.8812}\\
    22 & 28.80 0.7220 & 29.07 0.7521 & 24.05 0.6026 & 29.49 0.7734 & 28.23 0.7015 & 28.08 0.6953 & 30.47 0.8101 & 30.51 0.8102 & 30.69 0.8186 & \textbf{31.11 0.8335}\\
    23 & 31.89 0.8194 & 32.08 0.8916 & 25.32 0.6907 & 32.50 0.8597 & 32.38 0.8776 & 30.50 0.8579 & 33.68 0.8776 & 34.01 0.8942 & 33.93 0.8880 & \textbf{34.49 0.9030}\\
    24 & 27.72 0.7739 & 27.99 0.8171 & 24.03 0.7087 & 27.44 0.8074 & 27.70 0.7796 & 27.16 0.7615 & 29.57 0.8531 & 29.76 0.8585 & 29.38 0.8543 & \textbf{30.40 0.8810}\\
    \hline
    Avg. & 29.31 0.7744 & 29.50 0.8133 & 24.34 0.6672 & 29.59 0.8062 & 29.30 0.7855 & 28.22 0.7672 & 31.19 0.8472 & 31.34 0.8533 & 31.37 0.8574 & \textbf{31.94 0.8720}\\
    Time  & 4.92 & 2.53 & \textbf{1.59} & 402.94 & 489.4 & 581.63 & 766.43 & 973.67 & 583.24 & 708.87\\
    \bottomrule
\end{tabularx}
\label{tab_syn_315}
\end{table*}
\begin{table*}[tbp] 
	\centering\scriptsize
	\caption{Average improvements (PSNR and SSIM) of DtNFM over other methods.}
	\begin{tabularx}{\textwidth}{p{2.5cm}YYYYYYYYY}
\toprule
 & CBM3D & DRUNet & Restormer & HLTA-GN & NGMeet & DLRQP & MCWNNM & MCWSNM & NNFNM\\
\hline
(a) $[20, 35, 5]$ & 2.59 0.0861 & 1.64 0.0368 & 8.47 0.196 & 2.84 0.0744 & 2.47 0.0721 & 3.34 0.0821 & 0.41 0.0194 & 0.17 0.0074 & 0.12 0.0064\\
(b) $[30, 10, 50]$ & 2.63 0.0975 & 2.44 0.0586 & 7.6 0.2047 & 2.35 0.0658 & 2.64 0.0864 & 3.73 0.1048 & 0.75 0.0247 & 0.60 0.0187 & 0.57 0.0146\\
(c) Spatially variant noise& 2.81 0.0274 & 2.69 0.0323 & 1.41 0.0115 & 2.86 0.0345 & 0.89 0.0141 & 2.8 0.0259 & 2.34 0.0269 & 3.03 0.0578 & 2.94 0.0371\\
(d) Real-world noise & 3.71 0.0723 & 0.35 0.0033 & 2.02 0.0232 & 1.4 0.0129 & 0.87 0.0088 & 1.41 0.0165 & 0.56 0.0061 & 0.72 0.0076 & \!\!\!0.06 2.03E-5\\
\bottomrule
\end{tabularx}
\label{tab_improvements}
\end{table*}
%
%
\begin{table*}[tb]
	\centering\scriptsize
	\caption{Parameters of the proposed DtNFM method with respect to four experiments.}
    \begin{tabularx}{\textwidth}{p{3cm}YYYYYYYYYY}
		\toprule
        & $\theta$ & $N$ & $d$ & $s$ & $K$ & $t$ & $\alpha$ & $\lambda$ & $\rho_{0}$ & $\mu$ \\
		\midrule
		(a) $[20, 35, 5]$ & 3 & 60 & 6 & 5 & 10 & 2 & 1.80 & 0.80 & 0.30 & 1.002\\
		(b) $[30, 10, 50]$ & 2 & 60 & 6 & 5 & 10 & 2 & 1.80 & 1.00 & 0.50 & 1.002\\
		(c) Spatially variant noise & -- & 60 & 4 & 3 & 10 & 2 & 1.50 & 0.80 & 0.45 & 1.002\\
		(d) Real-world noise & -- & 60 & 6 & 5 & 10 & 0 & 2.00 & 2.30 & 0.90 & 1.002\\
		\bottomrule
	\end{tabularx}
	\label{tab_par_set}
\end{table*}
\begin{figure}
    \centering
    \captionsetup[subfigure]{captionskip=0pt, farskip=0pt}
    \subfloat[Kodak24]{
        \includegraphics[width=0.5\linewidth]{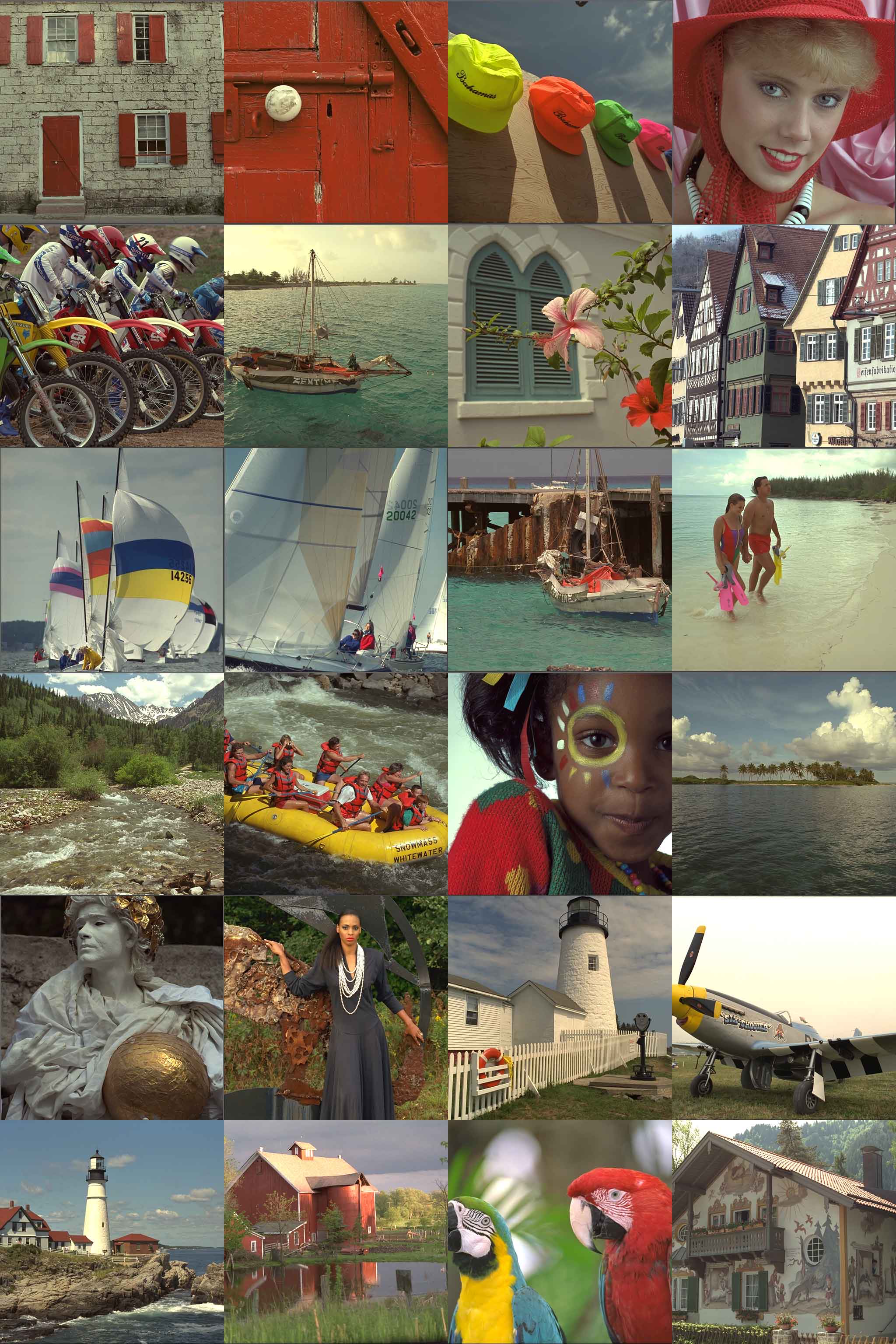}
    }%
    \subfloat[CC15]{
        \includegraphics[width=0.45\linewidth]{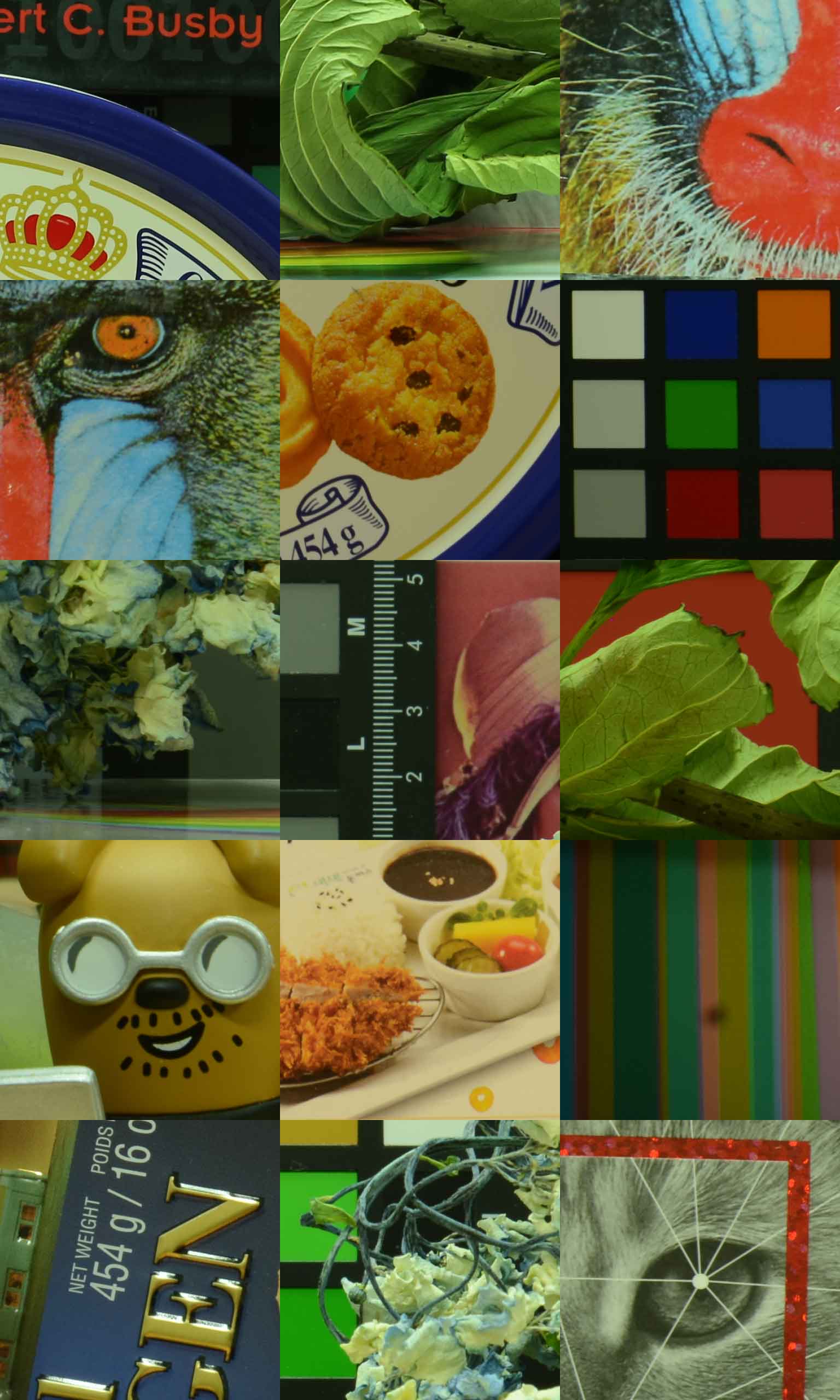}
    }%
    \caption{{{The ground truth images of the two dataset (enumerated from left-to-right and top--to-bottom).}}}
    \label{fig_dataset}
\end{figure}
\subsection{Spatially Invariant Noise Removal}
\label{sec_exp1}
{{%
The spatially invariant noise removal experiments are carried out on the Kodak24 dataset\footnote{{{http://r0k.us/graphics/kodak/}}}. 
It includes 24 noise-free color images, as shown in Fig. \ref{fig_dataset}(a).
}}%
The noise is synthetic, spatially invariant, but has cross-channel difference, as shown in Fig. \ref{fig_spatial_var}(b). 
The corrupted images are generated by zero-mean Gaussian noise with standard deviation $[\sigma_{r\_0}; \sigma_{g\_0}; \sigma_{b\_0}] \in \lbrace [20; 35; 5], [30; 10; 50] \rbrace$. 
For CBM3D, NGmeet and DLRQP, a single noise standard deviation should be given for denoising, which is set as 
\begin{equation}
	\sigma = \sqrt{(\sigma_{r\_0}^2 + \sigma_{g\_0}^2 + \sigma_{b\_0}^2)/3}. \label{eq_single_noise}
\end{equation}
Formula \eqref{eq_single_noise} is also used to calculate the training noise levels for the Restormer. 
The parameter settings of the proposed DtNFM method are listed in Table \ref{tab_par_set}(a) and Table \ref{tab_par_set}(b). 
\par
Table \ref{tab_syn_235} and Table \ref{tab_syn_315} display the PSNR and SSIM results of all competing methods. 
The best results are accentuated in bold. 
Under the noise $[\sigma_{r\_0}; \sigma_{g\_0}; \sigma_{b\_0}] = [20; 35; 5]$, the proposed DtNFM method achieves the highest PSNR on 22 out of 24 images, and the highest SSIM on 23 images. 
Under the noise $[\sigma_{r\_0}; \sigma_{g\_0}; \sigma_{b\_0}] = [30; 10; 50]$, 
The proposed DtNFM achieves the highest PSNR and SSIM on all of the 24 images. 
Focusing on the averages, we list the improvements of DtNFM over other methods at Table \ref{tab_improvements}(a) and Table \ref{tab_improvements}(b). 
As can be seen, DtNFM achieves higher PSNR and SSIM over the state-of-the-art deep learning methods, i.e., DRUNet and Restormer. 
DtNFM also outperforms its state-of-the-art counterparts, i.e., MCWNNM, MCWSNM and NNFNM. 
In terms of running times, the deep learning methods are significantly faster than the proposed DtNFM. 
However, please note that only their testing time is taken into account, and the training time is not considered. 
In contrast, the proposed DtNFM method does not have the process of training. 
\begin{figure}[tb] 
	\captionsetup[subfigure]{captionskip=0pt, farskip=0pt}
	\subfloat[]{
		\hspace{-2mm}
		\includegraphics[width=0.16\linewidth]{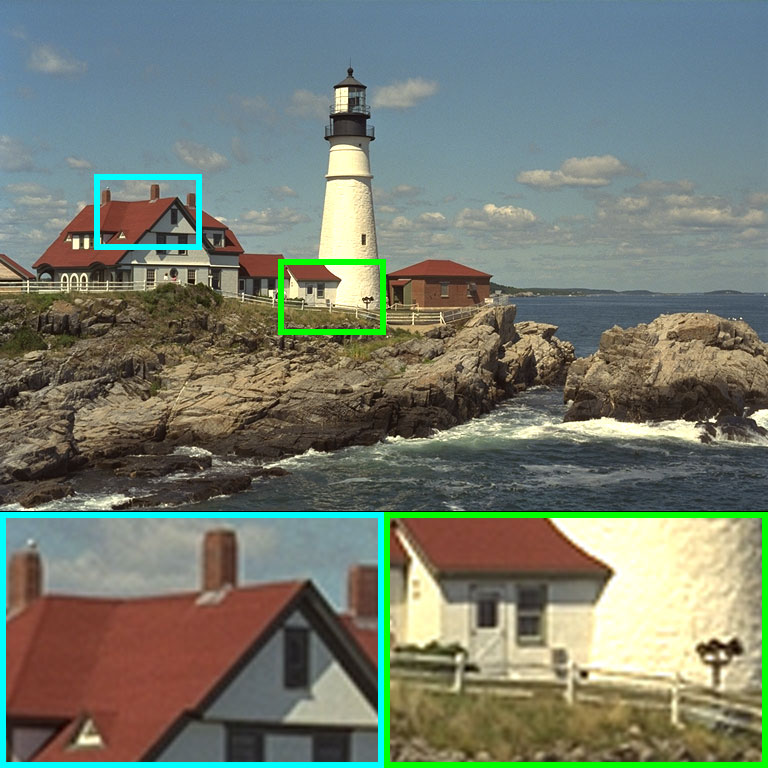}
		\hspace{-2mm}
	}%
	\subfloat[]{
		\hspace{-2mm}
		\includegraphics[width=0.16\linewidth]{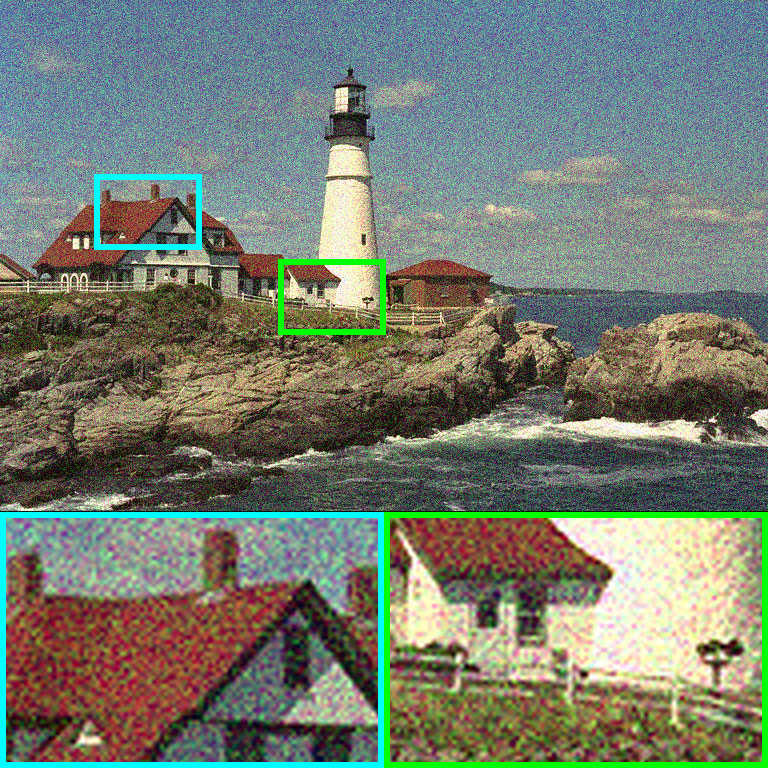}
		\hspace{-2mm}
	}%
	\subfloat[]{
		\hspace{-2mm}
		\includegraphics[width=0.16\linewidth]{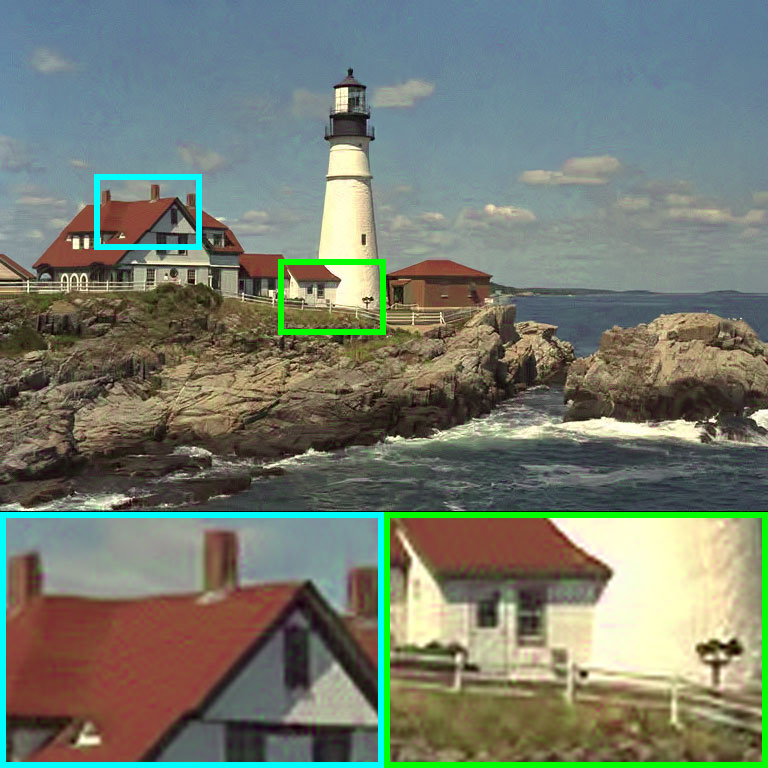}
		\hspace{-2mm}
	}%
	\subfloat[]{
		\hspace{-2mm}
		\includegraphics[width=0.16\linewidth]{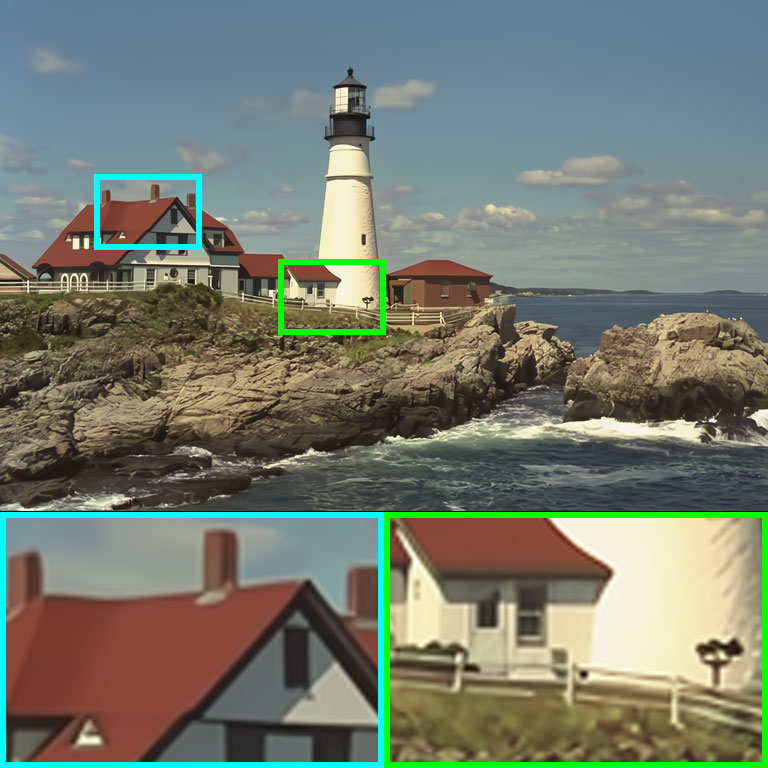}
		\hspace{-2mm}
	}%
	\subfloat[]{
		\hspace{-2mm}
		\includegraphics[width=0.16\linewidth]{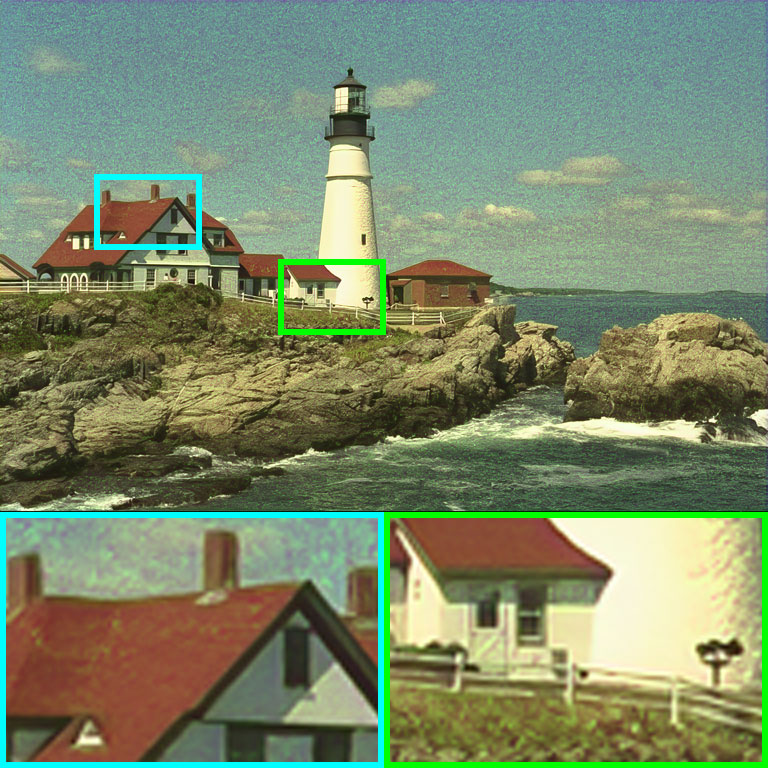}
		\hspace{-2mm}
	}%
	\subfloat[]{
		\hspace{-2mm}
		\includegraphics[width=0.16\linewidth]{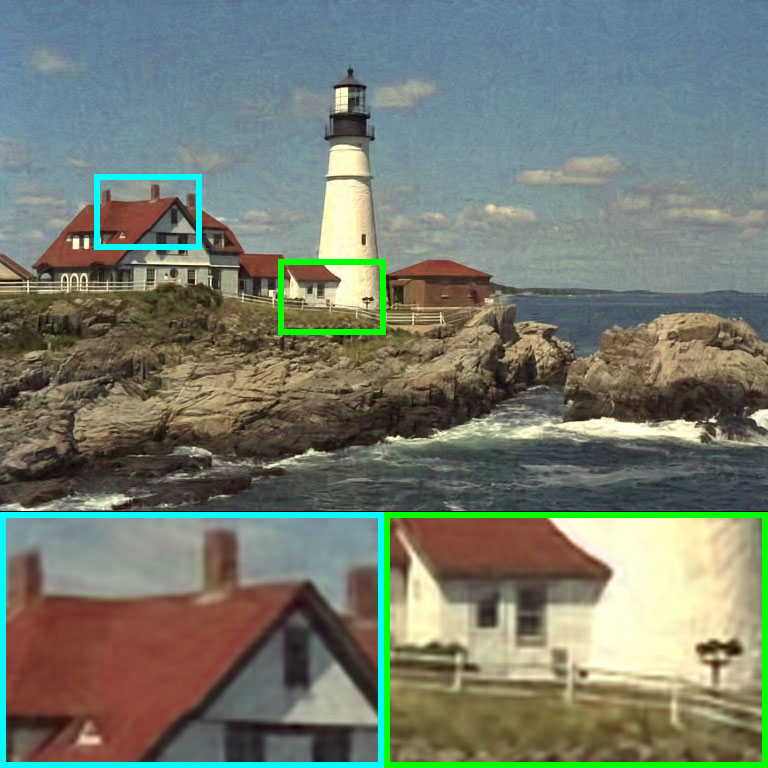}
		\hspace{-2mm}
	}%
	
	\subfloat[]{
		\hspace{-2mm}
		\includegraphics[width=0.16\linewidth]{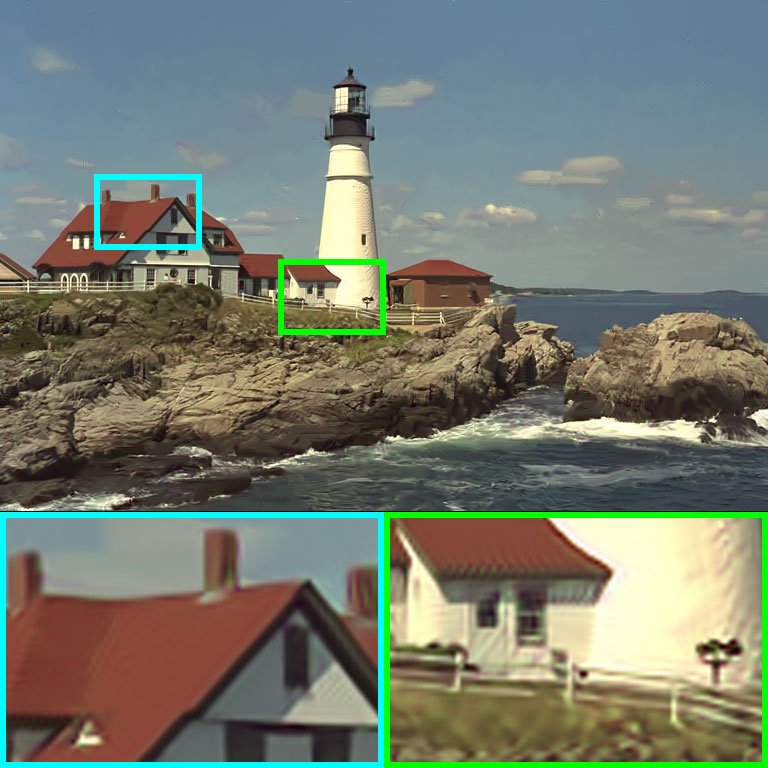}
		\hspace{-2mm}
	}%
	\subfloat[]{
		\hspace{-2mm}
		\includegraphics[width=0.16\linewidth]{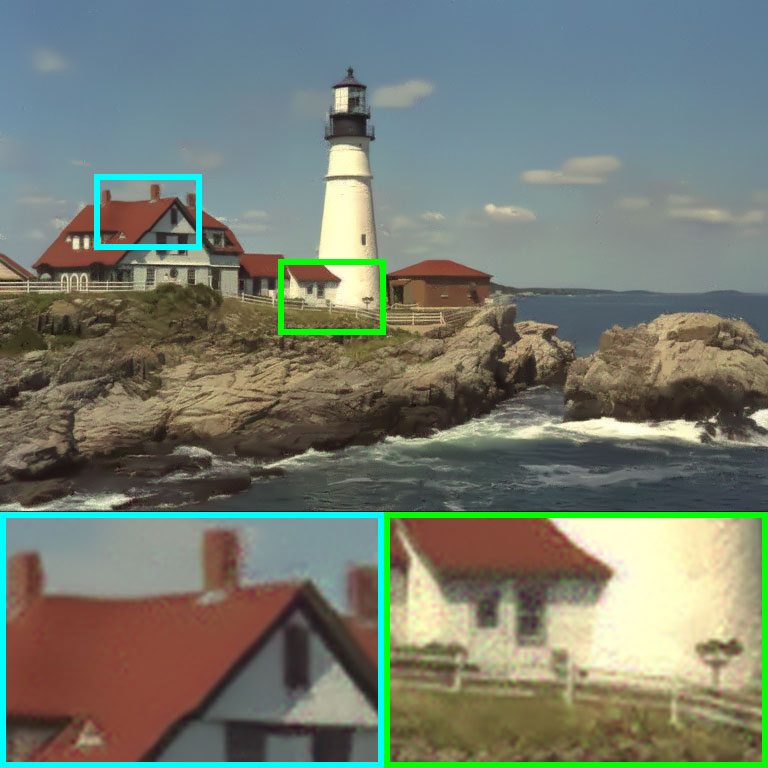}
		\hspace{-2mm}
	}%
	\subfloat[]{
		\hspace{-2mm}
		\includegraphics[width=0.16\linewidth]{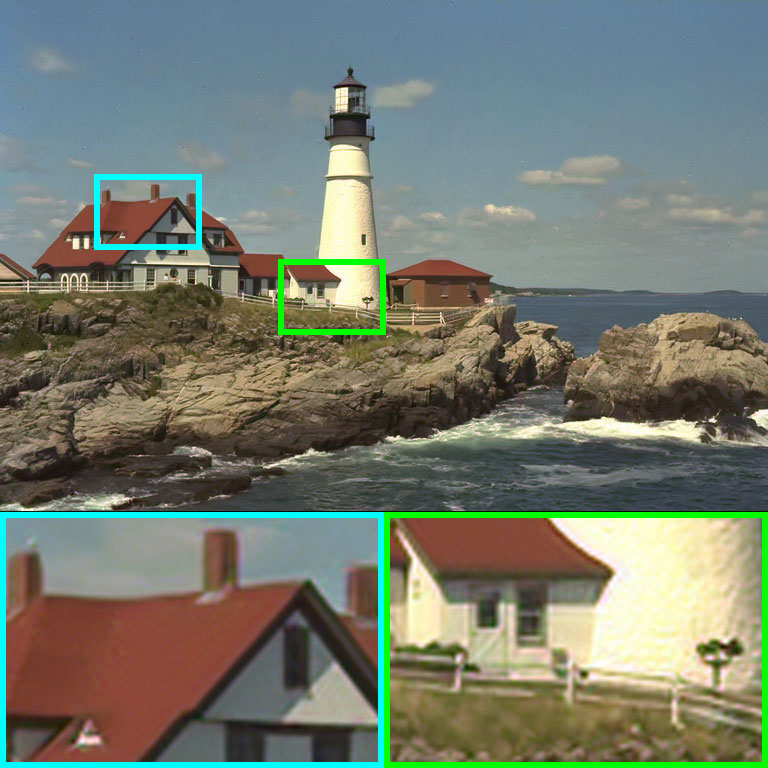}
		\hspace{-2mm}
	}%
	\subfloat[]{
		\hspace{-2mm}
		\includegraphics[width=0.16\linewidth]{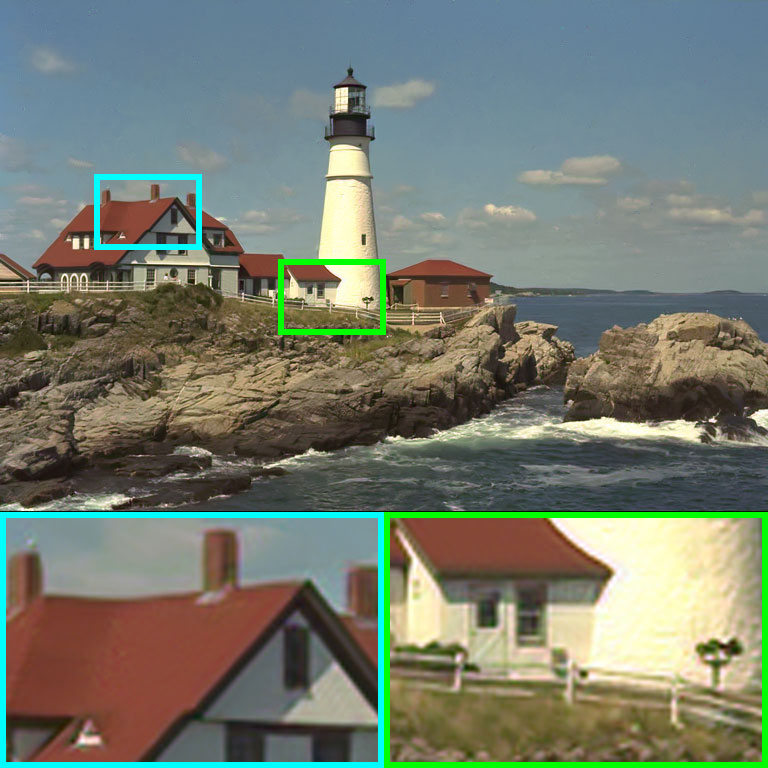}
		\hspace{-2mm}
	}%
	\subfloat[]{
		\hspace{-2mm}
		\includegraphics[width=0.16\linewidth]{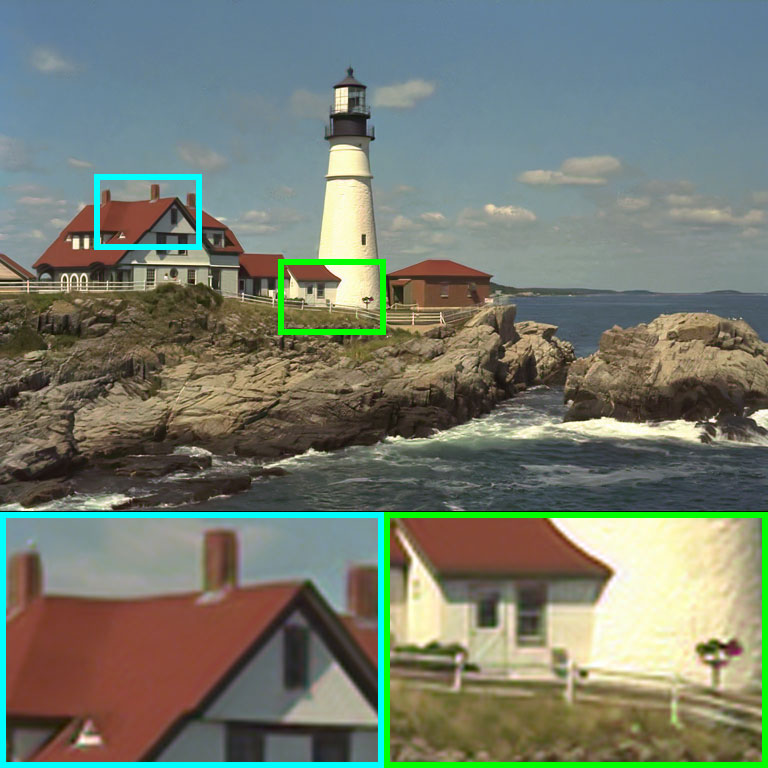}
		\hspace{-2mm}
	}%
	\subfloat[]{
		\hspace{-2mm}
		\includegraphics[width=0.16\linewidth]{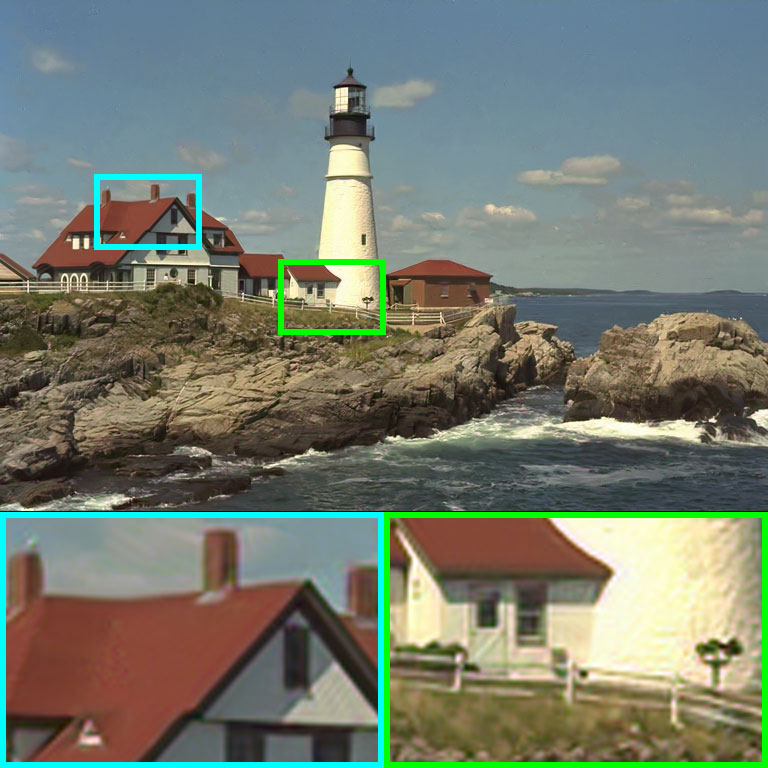}
		\hspace{-2mm}
	}%
	\captionsetup{format=plain} 
	\caption{Denoised results on ``kodim21'' ($[\sigma_{r\_0}; \sigma_{g\_0}; \sigma_{b\_0}] = [20;35;5]$). (a) Ground truth. (b) Noisy observation (20.73, 0.5010). (c) CBM3D (30.34, 0.8376). (d) DRUNet (31.34, 0.8904). (e) Restormer (24.46 0.7154). (f) HLTA-GN (30.08, 0.8501). (g) NGmeet (30.14, 0.8619). (h) DLRQP (28.18, 0.8264). (k) MCWNNM (32.48, 0.8974). (j) MCWSNM (32.67, 0.9106). (k) NNFNM (32.62, 0.9096). (l) DtNFM (\textbf{33.01}, \textbf{0.9162}).}
	\label{fig_235_8}
\end{figure}%
\begin{figure}[tb] 
	\captionsetup[subfigure]{captionskip=0pt, farskip=0pt}
	\subfloat[]{
		\hspace{-2mm}
		\includegraphics[width=0.16\linewidth]{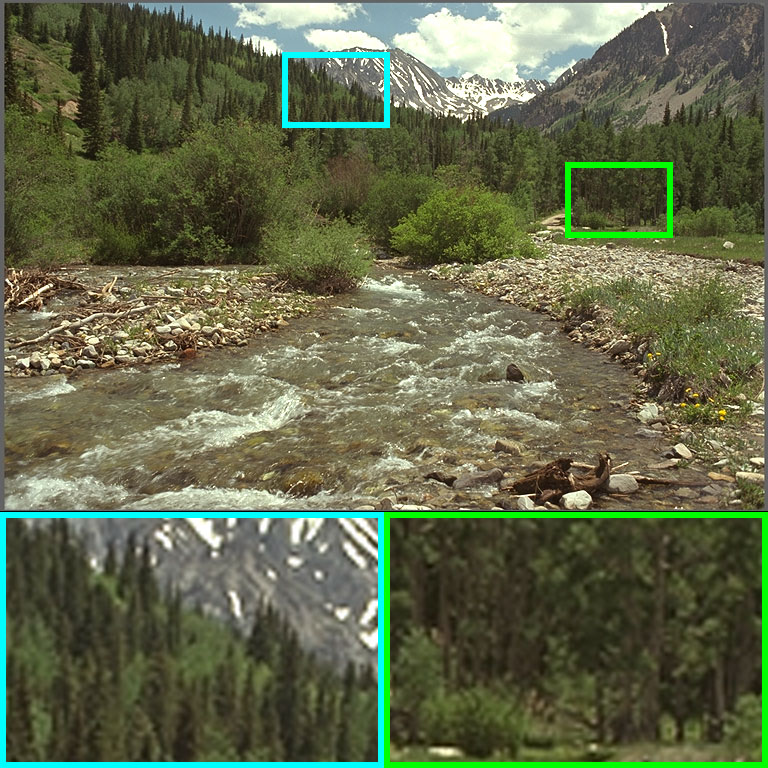}
		\hspace{-2mm}
	}%
	\subfloat[]{
		\hspace{-2mm}
		\includegraphics[width=0.16\linewidth]{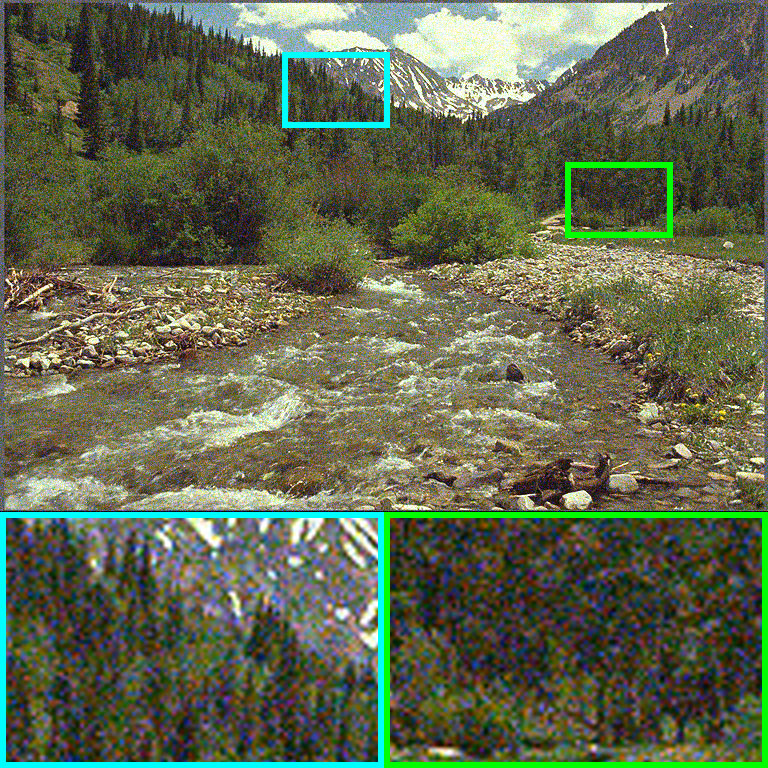}
		\hspace{-2mm}
	}%
	\subfloat[]{
		\hspace{-2mm}
		\includegraphics[width=0.16\linewidth]{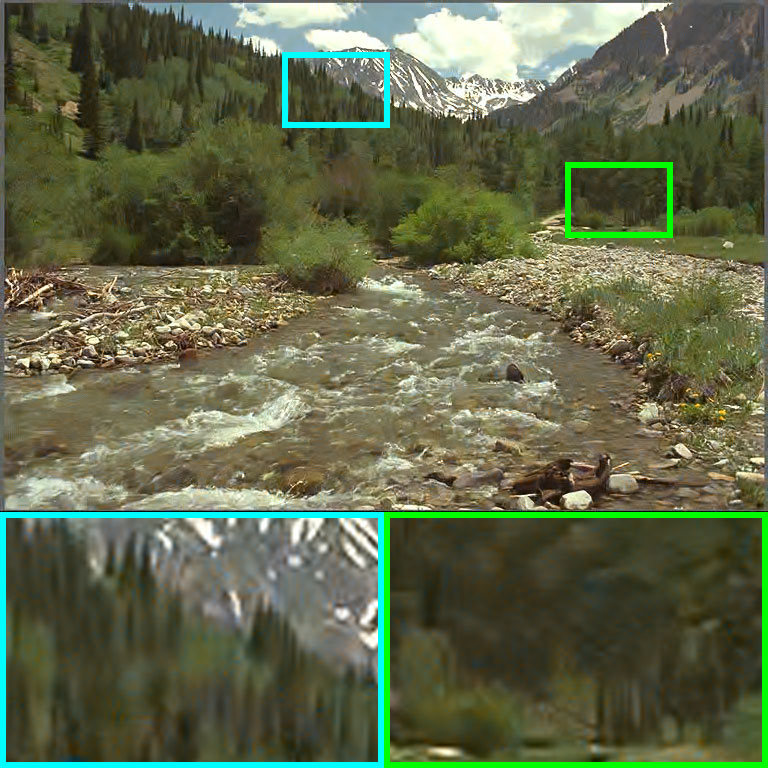}
		\hspace{-2mm}
	}%
	\subfloat[]{
		\hspace{-2mm}
		\includegraphics[width=0.16\linewidth]{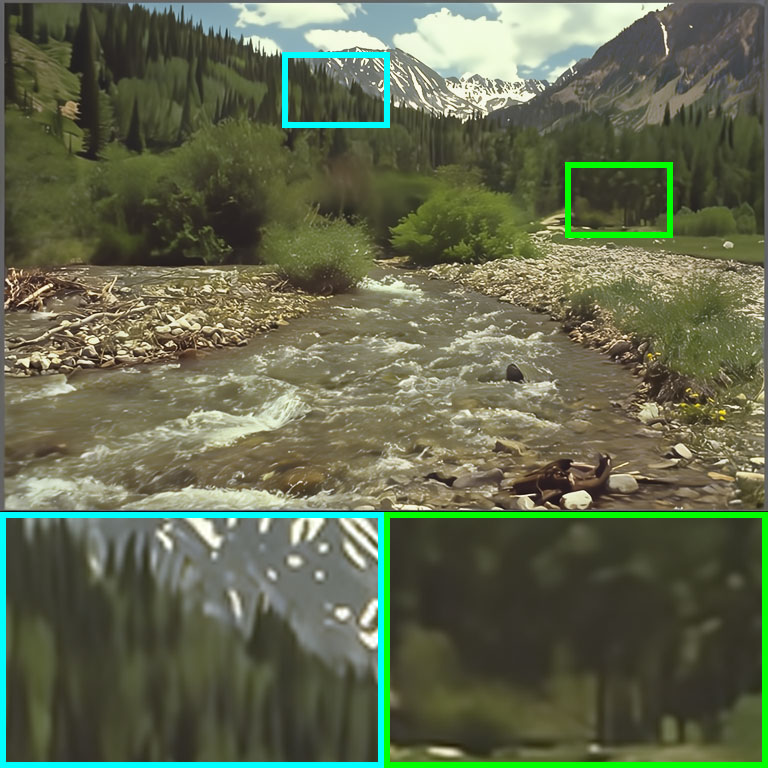}
		\hspace{-2mm}
	}%
	\subfloat[]{
		\hspace{-2mm}
		\includegraphics[width=0.16\linewidth]{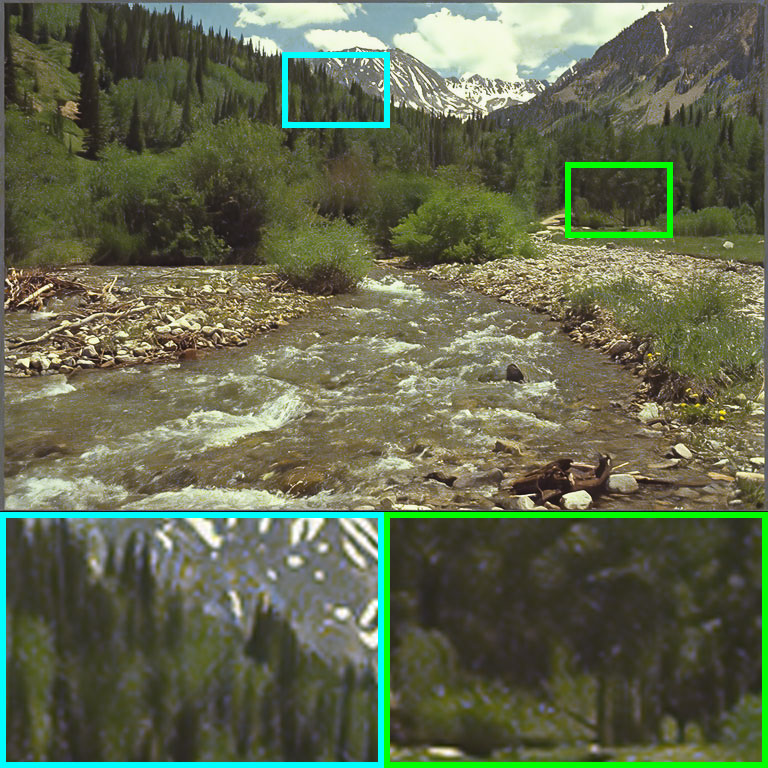}
		\hspace{-2mm}
	}%
	\subfloat[]{
		\hspace{-2mm}
		\includegraphics[width=0.16\linewidth]{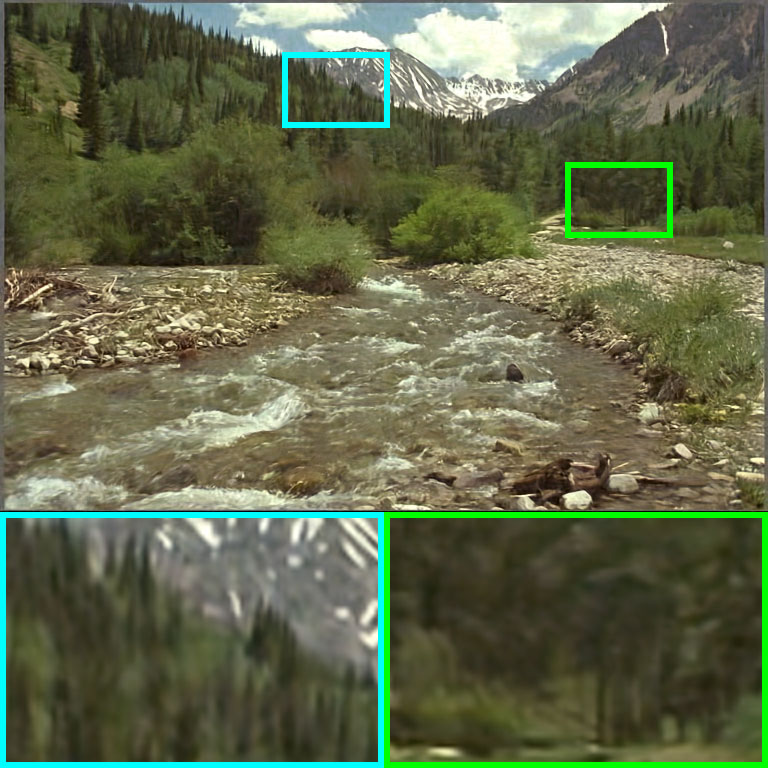}
		\hspace{-2mm}
	}%
	
	\subfloat[]{
		\hspace{-2mm}
		\includegraphics[width=0.16\linewidth]{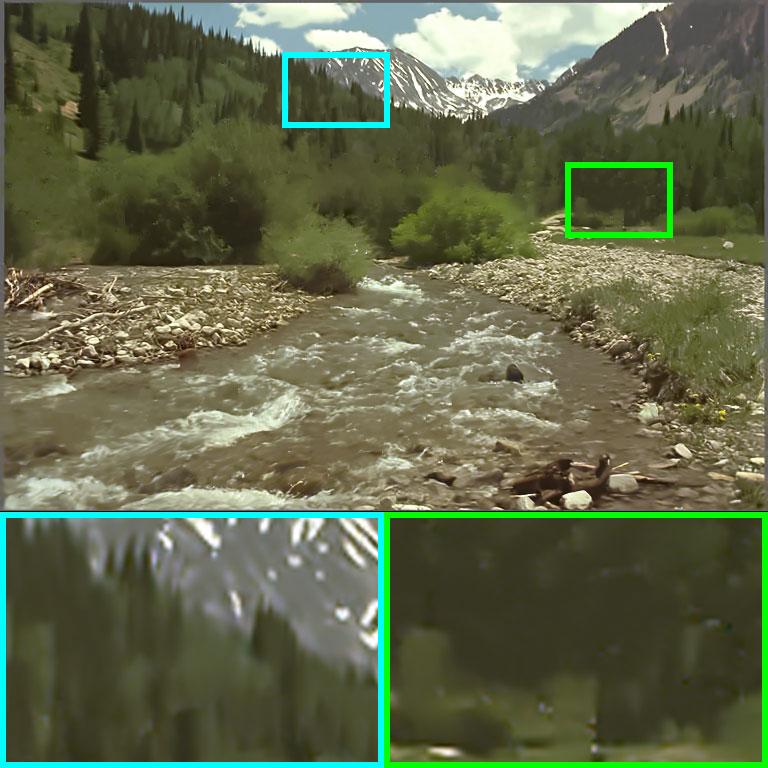}
		\hspace{-2mm}
	}%
	\subfloat[]{
		\hspace{-2mm}
		\includegraphics[width=0.16\linewidth]{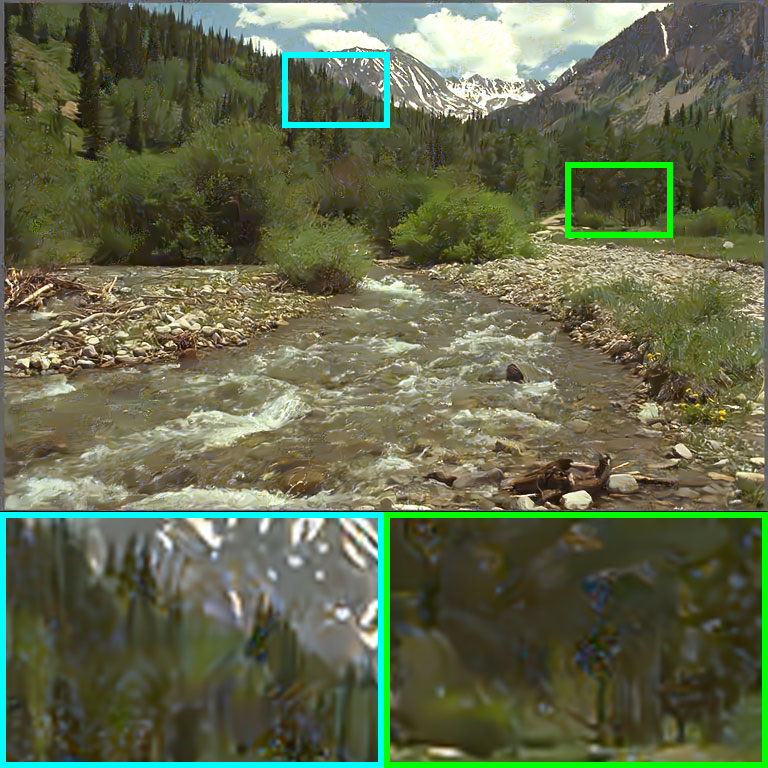}
		\hspace{-2mm}
	}%
	\subfloat[]{
		\hspace{-2mm}
		\includegraphics[width=0.16\linewidth]{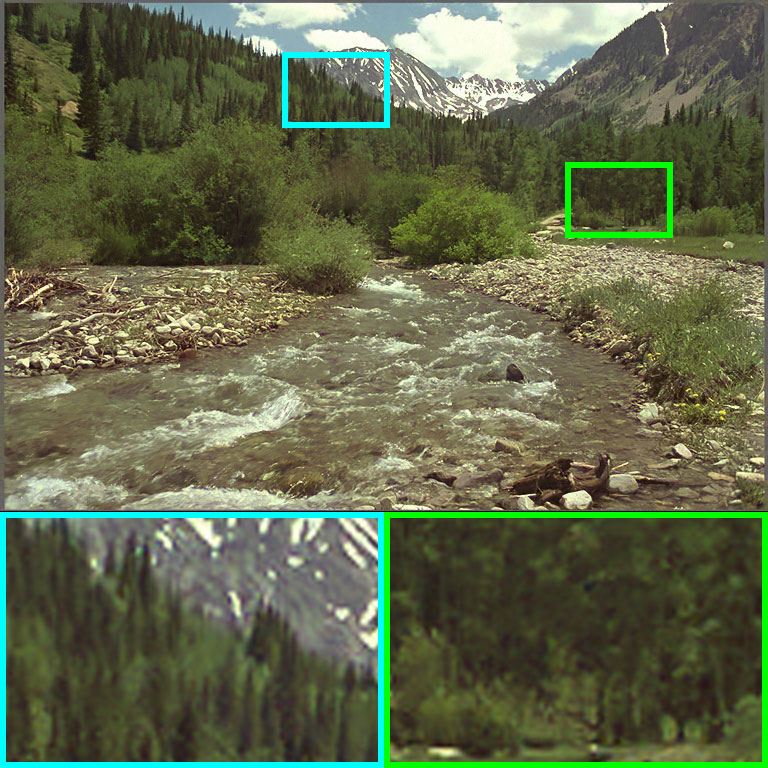}
		\hspace{-2mm}
	}%
	\subfloat[]{
		\hspace{-2mm}
		\includegraphics[width=0.16\linewidth]{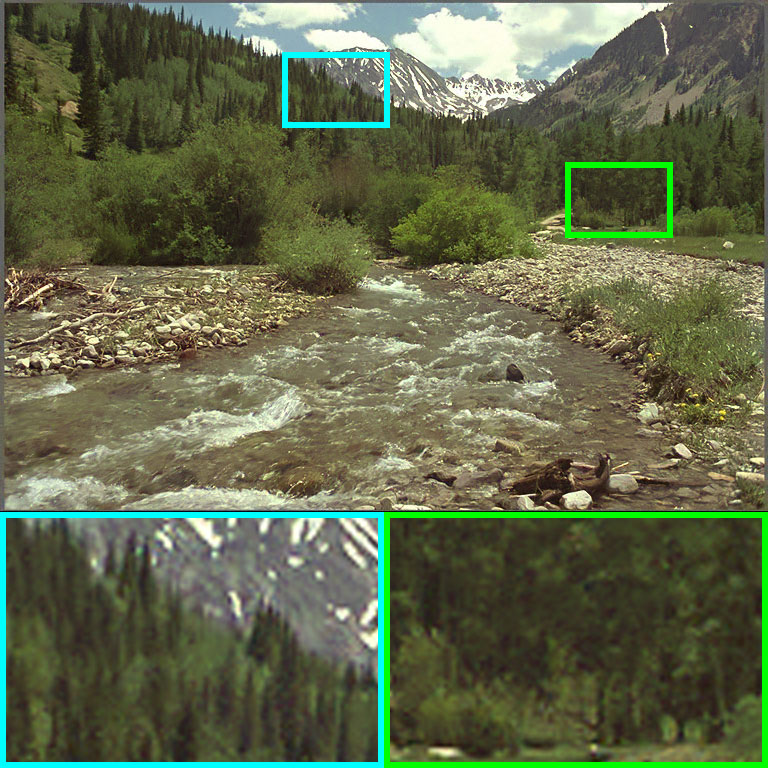}
		\hspace{-2mm}
	}%
	\subfloat[]{
		\hspace{-2mm}
		\includegraphics[width=0.16\linewidth]{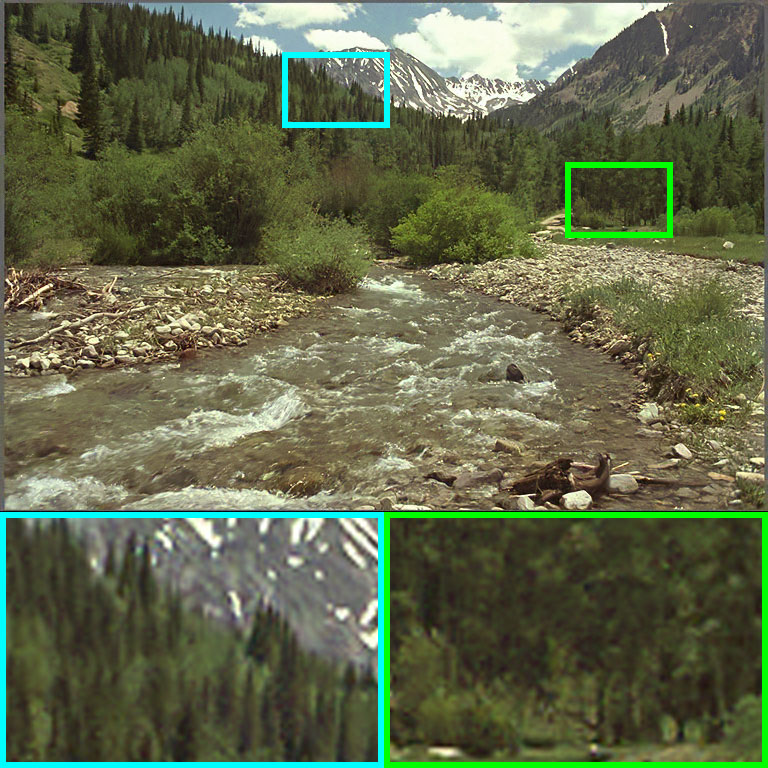}
		\hspace{-2mm}
	}%
	\subfloat[]{
		\hspace{-2mm}
		\includegraphics[width=0.16\linewidth]{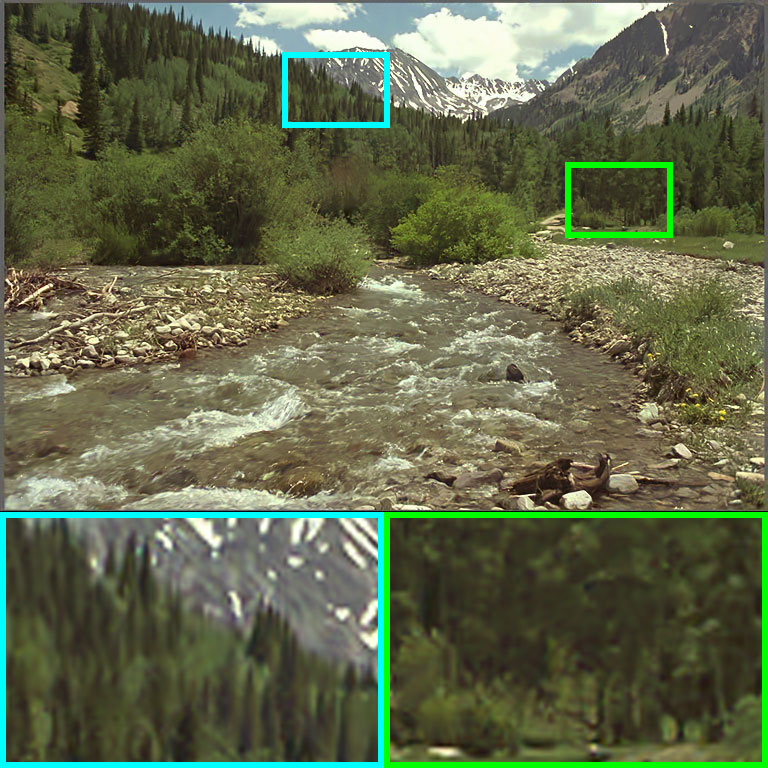}
		\hspace{-2mm}
	}%
	\captionsetup{format=plain} 
	\caption{ Denoised results on ``kodim13'' ($[\sigma_{r\_0}; \sigma_{g\_0}; \sigma_{b\_0}] = [30;10;50]$). (a) Ground truth. (b) Noisy observation (17.46, 0.5251). (c) CBM3D (25.45, 0.7100). (d) DRUNet (25.51, 0.6984). (e) Restormer (23.58, 0.6942). (f) HLTA-GN (25.05, 0.7228). (g) NGmeet (24.68, 0.6427). (h) DLRQP (25.38, 0.7233). (k) MCWNNM (27.07, 0.8094). (j) MCWSNM (27.25, 0.8049). (k) NNFNM (27.19, 0.8152). (l) DtNFM (\textbf{27.88} \textbf{0.8360}). }
	\label{fig_315_13}
\end{figure}%
\begin{figure}[tb] 
	\captionsetup[subfigure]{captionskip=0pt, farskip=0pt}
	\subfloat[]{
		\hspace{-2mm}
		\includegraphics[width=0.16\linewidth]{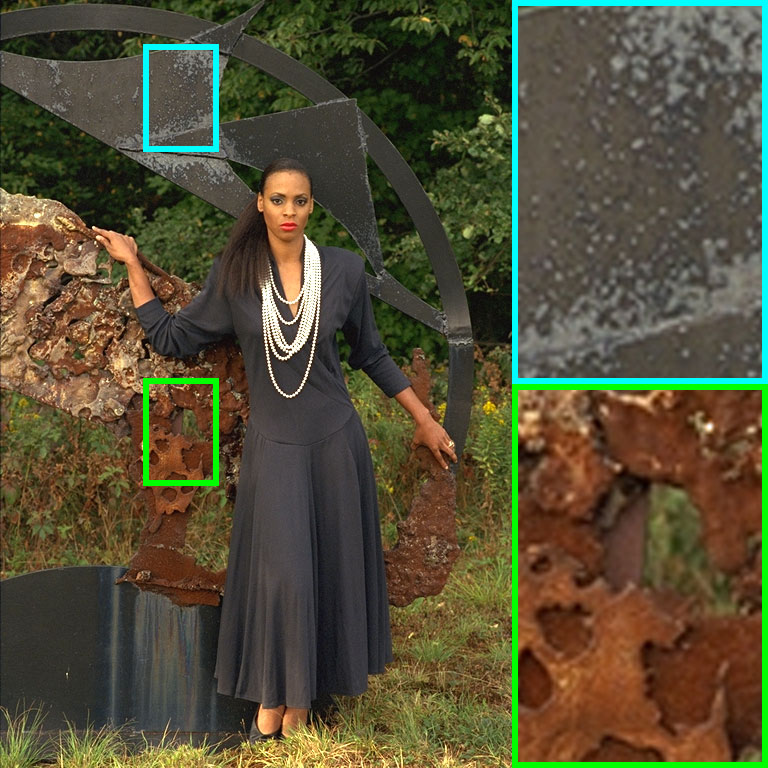}
		\hspace{-2mm}
	}%
	\subfloat[]{
		\hspace{-2mm}
		\includegraphics[width=0.16\linewidth]{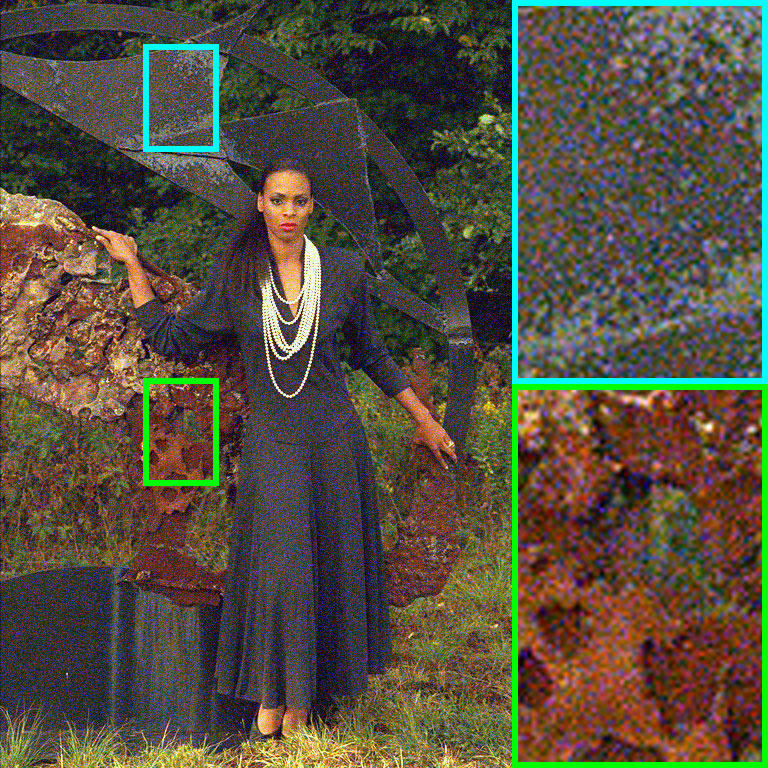}
		\hspace{-2mm}
	}%
	\subfloat[]{
		\hspace{-2mm}
		\includegraphics[width=0.16\linewidth]{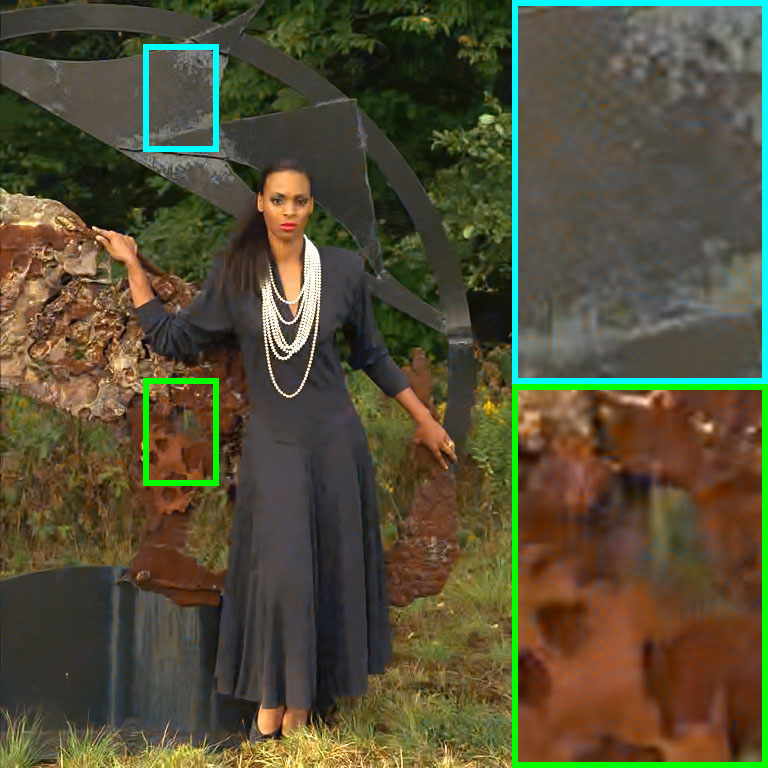}
		\hspace{-2mm}
	}%
	\subfloat[]{
		\hspace{-2mm}
		\includegraphics[width=0.16\linewidth]{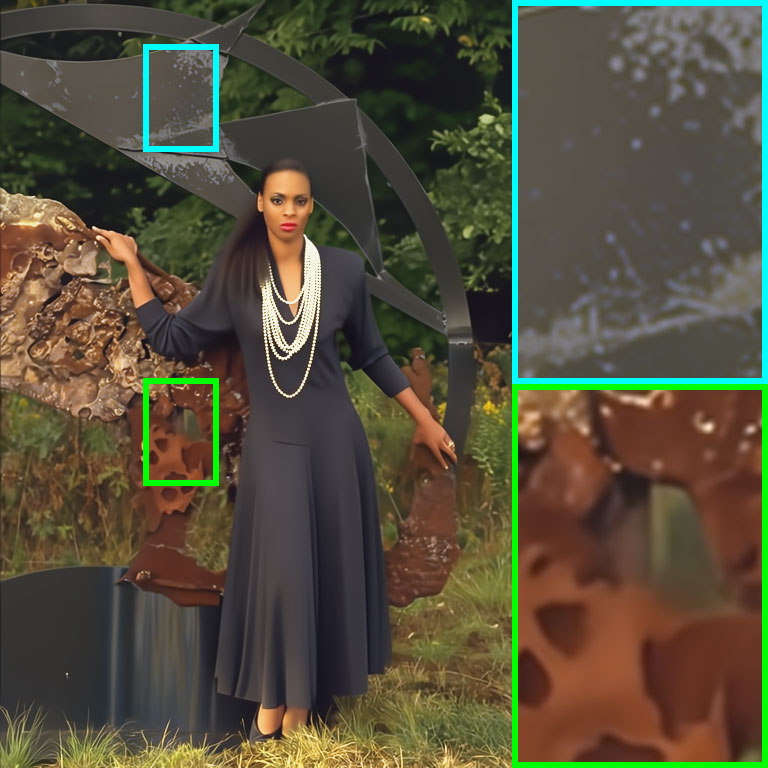}
		\hspace{-2mm}
	}%
	\subfloat[]{
		\hspace{-2mm}
		\includegraphics[width=0.16\linewidth]{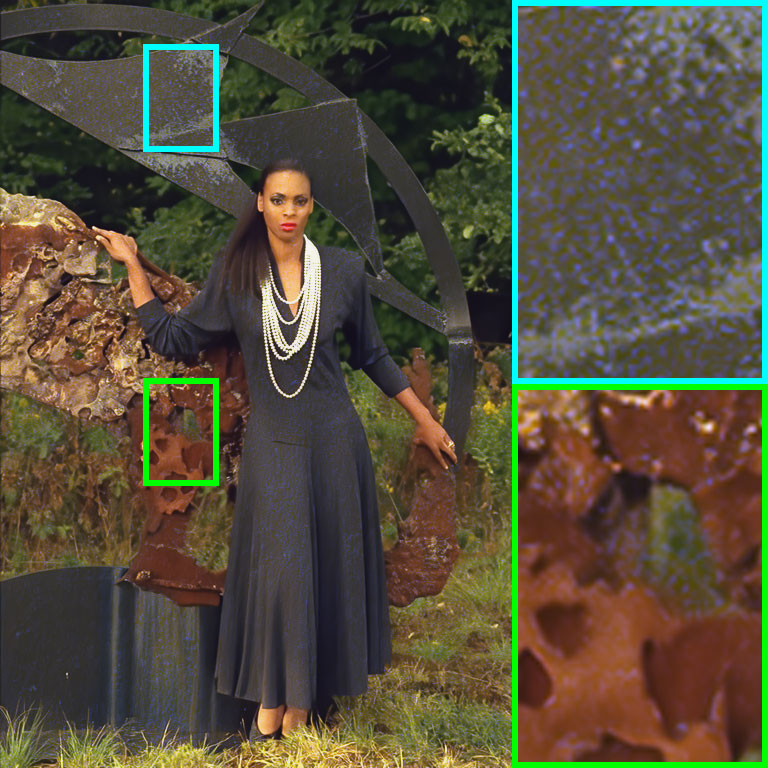}
		\hspace{-2mm}
	}%
	\subfloat[]{
		\hspace{-2mm}
		\includegraphics[width=0.16\linewidth]{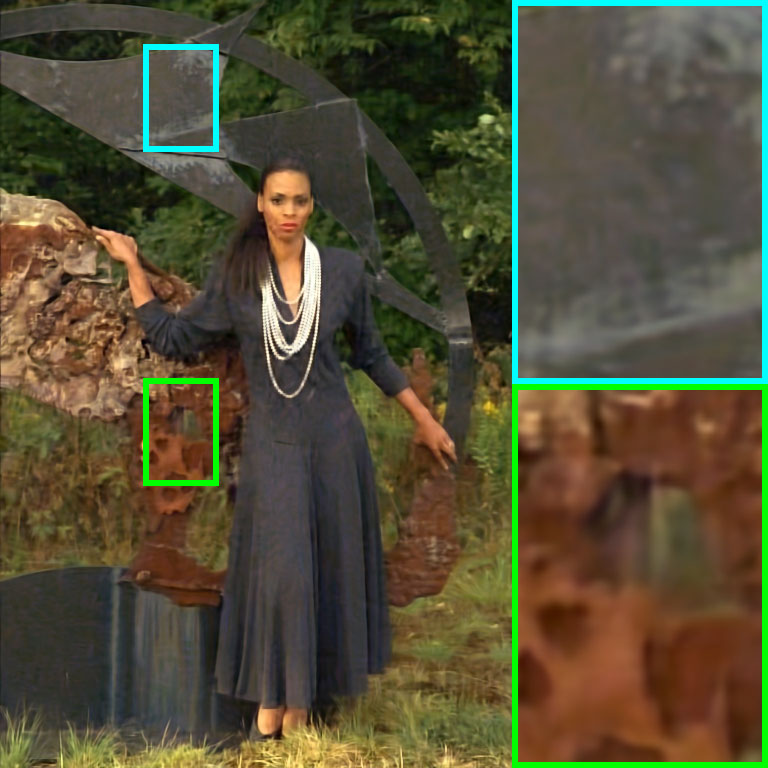}
		\hspace{-2mm}
	}%
	
	\subfloat[]{
		\hspace{-2mm}
		\includegraphics[width=0.16\linewidth]{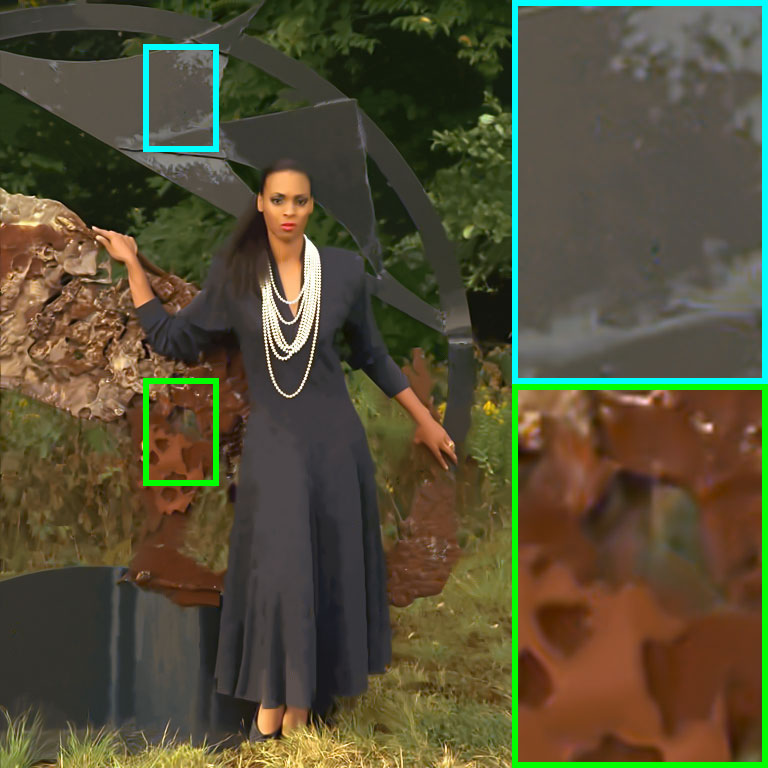}
		\hspace{-2mm}
	}%
	\subfloat[]{
		\hspace{-2mm}
		\includegraphics[width=0.16\linewidth]{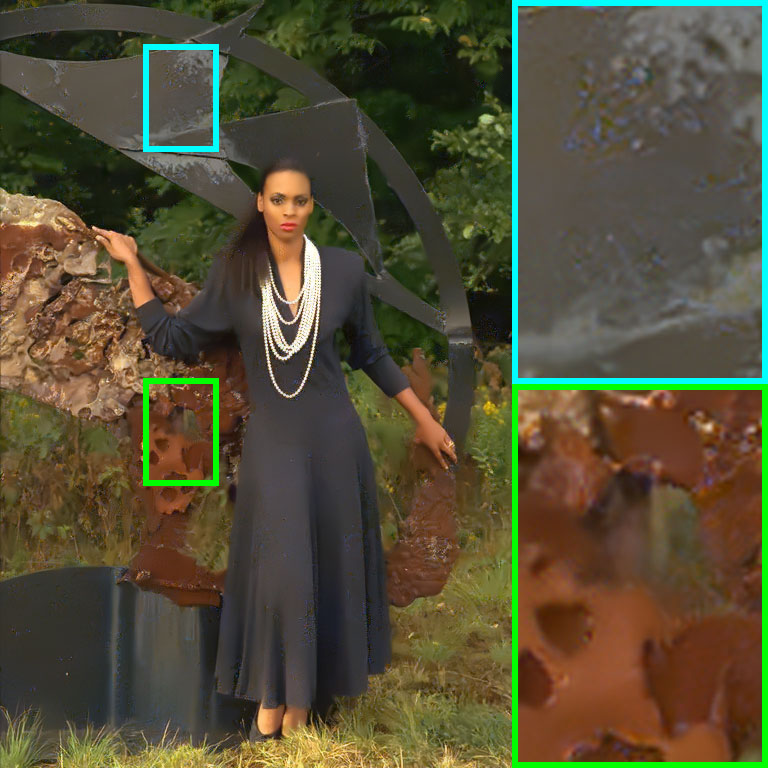}
		\hspace{-2mm}
	}%
	\subfloat[]{
		\hspace{-2mm}
		\includegraphics[width=0.16\linewidth]{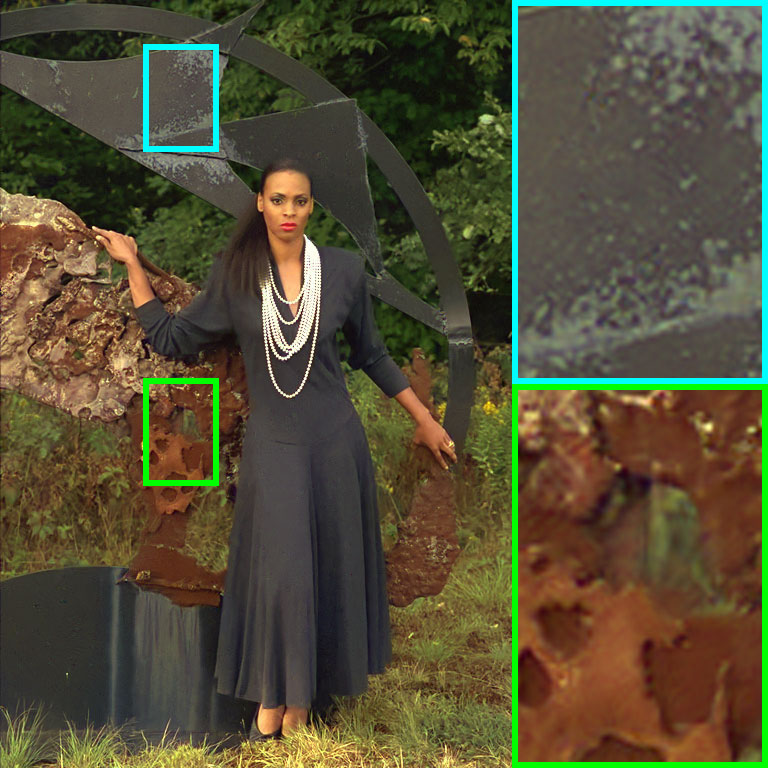}
		\hspace{-2mm}
	}%
	\subfloat[]{
		\hspace{-2mm}
		\includegraphics[width=0.16\linewidth]{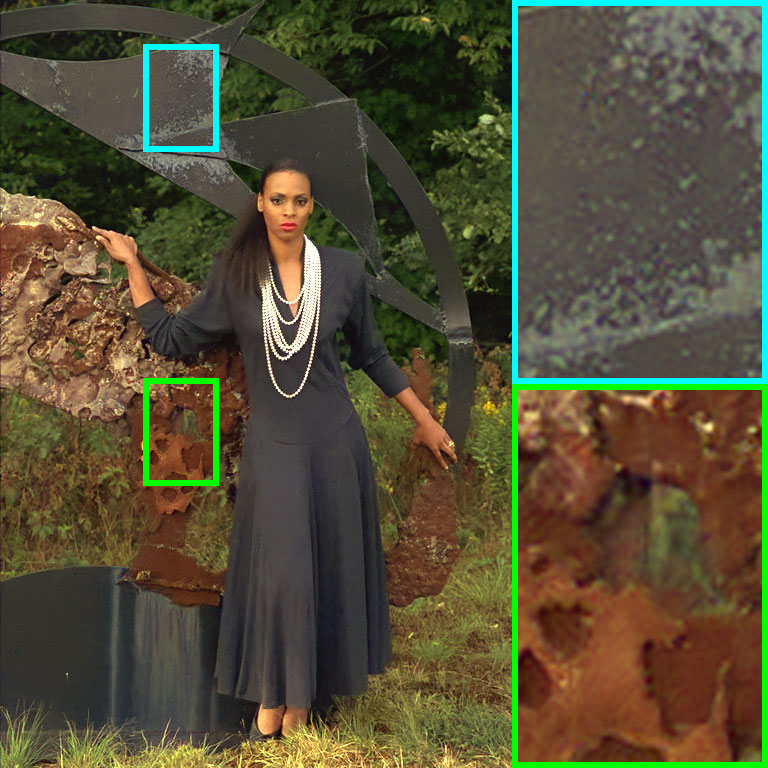}
		\hspace{-2mm}
	}%
	\subfloat[]{
		\hspace{-2mm}
		\includegraphics[width=0.16\linewidth]{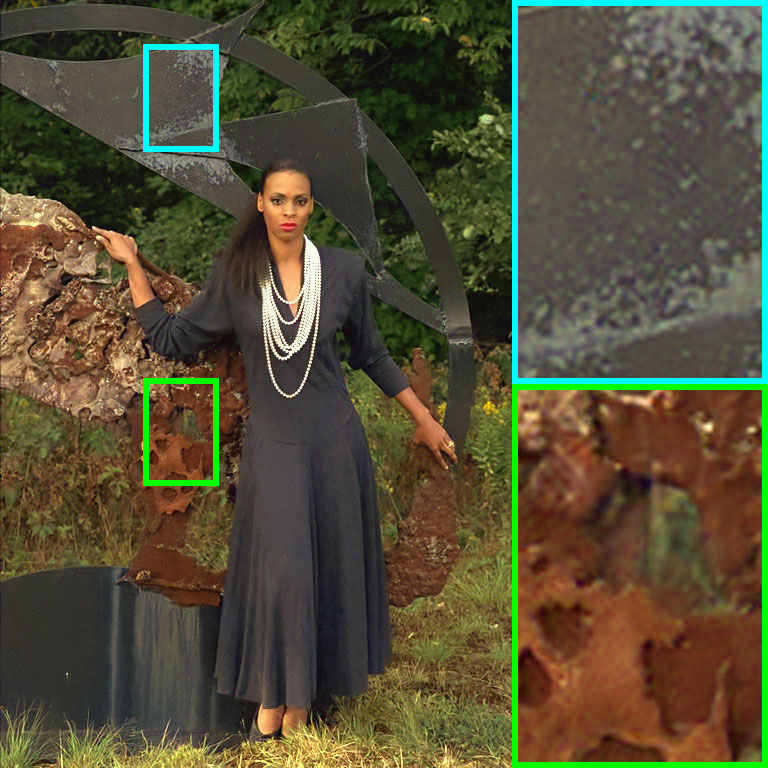}
		\hspace{-2mm}
	}%
	\subfloat[]{
		\hspace{-2mm}
		\includegraphics[width=0.16\linewidth]{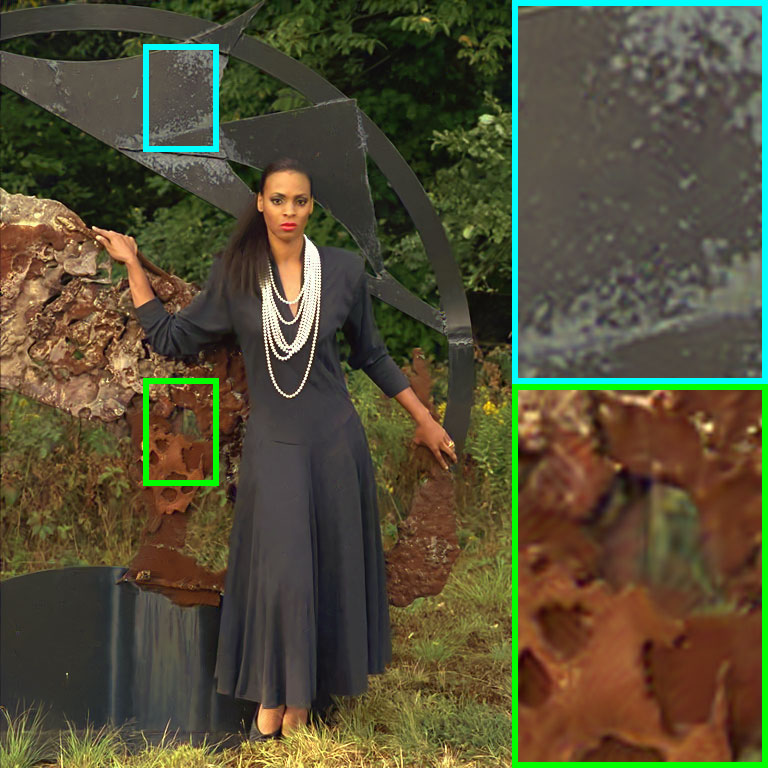}
		\hspace{-2mm}
	}%
	\captionsetup{format=plain} 
	\caption{ Denoised results on ``kodim18'' ($[\sigma_{r\_0}; \sigma_{g\_0}; \sigma_{b\_0}] = [30;10;50]$). (a) Ground truth. (b) Noisy observation (17.46, 0.3967). (c) CBM3D (27.54, 0.7386). (d) DRUNet (27.60, 0.7664). (e) Restormer (24.67, 0.6958). (f) HLTA-GN (26.30, 0.7210). (g) NGmeet (27.07, 0.7231). (h) DLRQP (27.01, 0.7209). (k) MCWNNM (29.33, 0.8202). (j) MCWSNM (29.42, 0.8230). (k) NNFNM (29.47, 0.8287). (l) DtNFM (\textbf{30.12}, \textbf{0.8483}). }
	\label{fig_315_18}%
\end{figure}
\par
The visual comparisons are shown in Fig. \ref{fig_235_8} $\sim$ Fig. \ref{fig_315_18}. 
{{%
We can find that the Restormer remains too much noise in the green channel (Fig. \ref{fig_235_8}(e)) or in the blue channel (Fig. \ref{fig_315_13}(e) and Fig. \ref{fig_315_18}(e)). 
That is because it is inadequate to handle the cross-channel difference of the noise. 
In Fig. \ref{fig_315_13}, DtNFM achieves better protection on the complex edges of trees and shrub. 
In contrast, the CBM3D, DRUNet and NGmeet over-smooth the image. 
The Restormer and DLRQP remains many blue spots in their denoised results. 
In Fig. \ref{fig_315_18}, both the textures of the sculpture and the detailed structure of the stone are well recovered by DtNFM. 
In contrast, the HLTA-GN generates some wrinkle texture, the NGmeet and DLRQP over-smooth the image. }}
In summary, the proposed DtNFM method achieves the best performance in both numerical results and visual quality. 
%
%
%
\subsection{Spatially Variant Noise Removal}
In the most previous section, the synthetic noise is spatially invariant. 
However, in practical applications, the noise does demonstrate spatial variation in most cases. 
Therefore, in this section we consider the noise possessing both cross-channel difference and spatial variation. 
To synthesize such kind of noise, we first choose the upper-bounds of the noise standard deviation, denoted as $[\sigma_{r\_0}; \sigma_{g\_0}; \sigma_{b\_0}]$. 
Then, we generate a map $\mathcal{M}(x,y)$ where $x,y \in \lbrace 1,\cdots, 512 \rbrace$, as shown in Fig. \ref{fig_spatial_var}(c). 
Note that $\mathcal{M}(\cdot, \cdot) \in [0,1]$. 
A point $(x,y,\mathcal{M}(x,y))$ means that for the pixel at $x$th row and $y$th column, the noise standard deviation imposed on it is $\mathcal{M}(x,y) \times [\sigma_{r\_0}; \sigma_{g\_0}; \sigma_{b\_0}]$. 
For all competing methods, the input noise standard deviation in channel $c \in \lbrace r,g,b \rbrace$ is given by
\begin{equation}
	\sigma_{c} = \frac{\sigma_{c\_0}}{512^2} \sum_{j=1}^{512} \sum_{i=1}^{512} \mathcal{M}(x,y). 
\end{equation}
\par
The ground truth images are still taken from Kodak24 data set. 
And they are cropped to $512\times 512$ pixels in order to match the size of the map $\mathcal{M}$. 
The upper-bounds of noise standard deviation $[\sigma_{r\_0}; \sigma_{g\_0}; \sigma_{b\_0}] = [30; 35; 40]$. 
Considering the noise distribution, i.e., the map $\mathcal{M}$, is hard to estimate, we make all the competing methods be unaware of it. 
Consequently, they have to assume that the noise is spatially invariant. 
Thus their flexibility on handling the noise can be fully tested. 
The parameters of DtNFM method are listed in Table \ref{tab_par_set}(c). 
\par
\begin{table*}[ptb] 
	\centering\scriptsize
	\caption{PSNR and SSIM results for all competing methods in the spatially variant noise experiments. Running times is in seconds.}
	\begin{tabularx}{\textwidth}{p{0.3cm}<{\centering}YYYYY YYYYY}
\toprule
& CBM3D & DRUNet & \!{{Restormer}} & \!{{HLTA-GN}} & \!{{NGMeet}} & \!{{DLRQP}} & MCWNNM & MCWSNM & NNFNM & DtNFM\\
\# & PSNR SSIM & PSNR SSIM & PSNR SSIM & PSNR SSIM & PSNR SSIM & PSNR SSIM & PSNR SSIM & PSNR SSIM & PSNR SSIM & PSNR SSIM\\
\hline
1 & 31.73 0.9231 & 31.79 0.9180 & 31.07 0.9231 & 30.46 0.8835 & 31.82 0.9121 & 32.00 0.9322 & 31.59 0.9205 & 29.41 0.8374 & 32.25 0.9277 & \textbf{33.91 0.9421}\\
2 & 32.39 0.8583 & 32.84 0.8557 & 34.10 0.8917 & 32.17 0.8506 & 34.01 0.8835 & 32.81 0.8800 & 34.87 0.8849 & 32.59 0.8364 & 33.43 0.8775 & \textbf{35.65 0.9081}\\
3 & 33.95 0.9175 & 33.85 0.9059 & 36.19 0.9207 & 35.54 0.9151 & 38.01 0.9438 & 33.63 0.8962 & 34.97 0.9186 & 35.44 0.9181 & 33.02 0.8749 & \textbf{38.58 0.9464}\\
4 & 32.70 0.8712 & 32.64 0.8616 & 34.00 0.8951 & 34.54 0.8867 & 35.58 0.9097 & 33.15 0.8838 & 33.09 0.8697 & 33.41 0.8534 & 33.04 0.8654 & \textbf{36.86 0.9177}\\
5 & 32.14 0.9485 & 32.54 0.9465 & 33.82 0.9582 & 28.62 0.8948 & 33.18 0.9529 & 31.75 0.9473 & 32.58 0.9464 & 30.94 0.9116 & 32.09 0.9473 & \textbf{33.87 0.9602}\\
6 & 32.39 0.9017 & 32.50 0.8961 & 33.86 \textbf{0.9291} & 32.56 0.9054 & 33.53 0.9159 & 32.49 0.9088 & 33.98 0.9196 & 31.32 0.8557 & 32.48 0.9059 & \textbf{34.65} 0.9264\\
7 & 33.62 0.9352 & 33.59 0.9317 & 36.65 0.9517 & 33.81 0.9368 & 37.64 \textbf{0.9620} & 33.40 0.9214 & 33.87 0.9273 & 34.63 0.9362 & 32.97 0.9125 & \textbf{37.77} 0.9602\\
8 & 32.25 0.9295 & 32.21 0.9238 & 32.95 0.9376 & 31.00 0.9289 & 34.03 0.9446 & 31.75 0.9311 & 32.78 0.9355 & 31.35 0.9129 & 32.12 0.9277 & \textbf{34.19 0.9496}\\
9 & 33.39 0.9098 & 33.50 0.9068 & 36.14 0.9215 & 35.03 0.9245 & 35.96 0.9263 & 33.24 0.8962 & 34.55 0.9105 & 34.62 0.9137 & 32.82 0.8770 & \textbf{37.03 0.9294}\\
10 & 33.42 0.9013 & 33.33 0.8971 & 35.66 0.9122 & 34.83 0.9179 & 36.58 \textbf{0.9255} & 33.22 0.8866 & 32.78 0.8856 & 34.44 0.8999 & 32.22 0.8604 & \textbf{36.95} 0.9249\\
11 & 32.14 0.8993 & 32.24 0.8914 & 34.11 0.9226 & 31.44 0.8877 & 32.76 0.8986 & 32.32 0.9135 & 31.39 0.8932 & 31.51 0.8574 & 32.09 0.9040 & \textbf{34.22 0.9285}\\
12 & 33.12 0.8846 & 32.88 0.8750 & 34.44 0.8999 & 35.76 0.9140 & 36.56 0.9156 & 33.13 0.8877 & 31.79 0.8682 & 33.72 0.8645 & 33.16 0.8742 & \textbf{36.98 0.9246}\\
13 & 31.25 0.9263 & 31.40 0.9246 & \textbf{31.76} 0.9346 & 28.48 0.8706 & 30.20 0.8568 & 30.99 0.9389 & 30.95 0.9353 & 28.63 0.8141 & 31.29 0.9423 & 31.53 \textbf{0.9449}\\
14 & 31.98 0.9018 & 32.22 0.8963 & \textbf{33.95 0.9308} & 29.91 0.8727 & 32.66 0.9078 & 32.15 0.9118 & 32.48 0.9065 & 30.59 0.8389 & 32.01 0.9082 & 33.87 0.9258\\
15 & 32.45 0.8743 & 32.91 0.8704 & 34.20 0.8955 & 32.18 0.8601 & 33.69 0.8904 & 32.82 0.8894 & 33.86 0.8851 & 32.66 0.8504 & 32.71 0.8747 & \textbf{35.66 0.9109}\\
16 & 32.99 0.8971 & 32.73 0.8854 & 35.19 0.9223 & 34.25 0.8978 & 35.65 0.9157 & 33.05 0.8944 & 33.92 0.8888 & 32.55 0.8564 & 32.34 0.8755 & \textbf{36.57 0.9303}\\
17 & 32.72 0.9005 & 32.80 0.8945 & 34.63 0.9214 & 31.89 0.8944 & \textbf{35.16} 0.9194 & 32.73 0.9068 & 32.03 0.8870 & 32.95 0.8794 & 32.62 0.8900 & 34.71 \textbf{0.9253}\\
18 & 32.10 0.8949 & 32.20 0.8901 & 33.74 \textbf{0.9215} & 28.14 0.8185 & 33.39 0.9125 & 32.21 0.8977 & 32.83 0.9047 & 30.72 0.8403 & 32.53 0.8928 & \textbf{33.93} 0.9204\\
19 & 32.50 0.9263 & 32.78 0.9203 & 34.57 0.9314 & 33.83 0.9295 & 34.40 0.9273 & 32.68 0.9306 & 33.44 0.9208 & 32.67 0.9023 & 32.43 0.9177 & \textbf{35.77 0.9436}\\
20 & 33.33 0.9074 & 34.21 0.9211 & 29.94 0.9086 & 35.33 0.9266 & \textbf{36.25 0.9464} & 33.26 0.8982 & 36.21 0.9413 & 33.51 0.8898 & 32.77 0.8865 & 35.76 0.9374\\
21 & 32.50 0.9134 & 32.60 0.9080 & 34.24 \textbf{0.9333} & 32.36 0.9134 & 33.71 0.9243 & 32.45 0.9090 & 32.29 0.9021 & 31.68 0.8902 & 31.81 0.8903 & \textbf{34.60} 0.9309\\
22 & 32.23 0.8826 & 32.38 0.8790 & 33.28 0.9010 & 33.01 0.8934 & 32.78 0.8864 & 32.49 0.8942 & 32.81 0.8854 & 31.68 0.8402 & 32.76 0.8853 & \textbf{34.97 0.9182}\\
23 & 33.62 0.9006 & 33.34 0.8902 & 35.45 0.9139 & 35.74 0.9223 & 37.78 \textbf{0.9434} & 33.34 0.8847 & 33.18 0.8811 & 35.08 0.9063 & 32.75 0.8657 & \textbf{37.95} 0.9365\\
24 & 32.34 0.9200 & 32.63 0.9173 & 33.02 0.9291 & 31.21 0.9089 & 33.99 0.9234 & 32.48 0.9194 & 32.42 0.9170 & 31.98 0.8896 & 32.48 0.9078 & \textbf{34.78 0.9391}\\
\hline
Avg. & 32.64 0.9052 & 32.75 0.9003 & 34.04 0.9211 & 32.59 0.8981 & 34.56 0.9185 & 32.65 0.9067 & 33.11 0.9056 & 32.42 0.8748 & 32.51 0.8955 & \textbf{35.45 0.9326}\\
Time  & 3.71 & 1.73 & \textbf{1.02} & 114.30 & 225.54 & 414.44 & 845.19 & 254.30 & 509.85 & 834.60\\
\bottomrule
	\end{tabularx}
	\label{tab_sv}
\end{table*}
The PSNR and SSIM results are listed in Table \ref{tab_sv}. 
The proposed DtNFM achieves the highest PSNR on 20 images, and the highest SSIM on 16 images. 
And the average improvements of DtNFM over other methods are listed in Table \ref{tab_improvements}(c). 
It demonstrates the DtNFM model has more flexibility in dealing with the spatially variant noise. 
In contrast, the MCWNNM, MCWSNM and NNFNM achieve sub-optimal performance since they are inadequate to handle the noise difference between the patches in the patch group. 
%
\par
\begin{figure}[tb] 
	\captionsetup[subfigure]{captionskip=0pt, farskip=0pt}
	\subfloat[]{
		\hspace{-2mm}
		\includegraphics[width=0.16\linewidth]{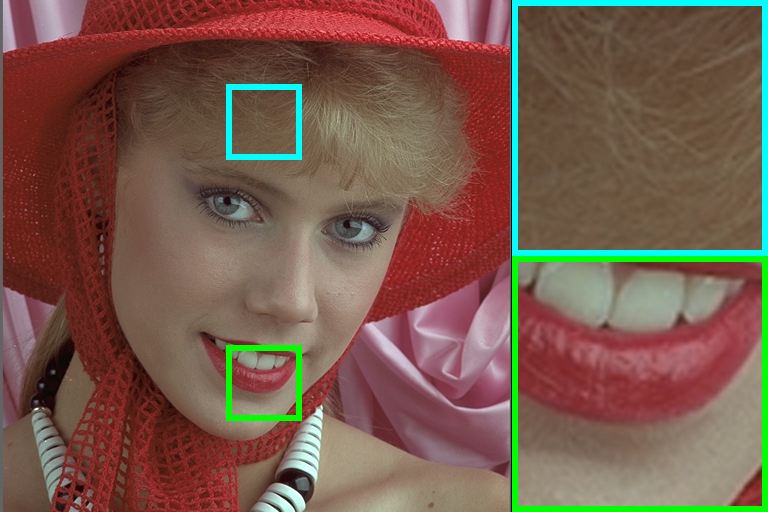}
		\hspace{-2mm}
	}%
	\subfloat[]{
		\hspace{-2mm}
		\includegraphics[width=0.16\linewidth]{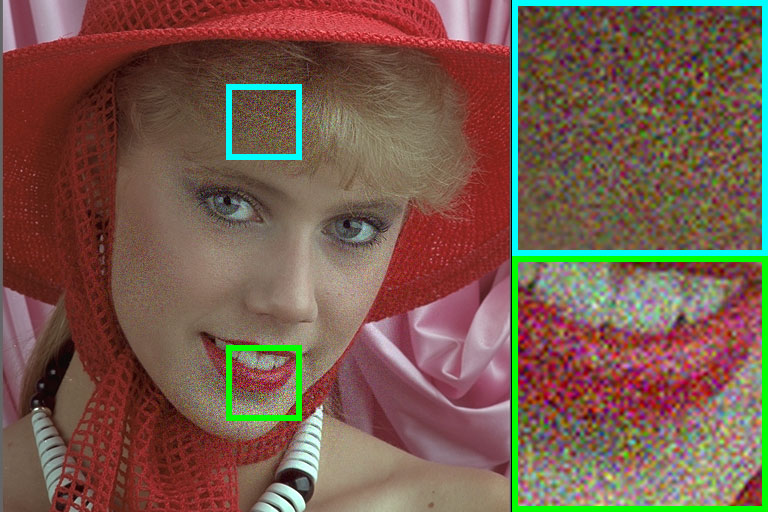}
		\hspace{-2mm}
	}%
	\subfloat[]{
		\hspace{-2mm}
		\includegraphics[width=0.16\linewidth]{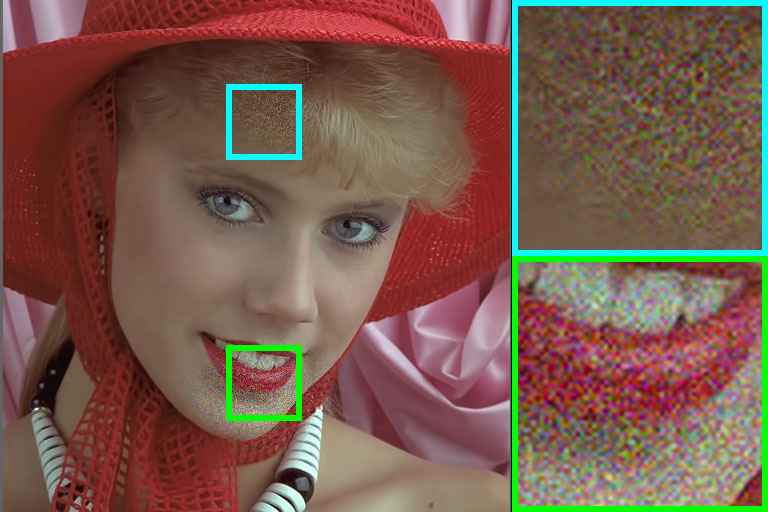}
		\hspace{-2mm}
	}%
	\subfloat[]{
		\hspace{-2mm}
		\includegraphics[width=0.16\linewidth]{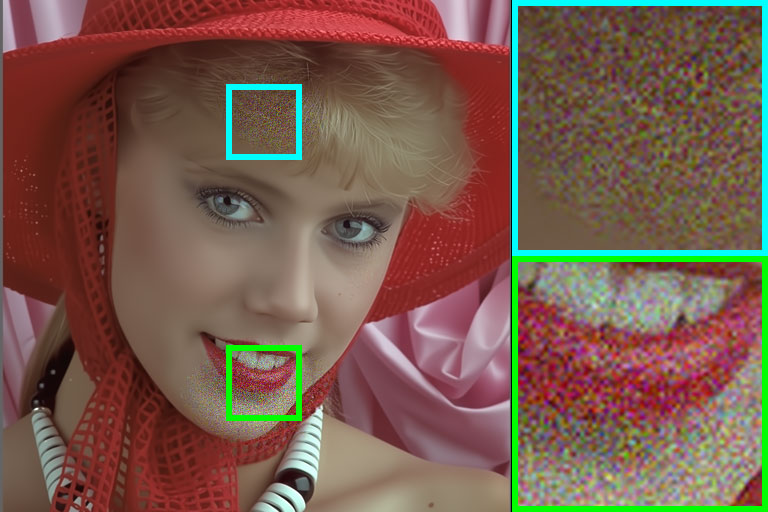}
		\hspace{-2mm}
	}%
	\subfloat[]{
		\hspace{-2mm}
		\includegraphics[width=0.16\linewidth]{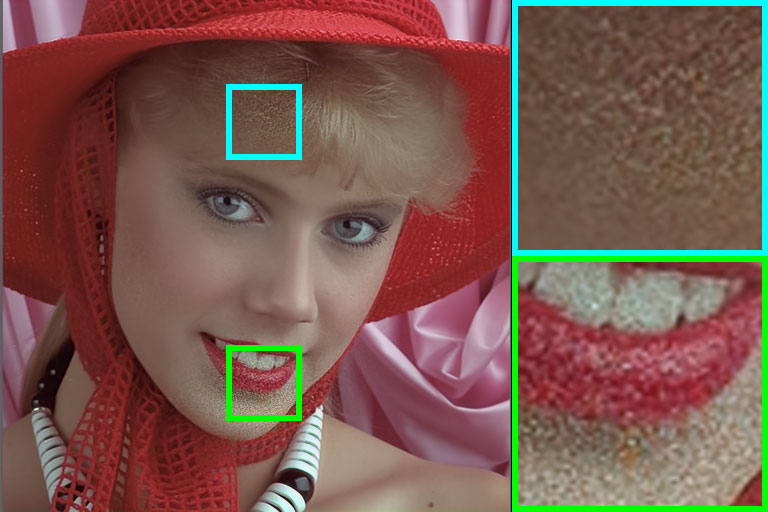}
		\hspace{-2mm}
	}%
	\subfloat[]{
		\hspace{-2mm}
		\includegraphics[width=0.16\linewidth]{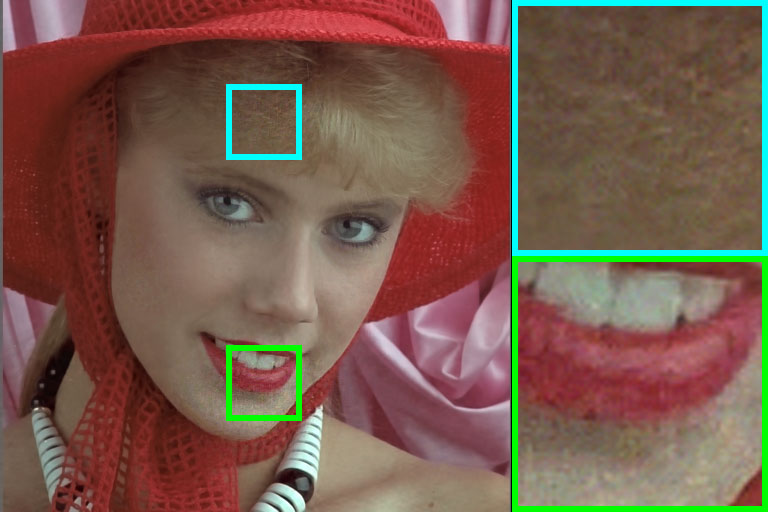}
		\hspace{-2mm}
	}%
	
	\subfloat[]{
		\hspace{-2mm}
		\includegraphics[width=0.16\linewidth]{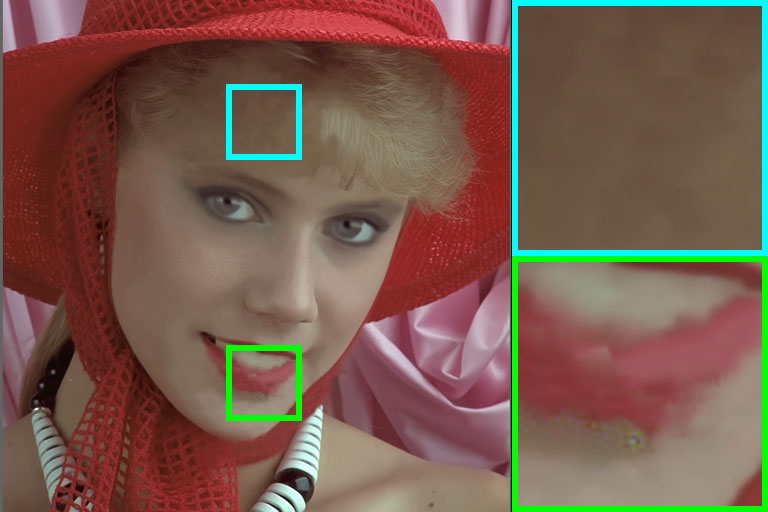}
		\hspace{-2mm}
	}%
	\subfloat[]{
		\hspace{-2mm}
		\includegraphics[width=0.16\linewidth]{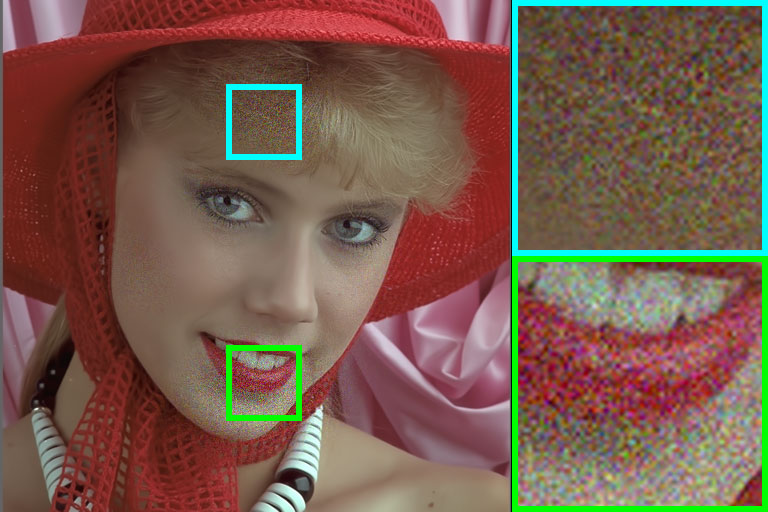}
		\hspace{-2mm}
	}%
	\subfloat[]{
		\hspace{-2mm}
		\includegraphics[width=0.16\linewidth]{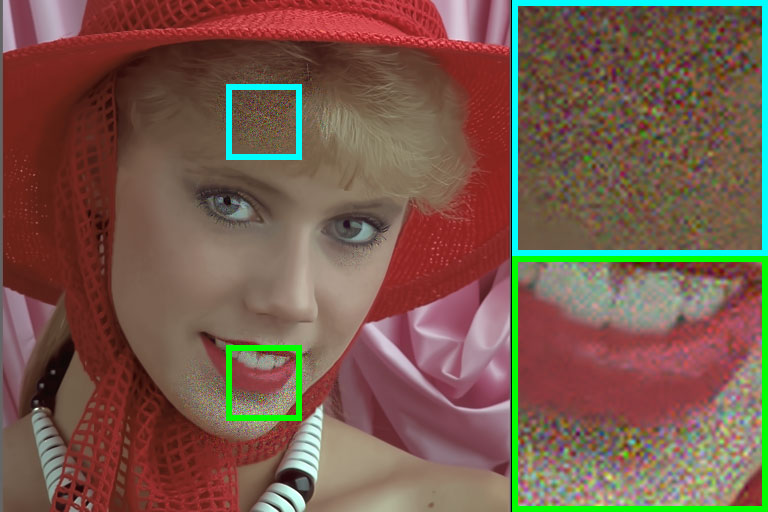}
		\hspace{-2mm}
	}%
	\subfloat[]{
		\hspace{-2mm}
		\includegraphics[width=0.16\linewidth]{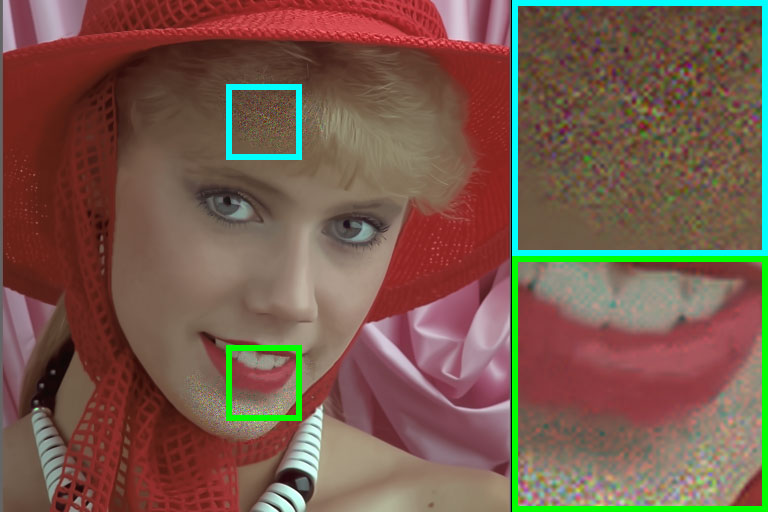}
		\hspace{-2mm}
	}%
	\subfloat[]{
		\hspace{-2mm}
		\includegraphics[width=0.16\linewidth]{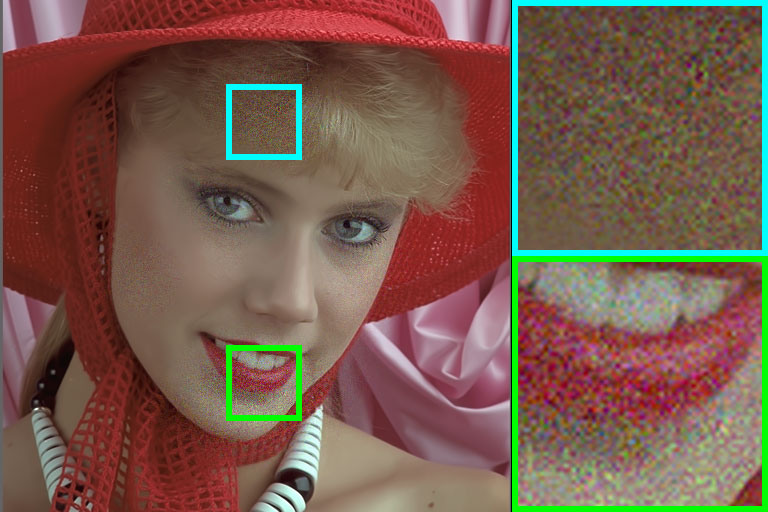}
		\hspace{-2mm}
	}%
	\subfloat[]{
		\hspace{-2mm}
		\includegraphics[width=0.16\linewidth]{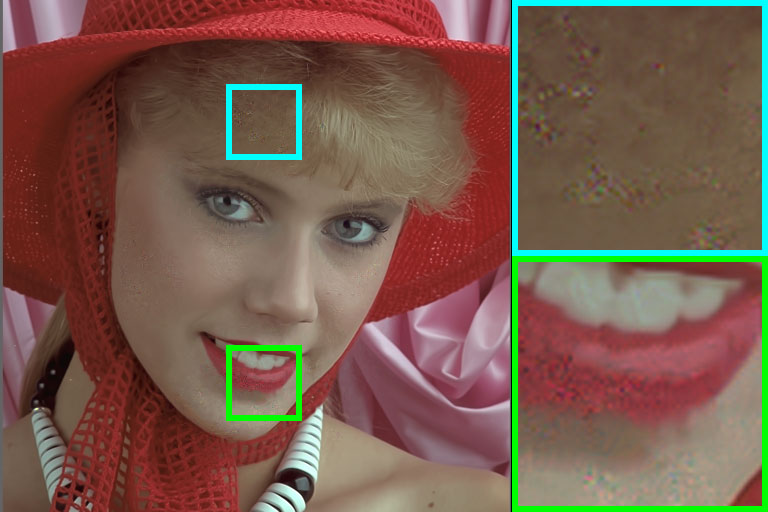}
		\hspace{-2mm}
	}%
	\captionsetup{format=plain} 
	\caption{Denoised results of ``kodim04'' (spatially variant noise). (a) Ground Truth. (b) Noisy observation (29.65, 0.8297). (c) CBM3D (32.70, 0.8712). (d) DRUNet (32.64, 0.8616). (e) Restormer (34.00, 0.8951). (f) HLTA-GN (34.54, 0.8867). (g) NGmeet (35.58, 0.9097). (h) DLRQP (33.15, 0.8838). (k) MCWNNM (33.09, 0.8697). (j) MCWSNM (33.41, 0.8534). (k) NNFNM (33.04, 0.8654). (l) DtNFM (\textbf{36.86}, \textbf{0.9177}). }
	\label{fig_sv_4}
\end{figure}%
\begin{figure}[tb] 
	\captionsetup[subfigure]{captionskip=0pt, farskip=0pt}
	\subfloat[]{
		\hspace{-2mm}
		\includegraphics[width=0.16\linewidth]{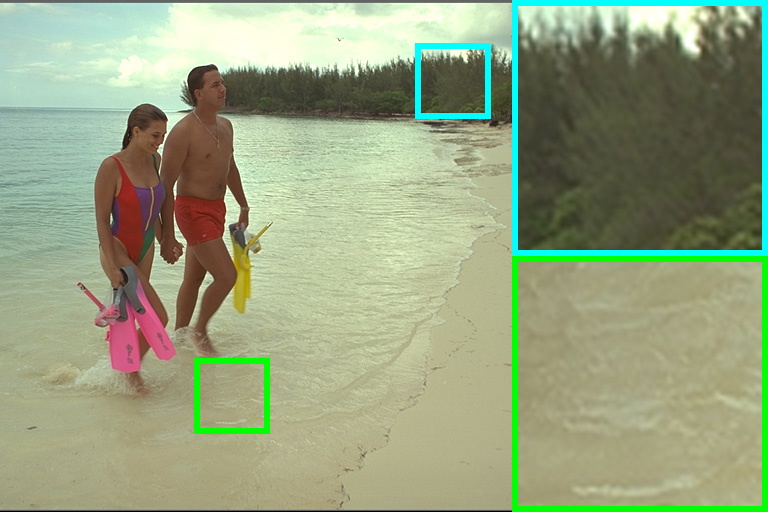}
		\hspace{-2mm}
	}%
	\subfloat[]{
		\hspace{-2mm}
		\includegraphics[width=0.16\linewidth]{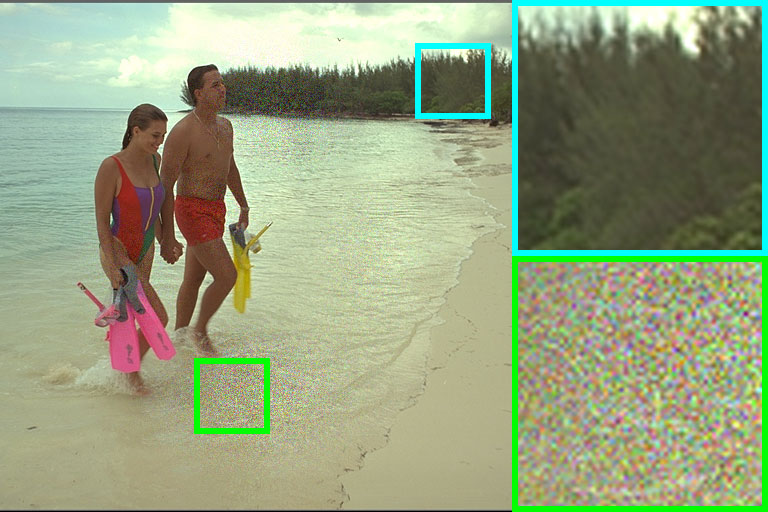}
		\hspace{-2mm}
	}%
	\subfloat[]{
		\hspace{-2mm}
		\includegraphics[width=0.16\linewidth]{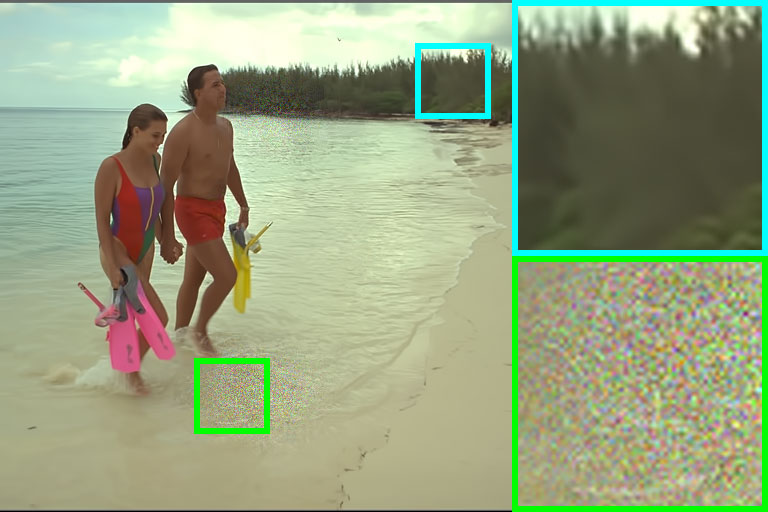}
		\hspace{-2mm}
	}%
	\subfloat[]{
		\hspace{-2mm}
		\includegraphics[width=0.16\linewidth]{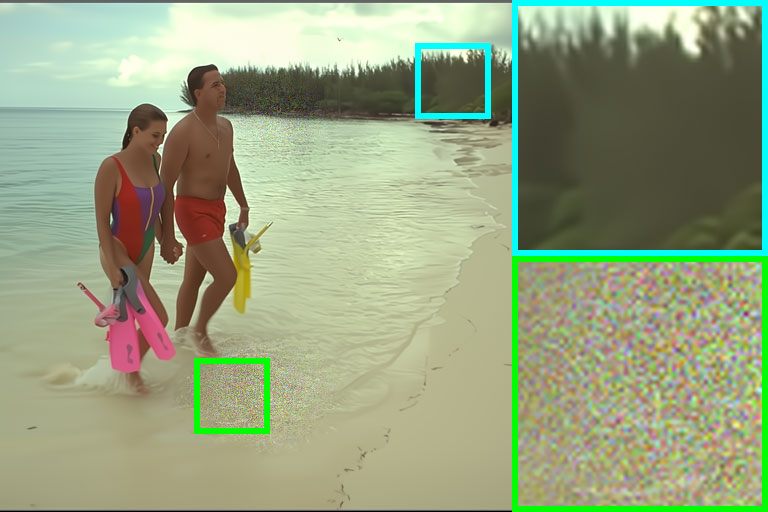}
		\hspace{-2mm}
	}%
	\subfloat[]{
		\hspace{-2mm}
		\includegraphics[width=0.16\linewidth]{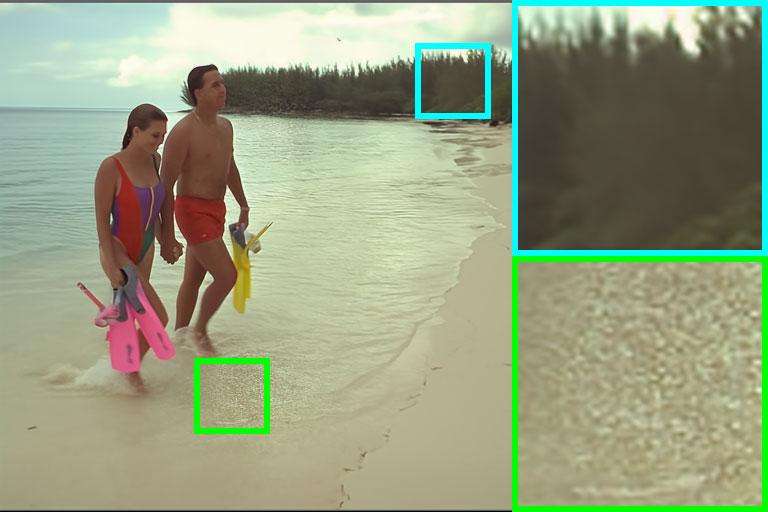}
		\hspace{-2mm}
	}%
	\subfloat[]{
		\hspace{-2mm}
		\includegraphics[width=0.16\linewidth]{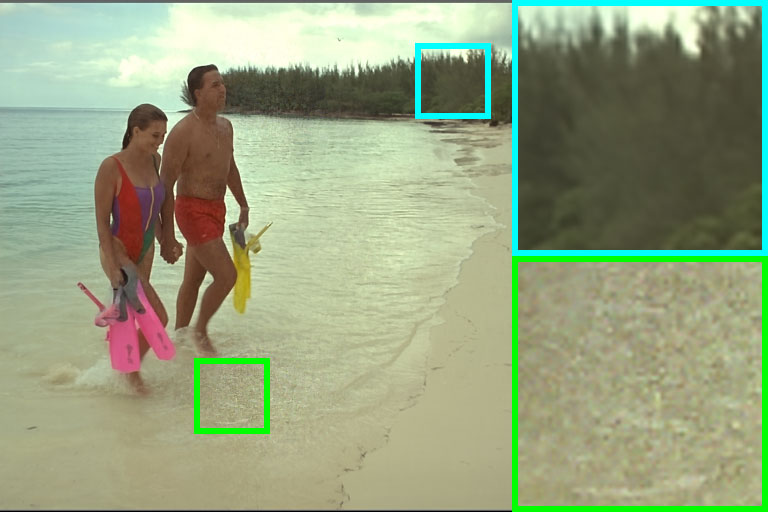}
		\hspace{-2mm}
	}%
	
	\subfloat[]{
		\hspace{-2mm}
		\includegraphics[width=0.16\linewidth]{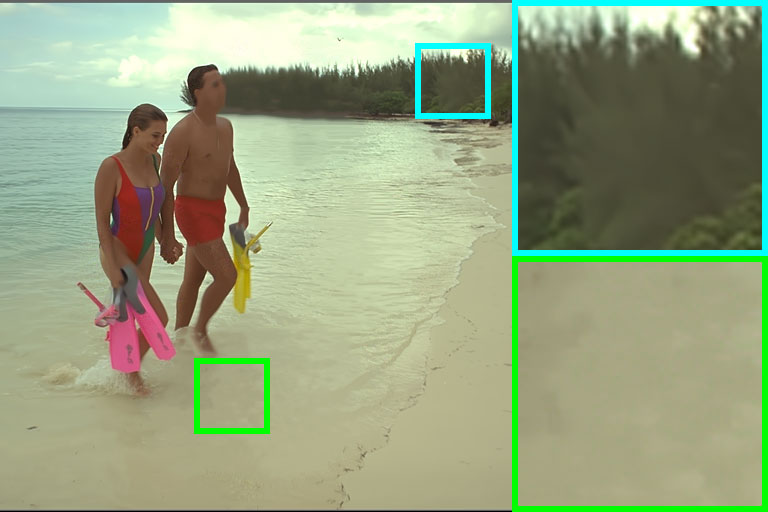}
		\hspace{-2mm}
	}%
	\subfloat[]{
		\hspace{-2mm}
		\includegraphics[width=0.16\linewidth]{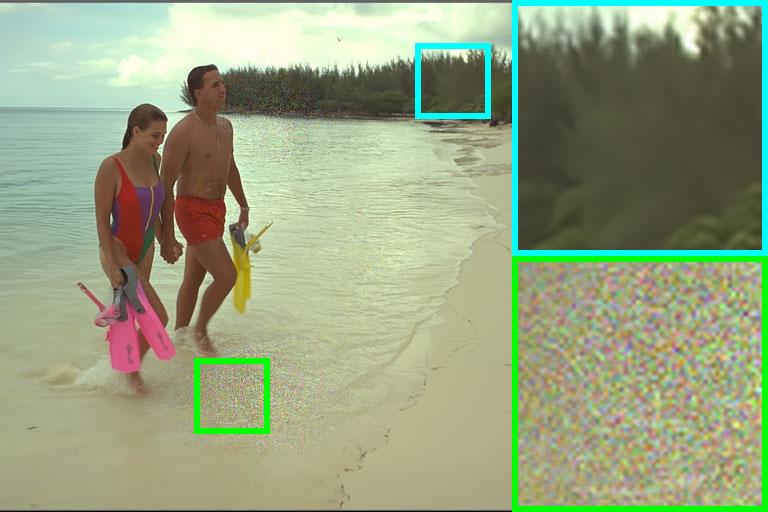}
		\hspace{-2mm}
	}%
	\subfloat[]{
		\hspace{-2mm}
		\includegraphics[width=0.16\linewidth]{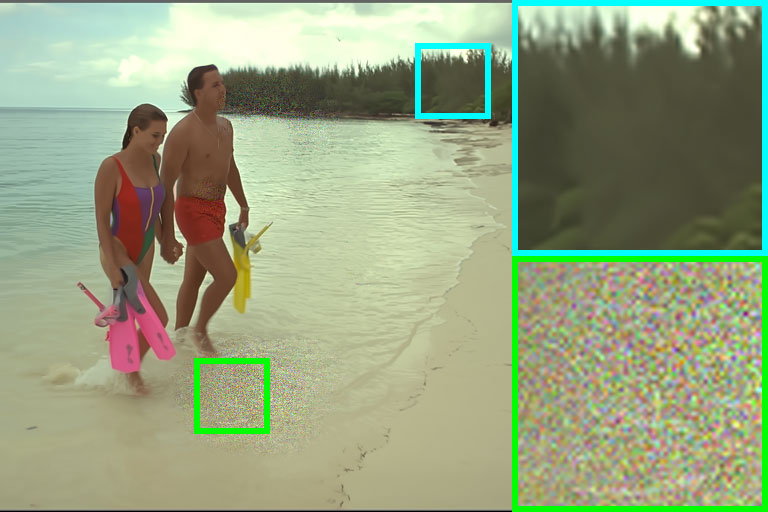}
		\hspace{-2mm}
	}%
	\subfloat[]{
		\hspace{-2mm}
		\includegraphics[width=0.16\linewidth]{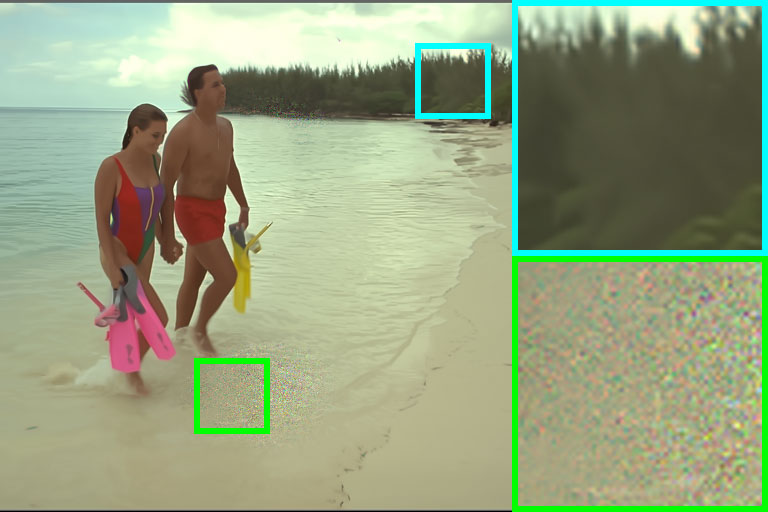}
		\hspace{-2mm}
	}%
	\subfloat[]{
		\hspace{-2mm}
		\includegraphics[width=0.16\linewidth]{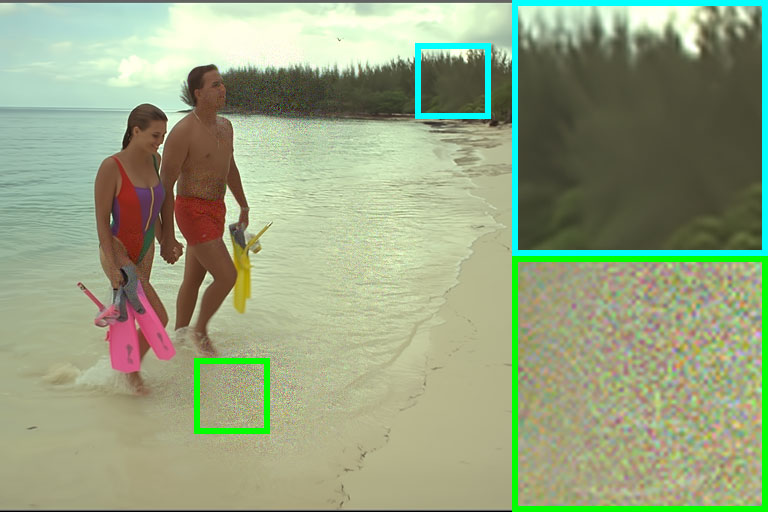}
		\hspace{-2mm}
	}%
	\subfloat[]{
		\hspace{-2mm}
		\includegraphics[width=0.16\linewidth]{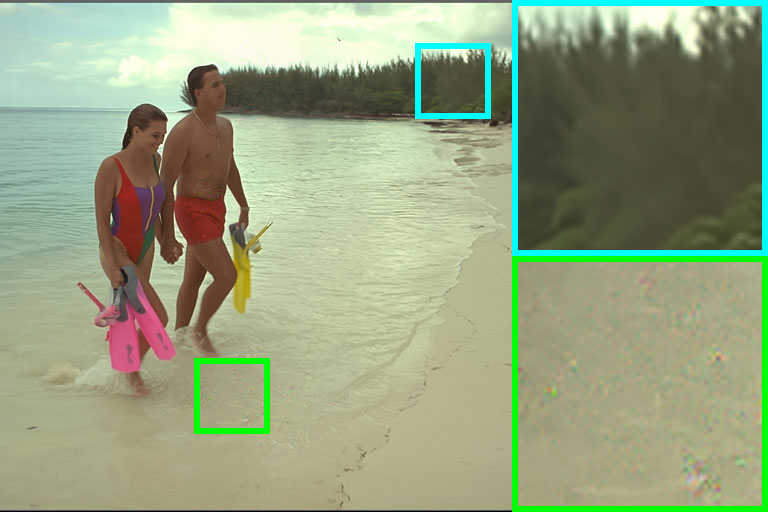}
		\hspace{-2mm}
	}%
	\captionsetup{format=plain} 
	\caption{Denoised results of ``kodim12'' (spatially variant noise). (a) Ground Truth. (b) Noisy observation (29.65, 0.8373). (c) CBM3D (33.12, 0.8846). (d) DRUNet (32.88, 0.8750). (e) Restormer (34.44, 0.8999). (f) HLTA-GN (35.76, 0.9140). (g) NGmeet (36.56, 0.9156). (h) DLRQP (33.13, 0.8877). (k) MCWNNM (31.79, 0.8682). (j) MCWSNM (33.72, 0.8645). (k) NNFNM (33.16, 0.8742). (l) DtNFM (\textbf{36.98}, \textbf{0.9246}). }
	\label{fig_sv_12}
\end{figure}%
The visual comparisons are shown in Fig. \ref{fig_sv_4} and Fig. \ref{fig_sv_12}. 
The proposed DtNFM not only removes the noise more completely, and preserves more details, such as the hair and mouth in Fig. \ref{fig_sv_4} and the trees at the top-right corner of Fig. \ref{fig_sv_12}.  
{{%
In contrast, the NGmeet over-smooth the images. 
Although the HLTA-GN removes the noise well, it is inadequate to preserve the textures. 
And the other seven methods remain too much noise. 
}}%
In summary, the proposed DtNFM method achieves a more rational balance between noise removal and detail protection. 
\begin{table*}[tb] 
	\centering\scriptsize
	\caption{ PSNR and SSIM results for all competing methods in the real-world noise experiments. Running times is in seconds.}
	\begin{tabularx}{\textwidth}{p{0.3cm}<{\centering}YYYYY YYYYY}
\toprule
& CBM3D & DRUNet & \!{{Restormer}} & \!{{HLTA-GN}} & \!{{NGMeet}} & \!{{DLRQP}} & MCWNNM & MCWSNM & NNFNM & DtNFM\\
\# & PSNR SSIM & PSNR SSIM & PSNR SSIM & PSNR SSIM & PSNR SSIM & PSNR SSIM & PSNR SSIM & PSNR SSIM & PSNR SSIM & PSNR SSIM\\
\hline
1 & 38.75 0.9688 & 40.88 0.9805 & 36.88 0.9718 & 38.80 0.9705 & 41.14 0.9830 & 40.15 0.9804 & 41.20 0.9829 & 41.22 0.9832 & 41.24 0.9835 & \textbf{41.43 0.9837}\\
2 & 35.52 0.9407 & 36.83 0.9486 & 36.71 0.9545 & 36.49 0.9517 & 37.32 0.9593 & 37.04 0.9585 & 37.25 0.9587 & \textbf{37.34 0.9608} & 37.25 0.9607 & 37.27 0.9592\\
3 & 35.69 0.9583 & 36.00 0.9542 & 35.65 0.9644 & 36.38 0.9640 & \textbf{37.08} 0.9691 & 36.66 0.9631 & 37.06 \textbf{0.9694} & 36.99 0.9670 & 36.96 0.9683 & 36.98 0.9665\\
4 & 33.84 0.9220 & 35.50 \textbf{0.9603} & 34.41 0.9545 & 34.81 0.9466 & 34.59 0.9463 & 34.89 0.9470 & 35.54 0.9598 & 35.28 0.9559 & \textbf{35.54} 0.9600 & 35.49 0.9591\\
5 & 34.66 0.9050 & 37.15 \textbf{0.9602} & 36.10 0.9560 & 36.07 0.9436 & 36.58 0.9490 & 36.22 0.9468 & 37.03 0.9568 & 36.66 0.9524 & 37.10 0.9584 & \textbf{37.18} 0.9590\\
6 & 36.22 0.9062 & 40.44 0.9810 & 38.09 0.9686 & 39.55 0.9735 & 40.29 0.9795 & 39.03 0.9692 & 39.56 0.9711 & 39.53 0.9710 & \textbf{41.27 0.9874} & 41.18 0.9873\\
7 & 36.63 0.9297 & \textbf{39.69 0.9705} & 38.70 0.9630 & 37.94 0.9546 & 38.77 0.9654 & 38.46 0.9605 & 39.26 0.9674 & 39.07 0.9668 & 39.35 0.9687 & 39.30 0.9676\\
8 & 37.32 0.9296 & \textbf{42.31 0.9832} & 38.93 0.9683 & 39.84 0.9730 & 40.86 0.9767 & 39.53 0.9655 & 41.45 0.9782 & 41.15 0.9768 & 41.84 0.9816 & 42.07 0.9822\\
9 & 36.31 0.9006 & \textbf{39.97} 0.9580 & 39.31 0.9481 & 37.96 0.9417 & 39.05 0.9532 & 38.60 0.9488 & 39.54 0.9561 & 39.39 0.9564 & 39.70 0.9556 & 39.65 \textbf{0.9585}\\
10 & 34.40 0.8647 & 39.30 0.9683 & 37.83 0.9434 & 38.78 0.9616 & 38.32 0.9639 & 38.48 0.9614 & 38.94 0.9636 & 38.89 0.9633 & 39.66 \textbf{0.9749} & \textbf{39.71} 0.9736\\
11 & 33.54 0.8743 & 36.73 0.9486 & 37.82 0.9584 & 37.22 0.9535 & 36.52 0.9426 & 36.47 0.9444 & 37.40 0.9524 & 37.23 0.9517 & \textbf{37.84 0.9586} & 37.65 0.9561\\
12 & 34.24 0.8352 & 41.10 0.9775 & 37.40 0.9177 & 39.93 0.9682 & 38.65 0.9572 & 39.45 0.9589 & 39.42 0.9591 & 39.52 0.9599 & 42.77 0.9832 & \textbf{42.99 0.9845}\\
13 & 30.64 0.7691 & \textbf{35.65 0.9398} & 31.94 0.8294 & 33.33 0.9097 & 35.50 0.9305 & 32.88 0.8841 & 34.85 0.9221 & 34.47 0.9200 & 35.18 0.9371 & 35.52 0.9381\\
14 & 30.88 0.8473 & \textbf{34.24} 0.9486 & 32.70 0.9208 & 33.12 0.9257 & 34.00 0.9392 & 32.51 0.9045 & 33.97 0.9397 & 33.56 0.9336 & 34.06 0.9471 & 34.20 \textbf{0.9490}\\
15 & 30.48 0.8146 & 33.83 0.9215 & 32.15 0.8839 & 33.66 0.9181 & 33.18 0.9037 & 33.30 0.9091 & 33.97 0.9215 & 33.78 0.9167 & 34.14 0.9245 & \textbf{34.25 0.9255}\\
\hline
Avg. & 34.61 0.8911 & 37.98 0.9601 & 36.31 0.9402 & 36.93 0.9504 & 37.46 0.9546 & 36.91 0.9468 & 37.76 0.9572 & 37.6 0.9557 & 38.26 0.9633 & \textbf{38.32 0.9633}\\
Time & 4.34 & \textbf{0.78} & 1.39 & 724.03 & 106.36 & 589.8 & 295.3 & 294.3 & 317.06 & 325.95\\
\bottomrule
	\end{tabularx}
	\label{tab_real}
\end{table*}
\subsection{Real-World Noise Removal}
{{%
The real-world noise removal experiments are carried out on the CC15 dataset \cite{CC15}, which is shown in Fig. \ref{fig_dataset}(b). 
It has 15 real-world noisy images and their noise-free versions. 
Those noise-free images are obtained by averaging each pixel from 500 images shot on the same scene under same camera settings. 
In practice, they can be used as the ground truth. 
With them, the PSNR and SSIM can be calculated for the quantitative comparison. }}%
For each corrupted image, the noise standard deviations $[\sigma_{r\_0}; \sigma_{g\_0}; \sigma_{b\_0}]$ are estimated by a state-of-the-art noise estimation method \cite{Noi_est}. 
Then, the competing methods denoise the images under the assumption that the noise is spatially invariant. 
The parameters of DtNFM method are listed in Table \ref{tab_par_set}(d). 
%
\par
\begin{figure}[tb] 
	\captionsetup[subfigure]{captionskip=0pt, farskip=0pt}
	\subfloat[]{
		\hspace{-2mm}
		\includegraphics[width=0.16\linewidth]{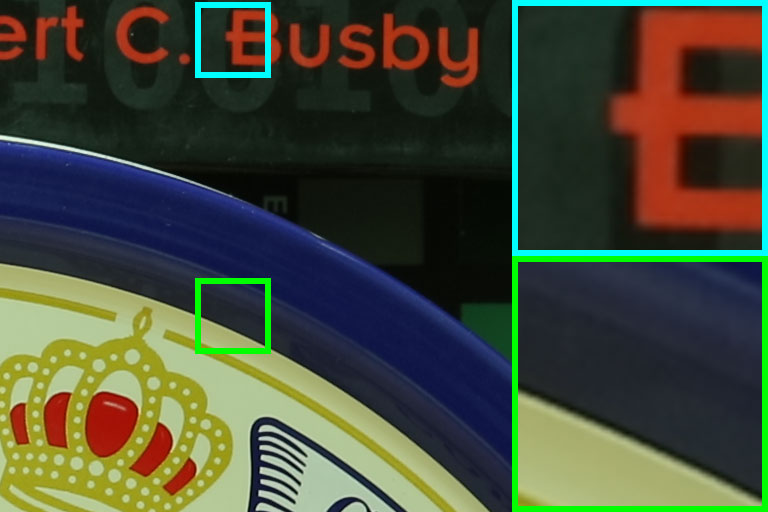}
		\hspace{-2mm}
	}%
	\subfloat[]{
		\hspace{-2mm}
		\includegraphics[width=0.16\linewidth]{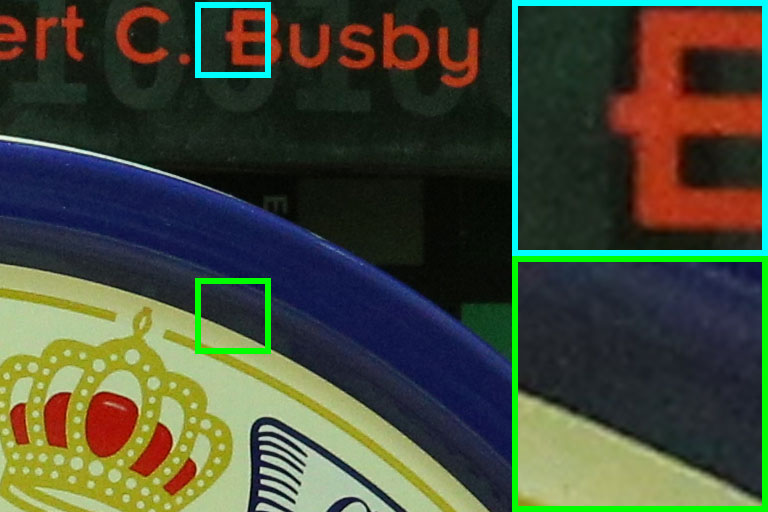}
		\hspace{-2mm}
	}%
	\subfloat[]{
		\hspace{-2mm}
		\includegraphics[width=0.16\linewidth]{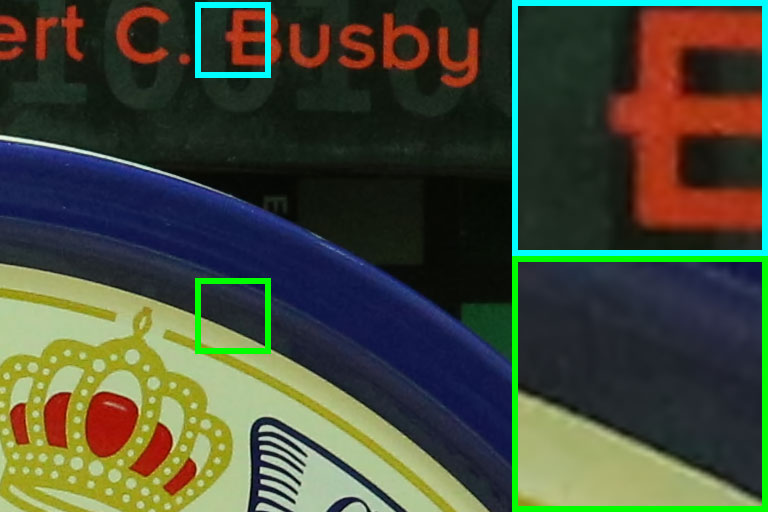}
		\hspace{-2mm}
	}%
	\subfloat[]{
		\hspace{-2mm}
		\includegraphics[width=0.16\linewidth]{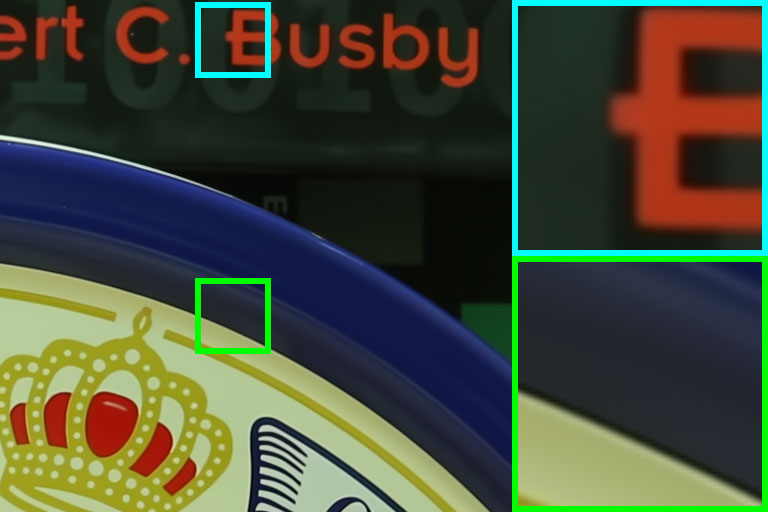}
		\hspace{-2mm}
	}%
	\subfloat[]{
		\hspace{-2mm}
		\includegraphics[width=0.16\linewidth]{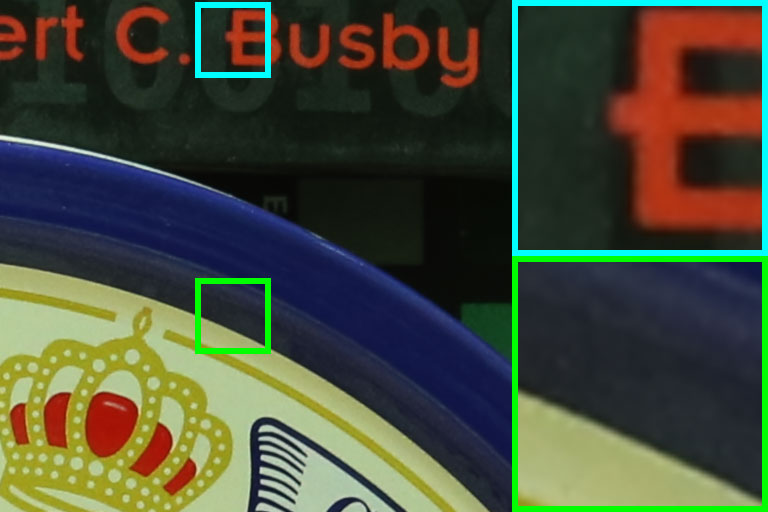}
		\hspace{-2mm}
	}%
	\subfloat[]{
		\hspace{-2mm}
		\includegraphics[width=0.16\linewidth]{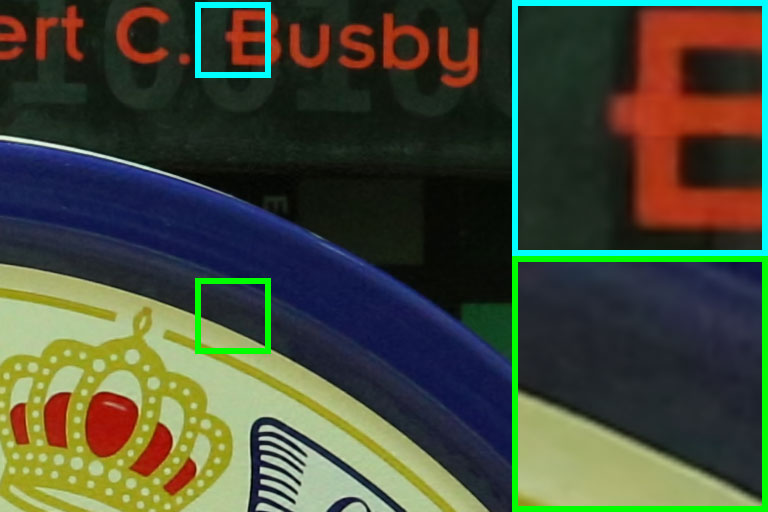}
		\hspace{-2mm}
	}%
	
	\subfloat[]{
		\hspace{-2mm}
		\includegraphics[width=0.16\linewidth]{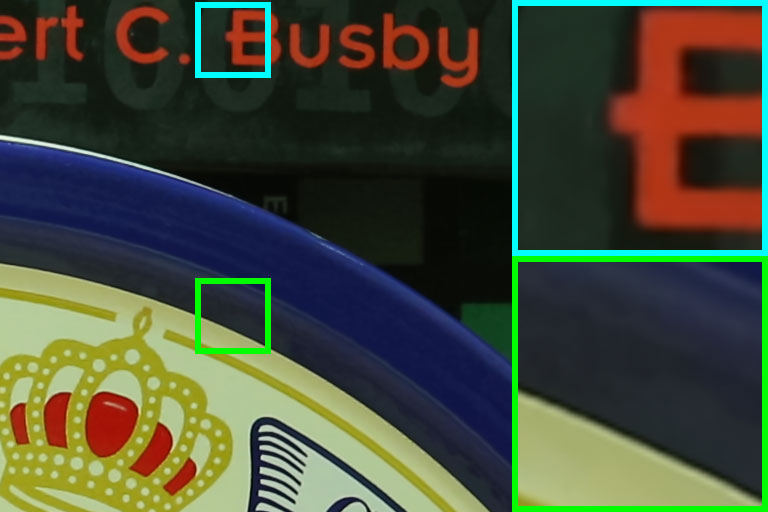}
		\hspace{-2mm}
	}%
	\subfloat[]{
		\hspace{-2mm}
		\includegraphics[width=0.16\linewidth]{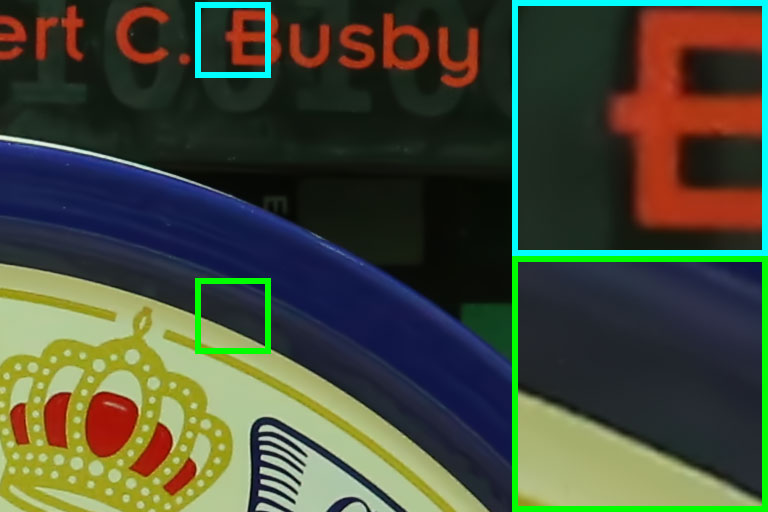}
		\hspace{-2mm}
	}%
	\subfloat[]{
		\hspace{-2mm}
		\includegraphics[width=0.16\linewidth]{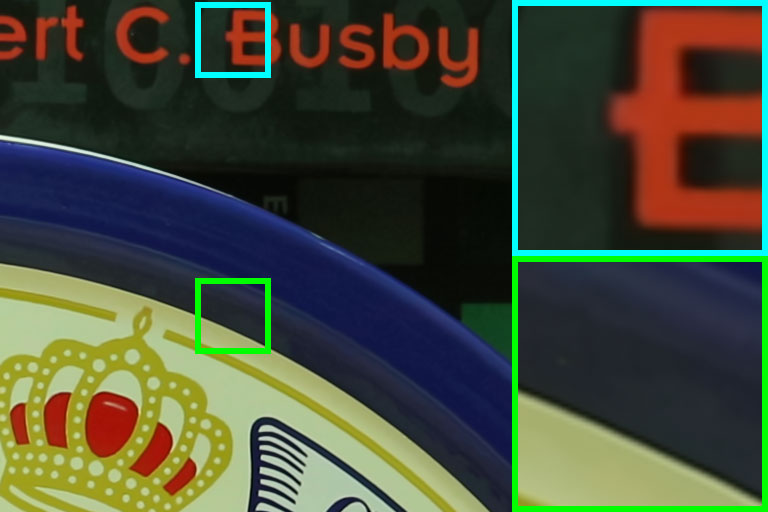}
		\hspace{-2mm}
	}%
	\subfloat[]{
		\hspace{-2mm}
		\includegraphics[width=0.16\linewidth]{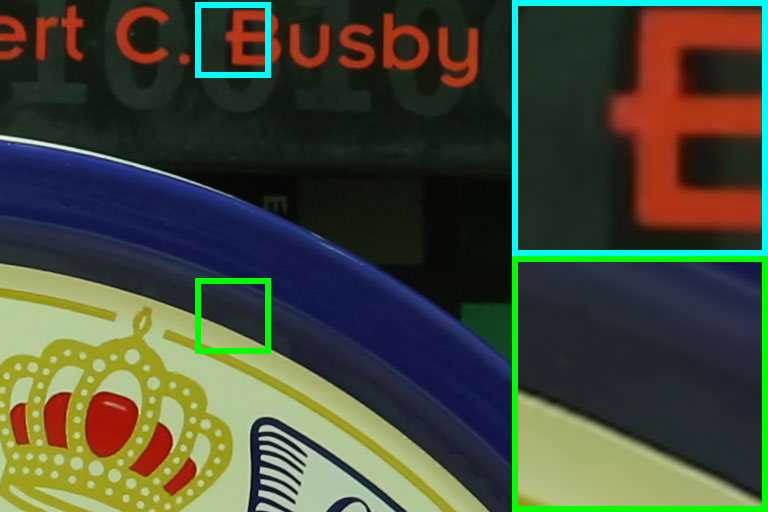}
		\hspace{-2mm}
	}%
	\subfloat[]{
		\hspace{-2mm}
		\includegraphics[width=0.16\linewidth]{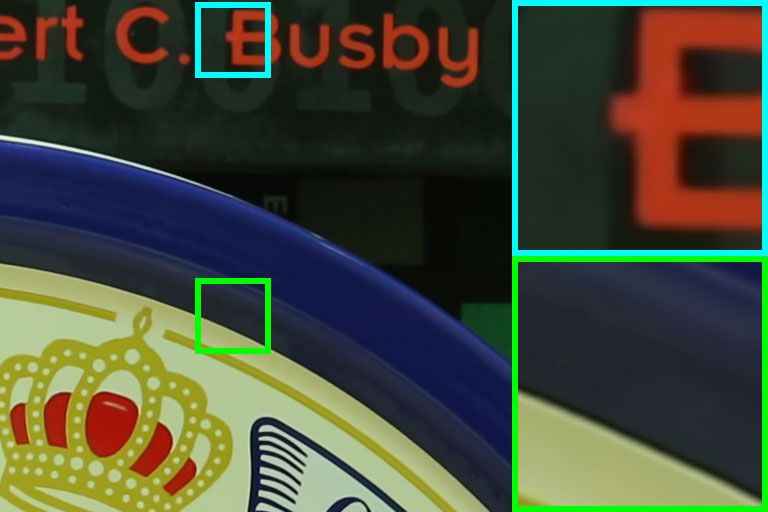}
		\hspace{-2mm}
	}%
	\subfloat[]{
		\hspace{-2mm}
		\includegraphics[width=0.16\linewidth]{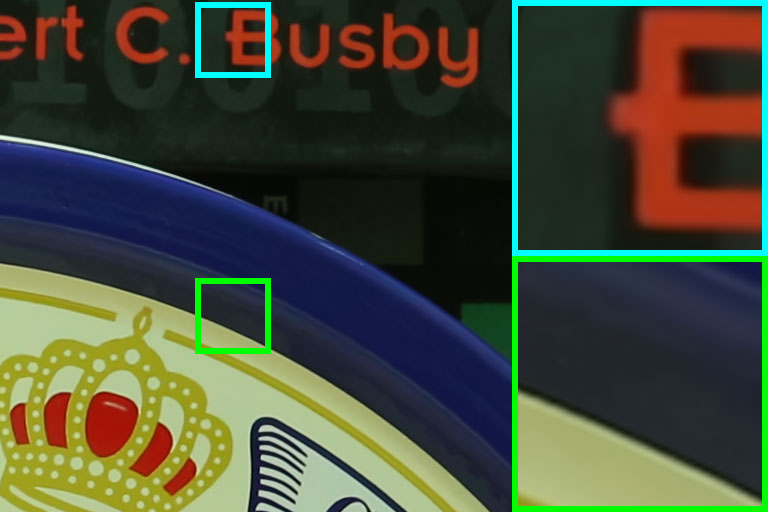}
		\hspace{-2mm}
	}%
	\captionsetup{format=plain} 
	\caption{Denoised results of image ``\#1'' (real-world noise). (a) Ground Truth. (b) Noisy observation (37.00, 0.9345). (c) CBM3D (38.75, 0.9688). (d) DRUNet (40.88, 0.9805). (e) Restormer (36.88, 0.9718). (f) HLTA-GN (38.80, 0.9705). (g) NGmeet (41.14, 0.9830). (h) DLRQP (40.15, 0.9804). (k) MCWNNM (41.20, 0.9829). (j) MCWSNM (41.22, 0.9832). (k) NNFNM (41.24, 0.9835). (l) DtNFM (\textbf{41.43}, \textbf{0.9837}). }
	\label{fig_real_1}
\end{figure}%
\begin{figure}[tb] 
	\captionsetup[subfigure]{captionskip=0pt, farskip=0pt}
	\subfloat[]{
		\hspace{-2mm}
		\includegraphics[width=0.16\linewidth]{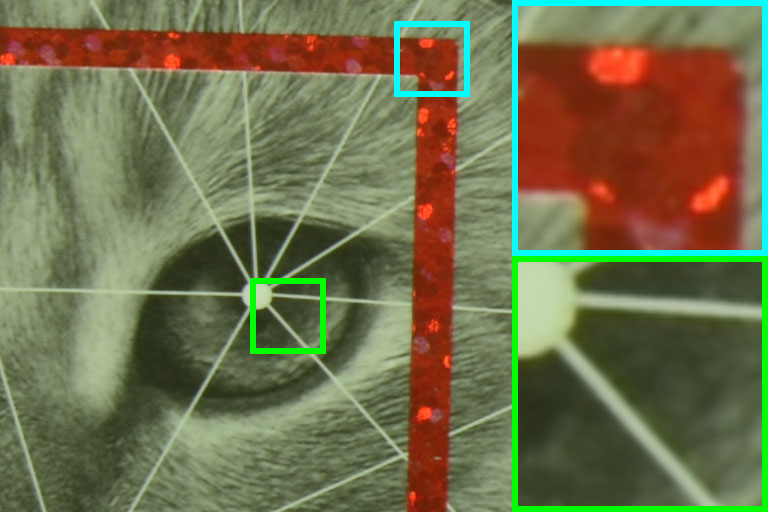}
		\hspace{-2mm}
	}%
	\subfloat[]{
		\hspace{-2mm}
		\includegraphics[width=0.16\linewidth]{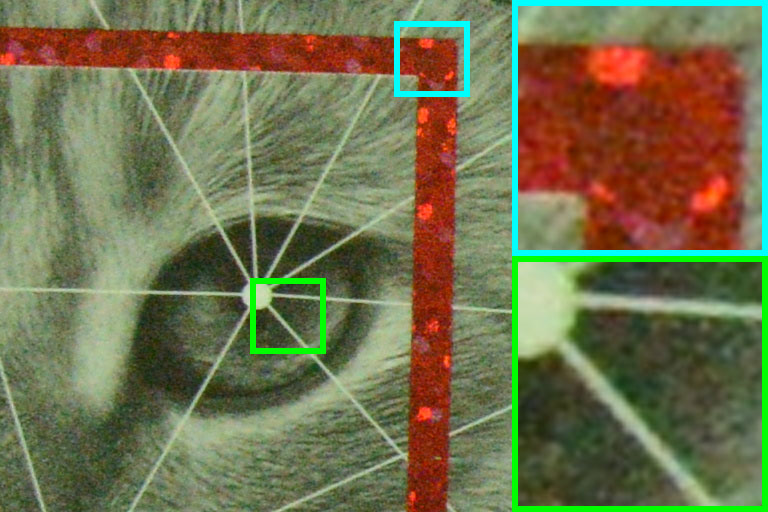}
		\hspace{-2mm}
	}%
	\subfloat[]{
		\hspace{-2mm}
		\includegraphics[width=0.16\linewidth]{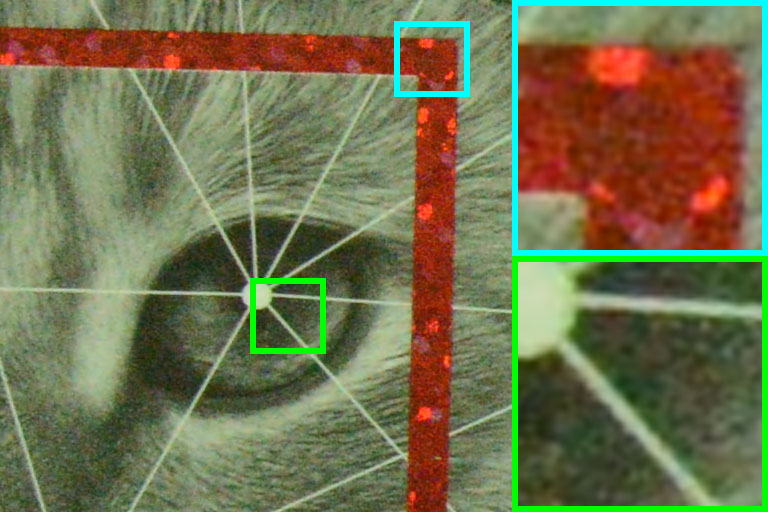}
		\hspace{-2mm}
	}%
	\subfloat[]{
		\hspace{-2mm}
		\includegraphics[width=0.16\linewidth]{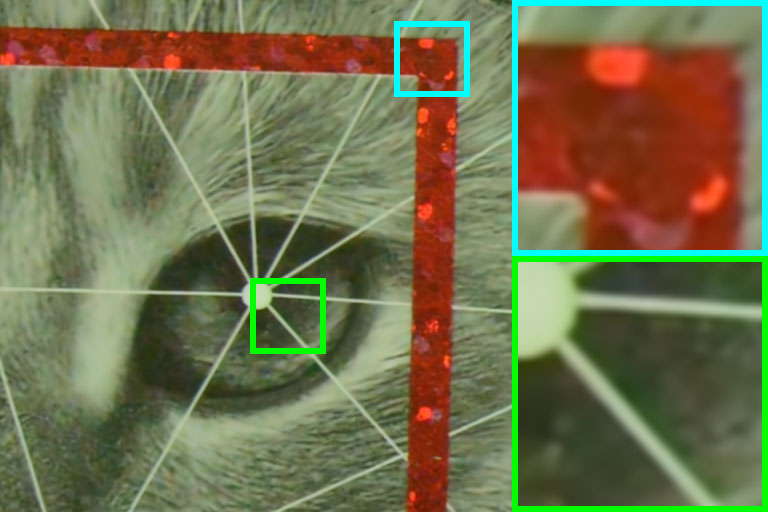}
		\hspace{-2mm}
	}%
	\subfloat[]{
		\hspace{-2mm}
		\includegraphics[width=0.16\linewidth]{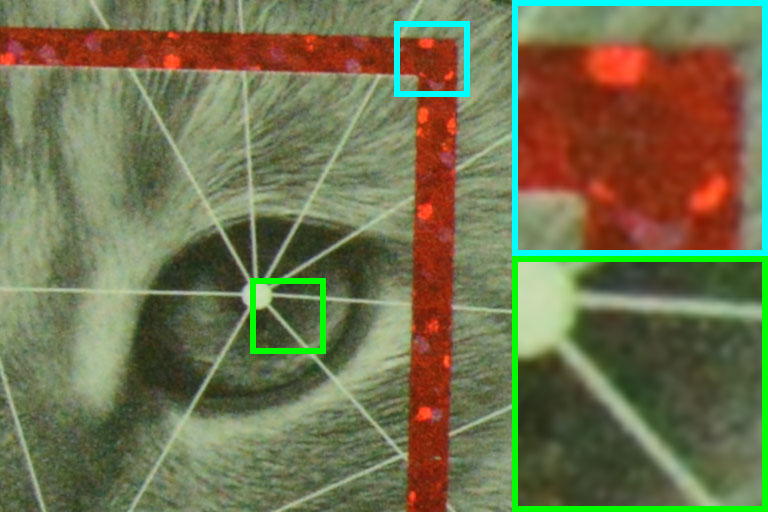}
		\hspace{-2mm}
	}%
	\subfloat[]{
		\hspace{-2mm}
		\includegraphics[width=0.16\linewidth]{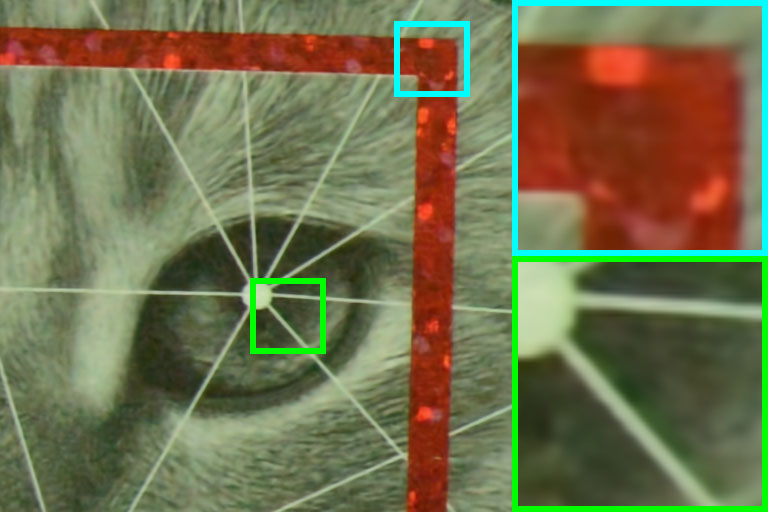}
		\hspace{-2mm}
	}%
	
	\subfloat[]{
		\hspace{-2mm}
		\includegraphics[width=0.16\linewidth]{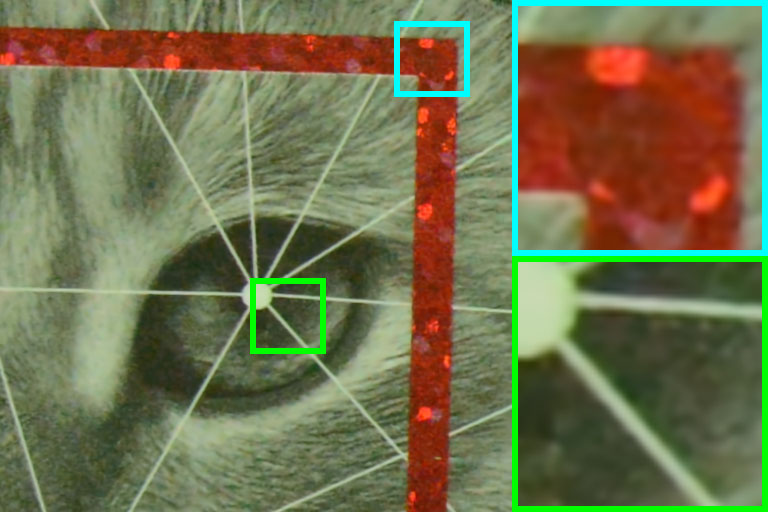}
		\hspace{-2mm}
	}%
	\subfloat[]{
		\hspace{-2mm}
		\includegraphics[width=0.16\linewidth]{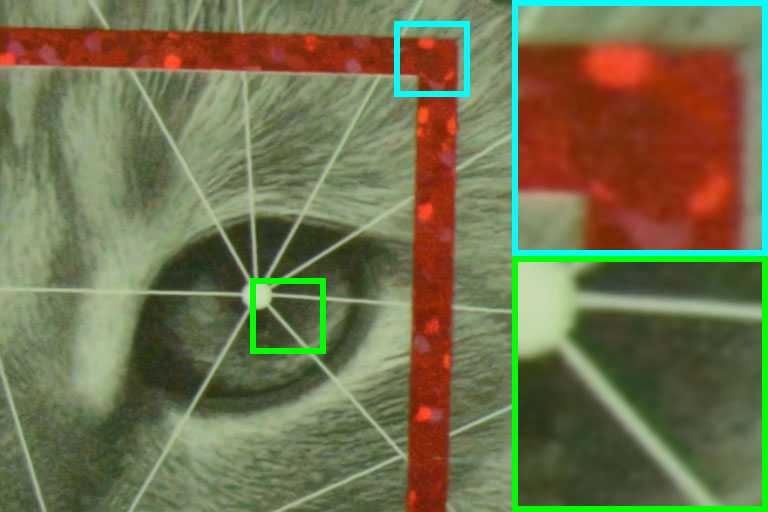}
		\hspace{-2mm}
	}%
	\subfloat[]{
		\hspace{-2mm}
		\includegraphics[width=0.16\linewidth]{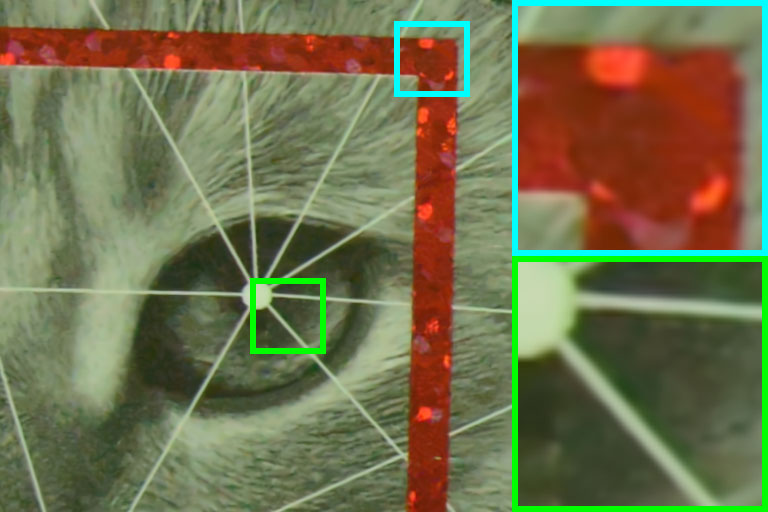}
		\hspace{-2mm}
	}%
	\subfloat[]{
		\hspace{-2mm}
		\includegraphics[width=0.16\linewidth]{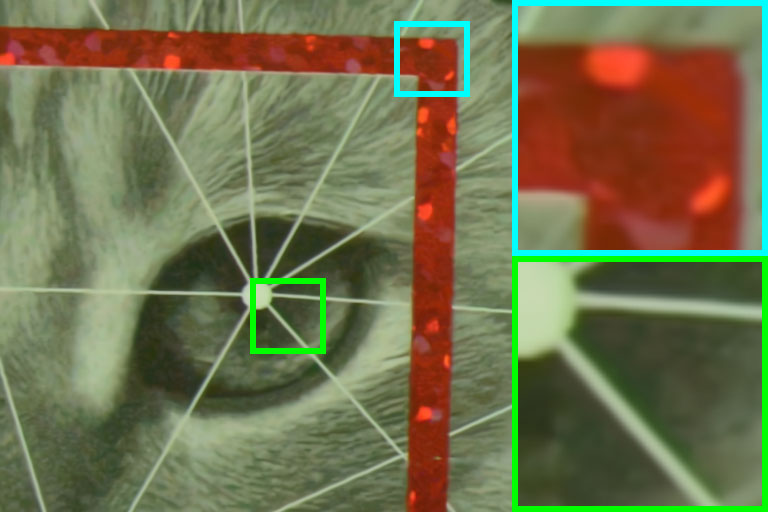}
		\hspace{-2mm}
	}%
	\subfloat[]{
		\hspace{-2mm}
		\includegraphics[width=0.16\linewidth]{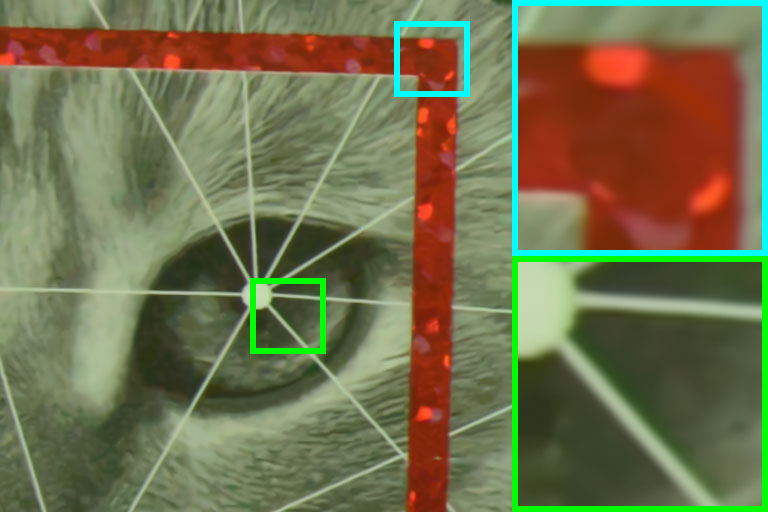}
		\hspace{-2mm}
	}%
	\subfloat[]{
		\hspace{-2mm}
		\includegraphics[width=0.16\linewidth]{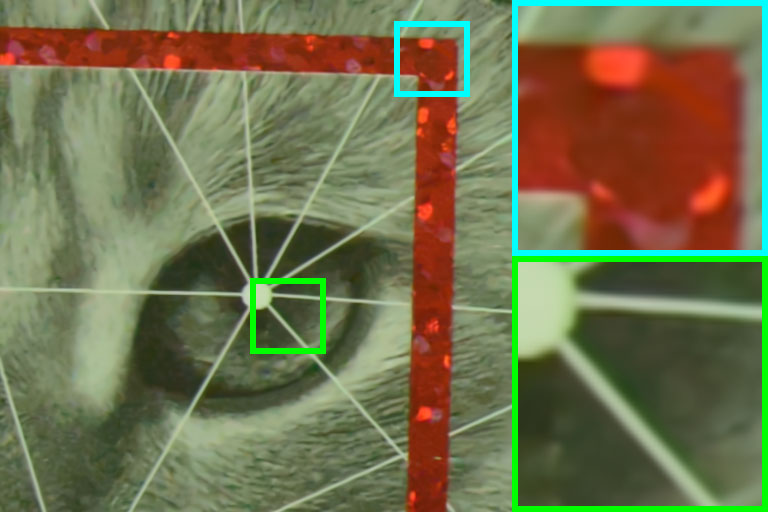}
		\hspace{-2mm}
	}%
	\captionsetup{format=plain} 
	\caption{Denoised results of image ``\#15'' (real-world noise). (a) Ground Truth. (b) Noisy observation (29.87, 0.7835). (c) CBM3D (30.48, 0.8146). (d) DRUNet (33.83, 0.9215). (e) Restormer (32.15, 0.8839). (f) HLTA-GN (33.66, 0.9181). (g) NGmeet (33.18, 0.9037). (h) DLRQP (33.30, 0.9091). (k) MCWNNM (33.97, 0.9215). (j) MCWSNM (33.78, 0.9167). (k) NNFNM (34.14, 0.9245). (l) DtNFM (\textbf{34.25}, \textbf{0.9255}). }
	\label{fig_real_15}
\end{figure}%
The PSNR and SSIM results are listed in Table \ref{tab_real}. 
As can be seen, DtNFM achieves the highest PSNR and SSIM on 5 images. 
And it achieves the highest average PSNR and SSIM. 
The average PSNR and SSIM improvements of DtNFM over other methods are listed in Table \ref{tab_improvements}(d). 
{{%
We also compare the DtNFM with other mainstream deep learning-based methods, including the Uformer \cite{Uformer}, DeamNet \cite{DeamNet}, DudeNet \cite{DudeNet}, SADNet \cite{SADNet}, and FFDNet \cite{FFDNet}. 
Due to space limit, the PSNR and SSIM results are shown in the supplementary material. 
}}
\par
The visual comparisons are shown in Fig. \ref{fig_real_1} and Fig. \ref{fig_real_15}. 
The DtNFM generates satisfactory visual quality, removing the noise fully and preserves image details well. 
In contrast, CBM3D and Restormer remain some noise, and the DLRQP over-smooth the images. 
To sum up, the proposed DtNFM has strong capacity on reducing the real-world noise, achieving higher PSNR and SSIM and producing promising visual quality. 
\begin{table}[t]
\centering
\caption{{{Ablation experiments for the DtNFM model.}} }
\begin{tabularx}{0.5\textwidth}{p{2.1cm}|Y|Y}
\bottomrule
\multirowcell{4}{Model} & \multicolumn{2}{c}{Noise}\\
\cline{2-3}
& Spatially invariant & Spatially variant\\
& $\mathbf{Std}=[30;10;50]$ & $\mathbf{Std} = [30;35;40]$ \\
\cline{2-3}
& PSNR SSIM & PSNR SSIM\\
\hline
\!\!(a) DtNFM & \textbf{31.94 0.8720} & \textbf{35.45 0.9326}\\
\hline
\!\!(b) DtNFM$-\mathbf{C}$ & 27.53 0.7042 & 32.78 0.9079\\
\hline
\!\!(c) DtNFM$-\mathbf{S}$ & 31.87 0.8701 & 35.27 0.9313\\
\toprule
\end{tabularx}
\label{tab_ablation_dtnfm}
\end{table}
\subsection{{{Ablation Studies}}}
{{%
In the proposed DtNFM model \eqref{eq_DtNFM}, the weight matrices $\mathbf{C}$ and $\mathbf{S}$ are used at once. 
To find their contributions, the ablation experiments are carried out. 
For the DtNFM model (baseline), the following two models can be generated by it: 
\begin{align}
    &\arg \min_{\mathbf{X} \in \mathbb{R}^{3d^2 \times N}} \Vert (\mathbf{Y} - \mathbf{X}) \mathbf{S} \Vert_F^2 + \lambda \Vert \mathbf{X} \Vert_{t,*-F}, \label{eq_wtnfm}\\
    &\arg \min_{\mathbf{X} \in \mathbb{R}^{3d^2 \times N}} \Vert \mathbf{C}(\mathbf{Y} - \mathbf{X}) \Vert_F^2 + \lambda \Vert \mathbf{X} \Vert_{t,*-F}, \label{eq_stnfm}
\end{align}
where $\mathbf{S} = \mathrm{Diag}([\sigma_{1}^{-1}, \ldots, \sigma_{N}^{-1}]) \in \mathbb{R}^{N \times N}$, $\mathbf{C} = \mathrm{Diag}([\sigma_{r}^{-1}\mathbf{1}; \sigma_{g}^{-1}\mathbf{1}; \sigma_{b}^{-1}\mathbf{1}]) \in \mathbb{R}^{3d^2 \times 3d^2}$, $\mathbf{1} \in \mathbb{R}^{d^2}$ is a column vector of ones, and the definition of $\sigma_{c}$ $(c\in \lbrace r,g,b \rbrace)$, $\sigma_{j}$ $(j \in \lbrace 1,\ldots, N \rbrace)$ can be found in Section III.A. 
And the models in \eqref{eq_wtnfm} and \eqref{eq_stnfm} are respectively denoted as ``DtNFM$-\mathbf{C}$'' and ``DtNFM$-\mathbf{S}$''. 
Two kinds of noise are tested, as shown in the last two columns of Table \ref{tab_ablation_dtnfm}. 
\par
The results of the ablation experiments are reported in Table \ref{tab_ablation_dtnfm}. 
Note that the PSNR and SSIM are averaged by the 24 images in Kodak24 dataset. 
And the parameters are all kept the same. 
We have the following observations:
\begin{itemize}
    \item Table \ref{tab_ablation_dtnfm}(b) demonstrates the weight matrix $\mathbf{C}$ provides a gain of $(4.41dB$, $0.1678)$ in the spatially invariant noise removal. And its gain reduces to $(2.67dB$, $0.0247)$ in the spatially variant noise removal. Hence the matrix $\mathbf{C}$ would provide more contribution when removing  the spatially invariant noise. 
    \item Table \ref{tab_ablation_dtnfm}(c) demonstrates the weight matrix $\mathbf{S}$ provides a gain of $(0.07dB, 0.0019)$ in the spatially invariant noise removal. And its contribution improves to $(0.18dB, 0.0013)$ in the spatially variant noise removal. Hence the matrix $\mathbf{S}$ would provide more contribution when removing  the spatially variant noise. 
\end{itemize}
However, one may point out that the contribution of matrix $\mathbf{S}$ is significantly smaller than that of matrix $\mathbf{C}$ in both two experiments. 
We believe this will be alleviated if a more reasonable scheme is used to determine the relative weight $p\in [0,1]$ between $\mathbf{C}$ and $\mathbf{S}$. }}
\subsection{The Impact of the Hyper-parameters $\lambda$ and $t$}
To give empirical schemes for determining the hyper-parameters $\lambda$ and $t$, we analysis their impacts on the model performance. 
All the test images are taken from Kodak24 data set. 
The corrupted images are generated by the spatially invariant noise with $[\sigma_{r\_0}; \sigma_{g\_0}; \sigma_{b\_0}] = [30; 10; 50]$. 
All the parameters are fixed but $\lambda$ and $t$. 
\par
\begin{figure}[tb] 
	\centering
	\captionsetup[subfigure]{captionskip=0pt, farskip=0pt}
	\begin{tabular}{p{1cm}l}
		& $\ \ \lambda=0.001 \quad\quad\quad\quad\lambda=1 \quad\quad\quad\ \ \lambda=1000$ \\
		$t=0$ & 
		\begin{minipage}{0.4\textwidth}
			\subfloat[{\scriptsize22.19, 0.4347}]{
				\hspace{-5mm}
				\includegraphics[width=0.32\linewidth]{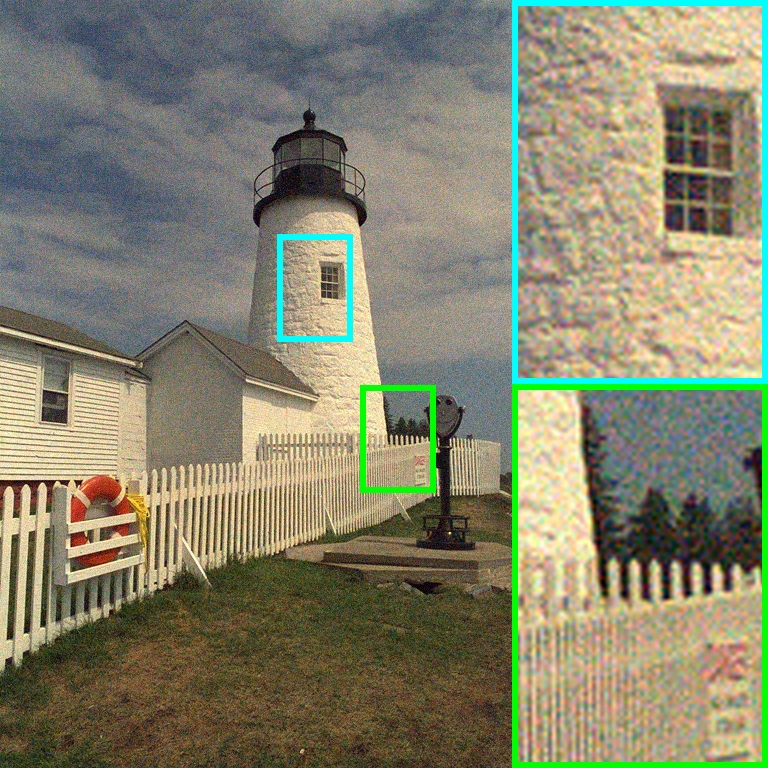} \label{fig_t0_0001}
			}%
			\subfloat[{\scriptsize\!\!31.42, 0.8501}]{
				\hspace{-2.5mm}
				\includegraphics[width=0.32\linewidth]{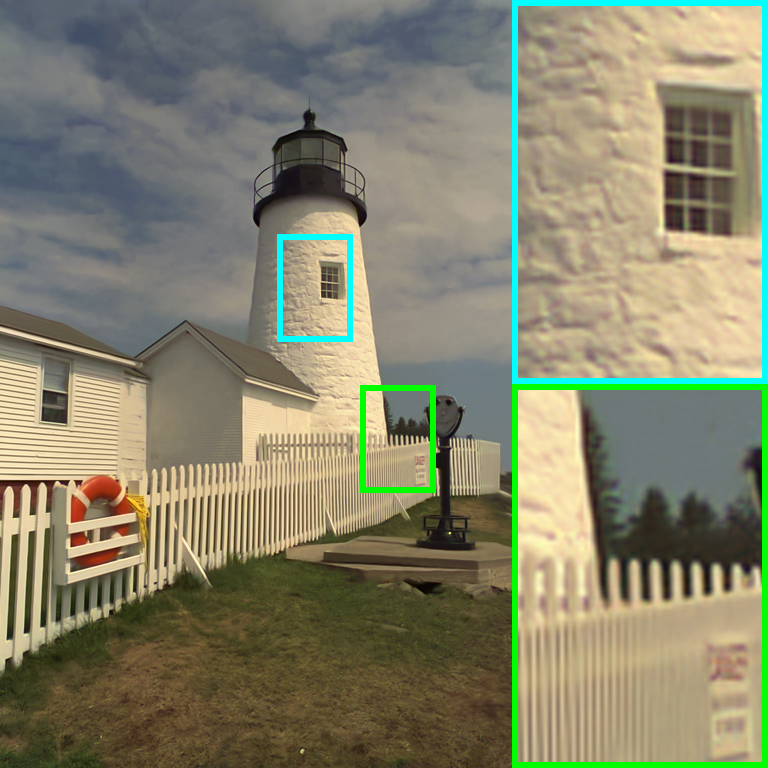}
			}%
			\subfloat[{\scriptsize\!\!27.68, 0.7281}]{
				\hspace{-2mm}
				\includegraphics[width=0.32\linewidth]{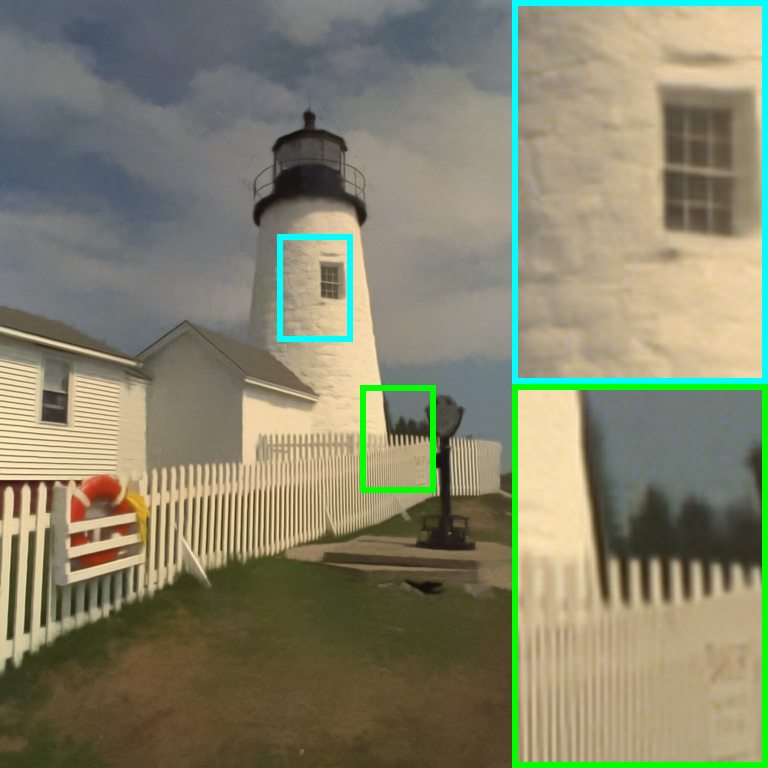} \label{fig_t0_1000}
			}
		\end{minipage}\\
		$t=1$ & 
		\begin{minipage}{0.4\textwidth}
			\subfloat[{\scriptsize22.19, 0.4347}]{
				\hspace{-5mm}
				\includegraphics[width=0.32\linewidth]{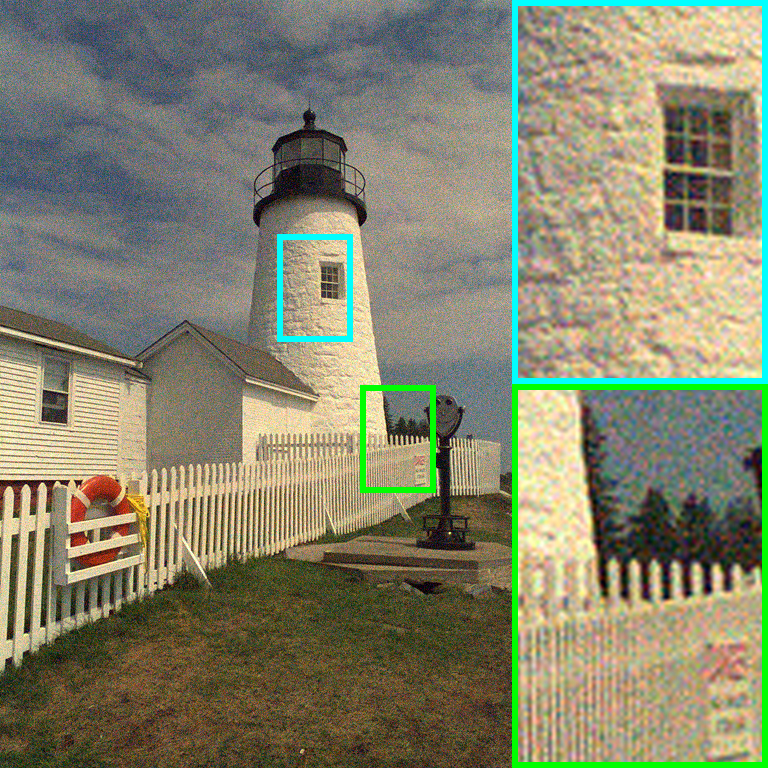} \label{fig_t1_0001}
			}%
			\subfloat[{\scriptsize31.67, 0.8561}]{
				\hspace{-3mm}
				\includegraphics[width=0.32\linewidth]{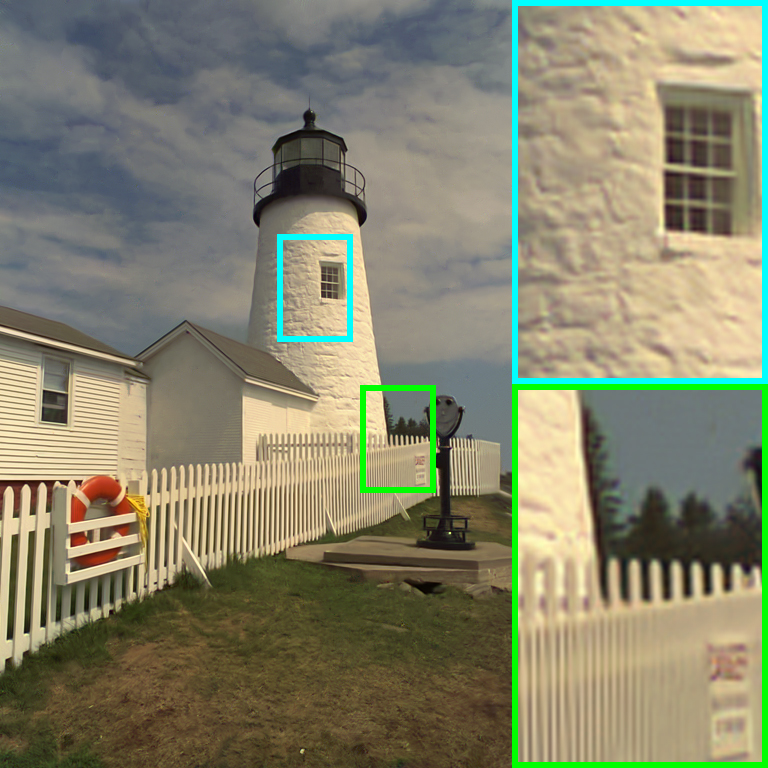}
			}%
			\subfloat[{\scriptsize28.68, 0.7475}]{
				\hspace{-2mm}
				\includegraphics[width=0.32\linewidth]{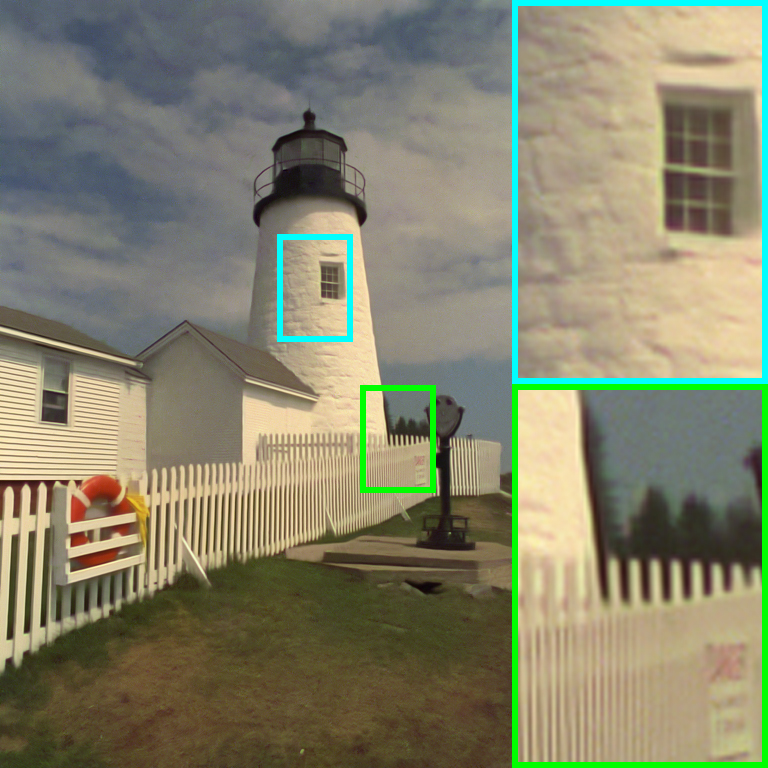} \label{fig_t1_1000}
			}
		\end{minipage}\\
		$t=2$ & 
		\begin{minipage}{0.4\textwidth}
			\subfloat[{\scriptsize22.19, 0.4347}]{
				\hspace{-5mm}
				\includegraphics[width=0.32\linewidth]{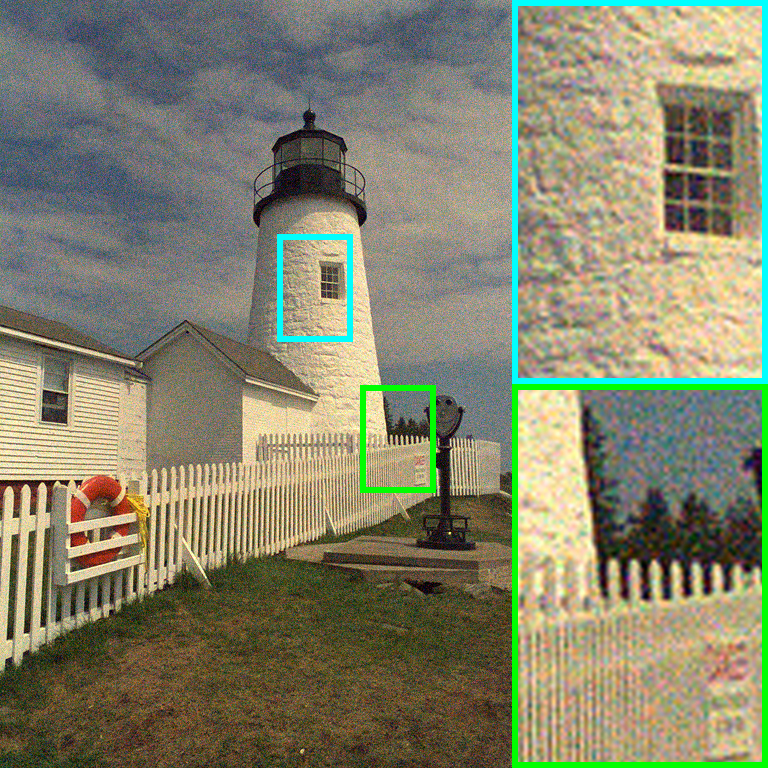} \label{fig_t2_0001}
			}%
			\subfloat[{\scriptsize31.82, 0.8597}]{
				\hspace{-3mm}
				\includegraphics[width=0.32\linewidth]{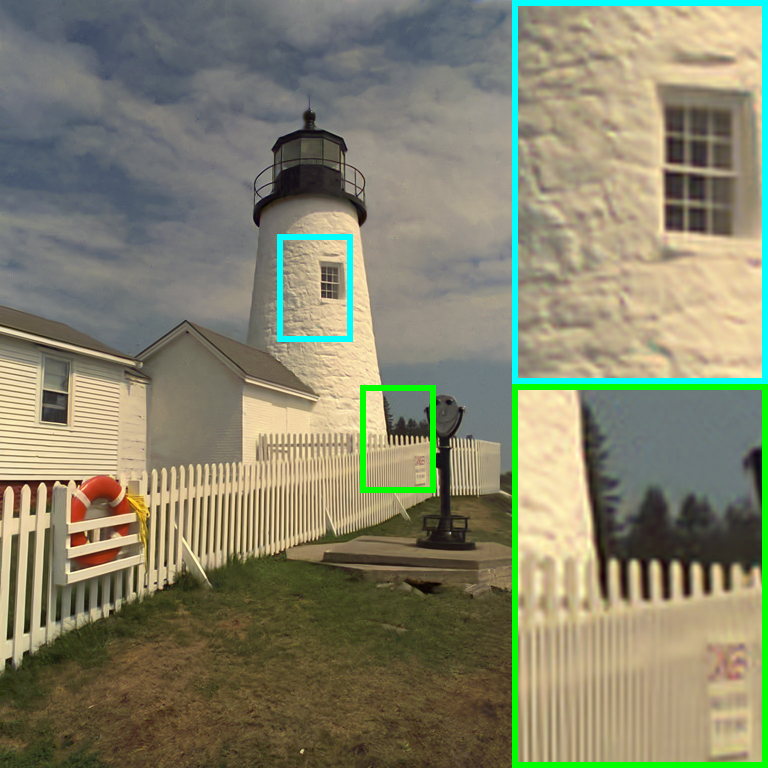}
			}%
			\subfloat[{\scriptsize29.36, 0.7665}]{
				\hspace{-2mm}
				\includegraphics[width=0.32\linewidth]{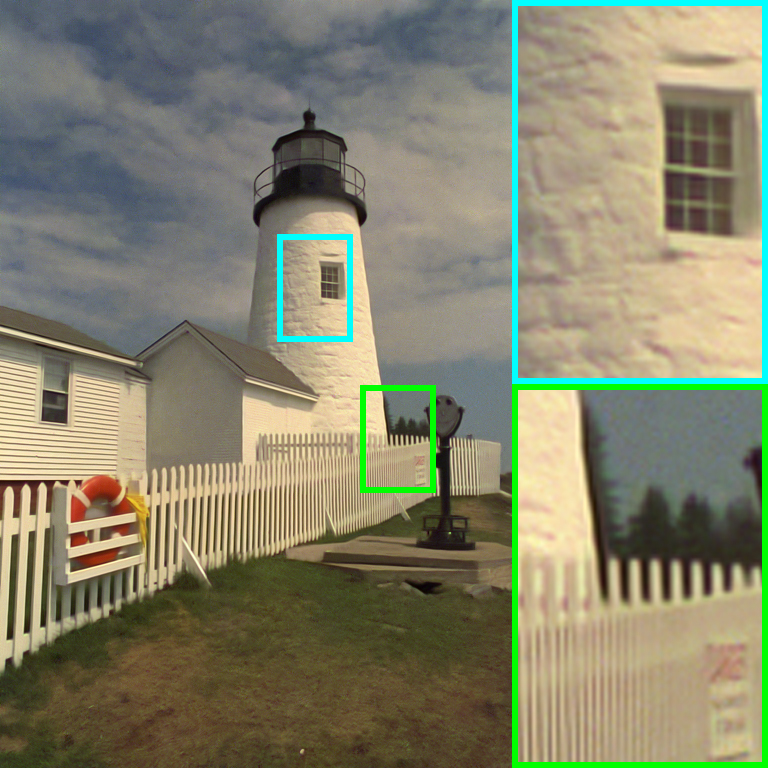} \label{fig_t2_1000}
			}
		\end{minipage}\\
	\end{tabular}
	\captionsetup{format=plain} 
	\caption{Denoised results of ``kodim19'' with different values of $\lambda$ and $t$.}
	\label{fig_lmd_19}
\end{figure}
The regularization parameter $\lambda$ is the most important parameter. 
It controls the regularization effects of the objective function in \eqref{eq_DtNFM}, and further controls the trade-off between noise removal and details preserving. 
Concretely, as it becomes too small, a resultant DtNFM model will suffer from overfitting since it mostly minimize the loss term in \eqref{eq_DtNFM}. 
Consequently, the output image will contain too much noise, as shown in Fig. \ref{fig_lmd_19}(a), Fig. \ref{fig_lmd_19}(d), and Fig. \ref{fig_lmd_19}(g). 
On the contrary, as $\lambda$ becomes too large, a resultant DtNFM model may suffer from underfitting since it minimizes the regularization term too much. 
Consequently, the output image may be over-smoothed, losing too many details, such as the Fig. \ref{fig_lmd_19}(c). 
However, the problem of over-smoothing will be alleviated as $t$ becomes larger, as shown in Fig. \ref{fig_lmd_19}(f) and Fig. \ref{fig_lmd_19}(i). 
That is because the $t$ most dominant rank components of the observed patch matrix $\mathbf{Y}$ are preserved without any condition. 
Therefore, the more information can be preserved. 
\par
\begin{figure}[tb] 
	\captionsetup[subfigure]{captionskip=0pt, farskip=0pt}
	\centering
	\subfloat[]{
		\includegraphics[width=0.45\linewidth]{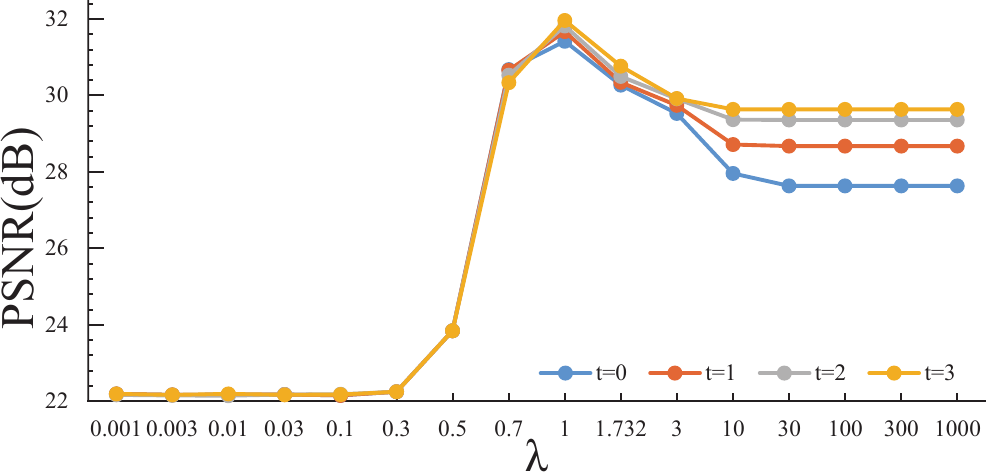}
	}%
	\subfloat[]{
		\includegraphics[width=0.45\linewidth]{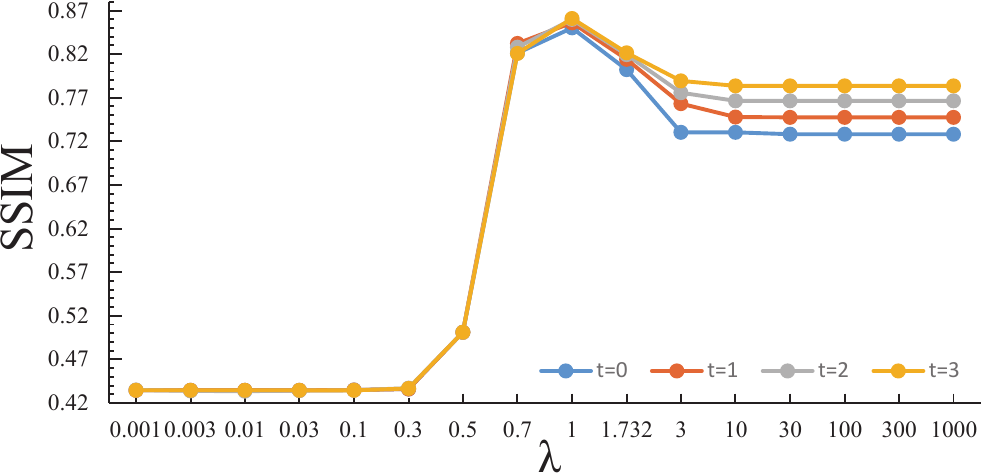}
	}%
	\captionsetup{format=plain} 
	\caption{PSNR and SSIM results of ``kodim19'' on the grid of $\lambda$ and $t$.}
	\label{fig_psnr_ssim_curves}
\end{figure}
Although a larger $t$ can protect the image from over-smoothing, the resultant DtNFM model is still inadequate to be performant. 
As shown in Fig. \ref{fig_psnr_ssim_curves}, suboptimal PSNR and SSIM will be obtained when $\lambda$ is too large for all $t$. 
While the optimal values of $\lambda$ dwell around 1.0. 
Therefore, determining the $\lambda$ needs at least a line search. 
\par
\begin{figure}[t]
	\centering
	\includegraphics[width=0.6\linewidth]{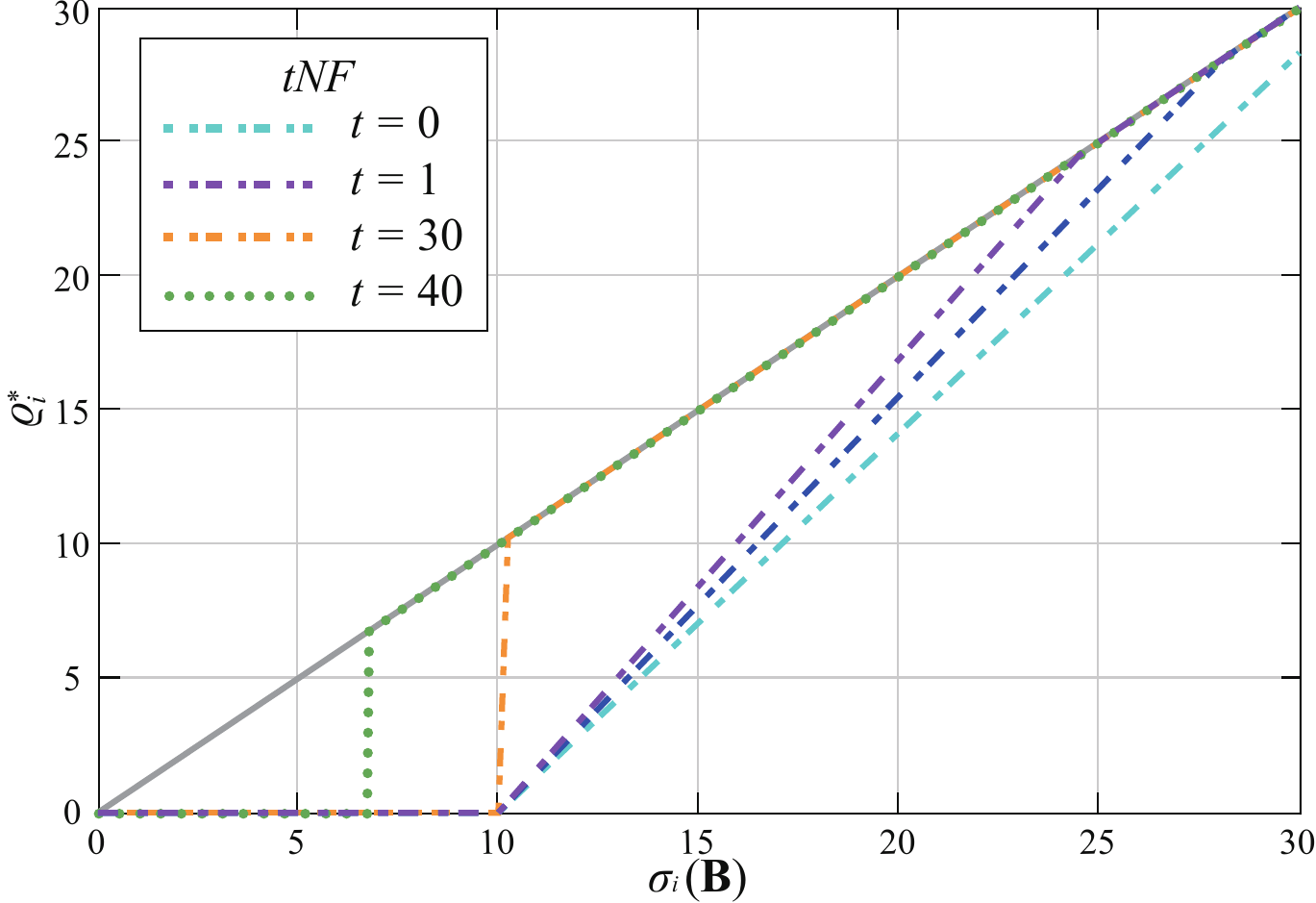}
	\caption{\centering The shrinkage functions of the proximal operator \eqref{eq_tnf_operator} with $\tau=10$.}
	\label{fig_tnf}
\end{figure}
\begin{figure}[t]
	\centering
	\subfloat[]{
		\includegraphics[width=0.45\linewidth]{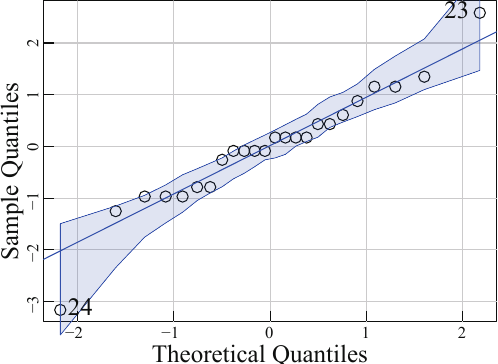}
		\label{fig_QQPlot}
	}%
	\subfloat[]{
		\includegraphics[width=0.45\linewidth]{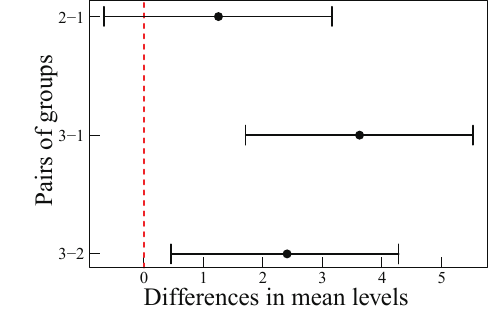}
		\label{fig_MultiComparison}
	}
	\caption{(a) Test the normality via the Q-Q plot. (b) Multiple comparisons between all pairs.}
\end{figure}
\begin{table*}[t] 
	\centering
	\caption{ Grouping results of the 24 test images.}
	\begin{tabularx}{\textwidth}{p{1.1cm}p{1.1cm}|YYYYYYYY|p{.8cm}<{\centering}p{.8cm}<{\centering}}
		\toprule
		\multirow{4}{*}{Group 1} & Image\# &23 &3 &12 &20 &2 &15 &9 &10 && \\
		&$SSIM_0$ &0.2523 &0.2578 &0.2687 &0.2754 &0.2763 &0.2785 &0.2885 &0.2930 & Avg($t$) & Std($t$)\\
		&{Best} $t$ &2 &2 &3 &3 &3 &4 &3 &5  &{3.125} &{0.991} \\
		\midrule
		\multirow{4}{*}{Group 2} & Image\# &4 &17 &16 &7 &22 &11 &21 &19 &&\\
		&$SSIM_0$ &0.2962 &0.3060 &0.3123 &0.3320 &0.3390 &0.3541 &0.3592 &0.3636 & Avg($t$) & Std($t$)\\
		&{Best} $t$ &3 &3 &4 &3 &5 &6 &6 &5  &{4.375} &{1.303}\\
		\midrule
		\multirow{4}{*}{Group 3} &Image\# &6 &18 &24 &14 &1 &5 &13 &8 &&\\
		&$SSIM_0$ &0.3895 &0.3967 &0.3989 &0.4037 &0.4783 &0.4888 &0.5251 &0.5418 & Avg($t$) & Std($t$)\\
		&{Best} $t$ &5 &7 &8 &7 &7 &7 &10 &3  &{6.750} &{2.053}\\
		\bottomrule
	\end{tabularx}
	\label{tab_groups}
\end{table*}
\begin{table}[t] 
	\caption{ The table of ANVOA.}
	\begin{tabularx}{\linewidth}{p{1.55cm} p{.9cm}<{\centering} p{1.3cm}<{\centering} p{0.6cm}<{\centering} p{0.5cm}<{\centering}Y}
		\toprule
		{Source} &\makecell[l]{Sum of\\Squares} &\makecell[l]{Degrees of\\Freedom} &\makecell[l]{Mean\\Square} &$F$ &$p$ \\
		\midrule
		Inter-group &67.13 &2 &33.56 &20.86 &$1.28\times 10^{-5}$ \\
		Within-group &32.18 &20 &1.61 && \\
		Total &99.30 &22 &&& \\
		\bottomrule
	\end{tabularx}
	\label{tab_ANOVA}
\end{table}
The parameter $t \in \mathbb{N}$ also impacts the model performance, since it controls the shrinkage on the singular values. 
Fig. \ref{fig_tnf} shows the shrinkage performed by the tNF regularizer, i.e., the proximal operator \eqref{eq_tnf_operator}. 
We can see that as $t$ becomes too small, the leading singular values may not be preserved well. 
Hence the model may over-smooth the image, as is the case in Fig. \ref{fig_lmd_19}(c). 
On the contrary, as $t$ becomes larger, more singular values would be preserved. 
Hence more noise would be preserved, since the inputted singular values come from the SVD of the corrupted data matrix. 
Therefore, the choice of $t$ should be judicious. 
\par
We give an empirical scheme in which a better $t$ can be determined by analyzing the SSIM of the corrupted image, denoted as $SSIM_0$. 
Given a corrupted image, the higher its $SSIM_0$ is, a larger $t$ is preferred. 
This is verified by the analysis of variance (ANOVA). 
Concretely, we first sort the $SSIM_0$ of 24 corrupted images in an ascending order. 
Based on the order, the 24 corrupted images are broke up into 3 groups, as shown in Table \ref{tab_groups}. 
Then, we search the best $t$ for each image, and list them in Table \ref{tab_groups}. 
As shown in the rightmost column, the average of ``Best $t$'' of Group 1 is the smallest, while that of Group 3 is the largest. 
This intuitive observation implies an connection between the $SSIM_0$ and the ``Best $t$''. 
Now we resort to the ANOVA to prove there exist a significant difference between the averages of ``Best $t$''. 
\par
To perform the ANOVA, the dependent variable $t$ should be normally distributed and have an equal variance in each group. 
The normality assumption can be assessed via the Q-Q plot, shown in Fig. \ref{fig_QQPlot}. 
As can be seen, the normality assumption is satisfied since all of the points fall within the 95\% confidence envelope, and the slope of the main diagonal is closed to 1.0. 
The equality of variances can be checked via the Barlett's test. 
And the result ($p=0.165$) suggests that the variances in three groups do not differ significantly. 
Hence the ANOVA can be carried out. 
\par
The results of ANOVA are shown in Table \ref{tab_ANOVA}. 
As $p < 0.05$, the average of ``best $t$'' does have a significant difference among three groups. 
Although the multiple comparisons in Fig. \ref{fig_MultiComparison} demonstrate that the averages of ``best $t$'' in group 1 and 2 are not significantly different, it is still easy to find the positive correlation between $SSIM_0$ and ``best $t$''. 
\section{Conclusion}
\label{sec_conclusion}
In this paper, the DtNFM model was proposed and applied to color image denoising via integrating with the NSS prior. 
The DtNFM model possesses two advantages. 
On the one hand, it can fully model and utilize the cross-channel difference and the spatial variation of noise. 
On the other hand, it can provide flexible treatments for different rank components, and further give a close approximation to the underlying low-rank matrix. 
To solve the resultant optimization problem, an accurate and effective algorithm was proposed by exploiting the framework of ADMM. 
Importantly, we mathematically proved the global optima of all subproblems can be obtained in closed-form. 
The convergence guarantee was established. 
Extensive experiments are carried out on the synthetic noise, spatially variant noise, and real-world noise images, respectively. 
The results demonstrated that the proposed method outperforms many state-of-the-art color image denoising methods. 
\par
{{In the future, we will try to extend our work to low-rank tensor approximation. 
If so, we would no longer have to stretch the similar patches to vectors. 
Hence the structure of patches would be preserved better. 
Moreover, we will try to devise heuristic schemes to search and update the $t$ and $\alpha$ adaptively. 
If so, the practicality of the proposed DtNFM model would be further improved. }}
\bibliographystyle{IEEEtran}
\bibliography{References}
%







\vfill

\end{document}